\documentclass[twocolumn]{aastex701}
\usepackage{graphicx}	
\usepackage{amsmath}	
\usepackage[table]{xcolor}
\usepackage{booktabs}
\usepackage{array}

\newcommand{\kms}{$\rm\ km\ s^{-1}$}
\newcommand{\vvir}{$V_{\rm vir}$}
\newcommand{\zzero}{$z=0$}
\newcommand{\zsix}{$z=6$}
 
\newcommand{\rvir}{$r_{\rm vir}$}

\newcommand{\mvir}{$M_{\rm vir}$}
\newcommand{\mhalo}{$M_{\rm halo}$}
\newcommand{\mhot}{$M_{\rm hot}$}

\newcommand{\fin}{$f^{\rm in}_{\rm CGM}$}
\newcommand{\sfe}{$\tau_{*}$}
\newcommand{\mstar}{$M_{\rm star}$}
\newcommand{\mcold}{$M_{\rm cold}$}
\newcommand{\msun}{$M_\odot$}
\newcommand{\rcool}{$r_{\rm cool}$}
\newcommand{\tcool}{$t_{\rm cool}$}
\newcommand{\zhot}{$Z_{\rm hot}$}
\newcommand{\zcold}{$Z_{\rm cold}$}
\newcommand{\zstar}{$Z_{\rm star}$}
\newcommand{\etahalo}{$\eta_{\rm CGM}$}
\newcommand{\etaism}{$\eta_{\rm ISM}$}
\newcommand{\zetaism}{$\zeta^{\rm in/out}_{\rm ISM}$}
\newcommand{\zetahalo}{$\zeta^{\rm in/out}_{\rm CGM}$}
\newcommand{\fire}{\textsc{FIRE}}

\newcommand{\mdotinism}{$\dot{M}^{\rm in}_{\rm ISM}$}

\newcommand{\zmdotinism}{$\dot{M}^{\rm in}_{\rm Z,ISM}$}
\newcommand{\zmdotoutism}{$\dot{M}^{\rm out}_{\rm Z,ISM}$}
\newcommand{\mdotincgm}{$\dot{M}^{\rm in}_{\rm CGM}$}

\newcommand{\zmdotincgm}{$\dot{M}^{\rm in}_{\rm Z,CGM}$}
\newcommand{\zmdotoutcgm}{$\dot{M}^{\rm out}_{\rm Z,CGM}$}

\definecolor{tablecolor}{HTML}{DBDBDB}
\defcitealias{oren_cosmic_2025}{YO25}
\defcitealias{henriques_galaxy_2015}{H15}
\defcitealias{mitchell_how_2022}{MS22}
\defcitealias{neistein_hydrodynamical_2012}{N12}

\begin{document}

\title{TNG SAM: Bridging Hydrodynamical Complexity and Semi-Analytic Efficiency to Model Galaxy Formation}

\author[orcid=0000-0002-3649-5362,sname='Omoruyi']{Osase Omoruyi}
\affiliation{Center for Computational Astrophysics, Flatiron Institute, 162 5th Avenue, New York, NY 10010, USA}
\affiliation{Department of Astrophysical Sciences, Peyton Hall, Princeton University, Princeton, NJ 08544, USA}
\affiliation{Center for Astrophysics $|$ Harvard \& Smithsonian, 60 Garden St., Cambridge, MA 02138, USA}
\email[show]{osase.omoruyi@gmail.com}  

\author[orcid=0000-0001-5529-7305,sname='Terrazas']{Bryan Terrazas}
\affiliation{Department of Physics \& Astronomy, Oberlin College, Oberlin, OH 44074, USA}
\email{bterraza@oberlin.edu}

\author[orcid=0009-0001-3658-6950,sname=Oren]{Yossi Oren}
\affiliation{School of Physics and Astronomy, Tel Aviv University, Ramat Aviv 69978, Israel}
\email{cohenyossi1@gmail.com}

\author[orcid=0000-0003-4295-3793,sname=Gabrielpillai]{Austen Gabrielpillai}
\affiliation{Columbia University, Department of Astronomy, New York, NY 10027, USA}
\affiliation{The City University of New York, 365 5th Ave, New York, NY 10016, USA}
\email{a.gabrielpillai@gmail.com}

\author[orcid=0000-0002-2499-9205,sname=Pandya]{Viraj Pandya}
\affiliation{Columbia University, Department of Astronomy, New York, NY 10027, USA}
\email{vgp2108@columbia.edu}

\author[orcid=0000-0002-6748-6821,sname=Somerville]{Rachel S. Somerville}
\affiliation{Center for Computational Astrophysics, Flatiron Institute, 162 5th Avenue, New York, NY 10010, USA}
\email{rsomerville@flatironinstitute.org}

\author[orcid=0000-0001-5065-9530,sname=Sternberg]{Amiel Sternberg}
\affiliation{School of Physics and Astronomy, Tel Aviv University, Ramat Aviv 69978, Israel}
\affiliation{Center for Computational Astrophysics, Flatiron Institute, 162 5th Avenue, New York, NY 10010, USA}
\affiliation{Max-Planck-Institut f\"ur extraterrestrische Physik (MPE), Giessenbachstr., 85748 Garching, Germany} 
\email{asternberg@flatironinstitute.org}

\author[orcid=0000-0001-6950-1629,sname=Hernquist]{Lars Hernquist}
\affiliation{Center for Astrophysics $|$ Harvard \& Smithsonian, 60 Garden St., Cambridge, MA 02138, USA}
\email{lhernquist@cfa.harvard.edu}

\begin{abstract}

All cosmological models of galaxy formation must navigate the trade-off between physical accuracy and computational efficiency. Hydrodynamical simulations provide spatially resolved predictions for the co-evolution of dark matter, gas, stars, and black holes, but rely on phenomenological subgrid models for small-scale processes (e.g., star formation).  Semi-analytic models (SAMs), by contrast, gain efficiency through simplified, analytic treatments of the same processes, at the cost of reduced predictive scope. In this work, we leverage the strengths of the Santa Cruz SAM and the IllustrisTNG hydrodynamical simulation to develop the TNG SAM. Calibrated to reproduce baryon cycling in stellar feedback–dominated TNG galaxies ($\sim 10^{10}M_\odot < M_{200} < 10^{12}M_\odot$), the TNG SAM introduces several key updates to the Santa Cruz framework regarding: 1) halo gas (re-)accretion efficiency, 2) a cooling model that moves beyond the traditional cold–hot mode dichotomy, 3) explicit treatment of both galactic- and halo-scale outflows, 4) star formation efficiency, and 5) the circulation of metals between galaxies and their surroundings. These changes enable the TNG SAM to reproduce TNG's flow of gas and metals from the scale of the galaxy to the halo, as well as global galaxy (e.g., stellar mass) and halo (e.g., hot halo gas mass) properties within $\lesssim 30\%$ accuracy out to $z=6$. This work demonstrates that, with appropriate calibration, SAMs can capture the complex physics of galaxy formation modeled in hydrodynamical simulations while providing a flexible framework for studying galaxy evolution across the large cosmological volumes targeted by future observational surveys.

\end{abstract}
\keywords{\uat{Galaxies}{573}}

\section{Introduction}
\label{sec:intro}

\hspace{1em} Observations across the electromagnetic spectrum have revealed a Universe rich in dark matter, stars, gas, and dust, sparking numerous questions regarding their formation and evolution. The $\Lambda$ Cold Dark Matter ($\Lambda$CDM) model of cosmological structure formation proposes that the Universe consists of a complex filamentary network molded by gravity and populated with dark matter halos. Within these halos, baryonic matter falls to the center of the gravitational potential well, condenses, and forms stars \citep{white_core_1978}. To study this process from cosmological to stellar scales, numerous studies have used numerical simulations to predict and correlate observable galaxy properties-- such as stellar mass, cold gas mass, and rate of star formation-- with dark matter halo characteristics (see \cite{somerville_physical_2015,naab_theoretical_2017, crain_hydrodynamical_2023, feldmann_cosmological_2025} for reviews). With the advent of powerful observatories like \textit{JWST}  and the upcoming \textit{Nancy Grace Roman Space Telescope}, our ability to observe galaxies in the earliest days of the Universe and across larger cosmic volumes is rapidly improving, necessitating the development of more sophisticated and flexible cosmological models of galaxy formation.

Modeling how galaxies form and evolve requires investigating how they interact with their surrounding environment. As dark matter halos grow, galaxies accrete gas from the intergalactic medium (IGM) into the circumgalactic medium (CGM), connecting the large-scale cosmic web with the galaxies at the centers of these halos. Gravity continues to drive flows of gas from the CGM into the interstellar medium (ISM), fueling star formation. As stars form, evolve, and eventually die, the resulting feedback from stellar winds, supernovae (SNe), and black hole activity heats and/or expels material back into the CGM or IGM. This matter can later be re-accreted, fueling new star formation. This cycle of gas accretion, star formation, feedback, and re-accretion—known as the ``baryon cycle”— is fundamental to shaping how galaxies evolve.

To model the baryon cycle within the $\Lambda$CDM paradigm, two primary techniques have emerged: numerical hydrodynamical simulations and semi-analytic models (SAMs). Numerical hydrodynamical simulations explicitly solve partial differential equations of gravity, (magneto)hydrodynamics, and thermodynamics. This comprehensive approach enables them to self-consistently model the coupled evolution of dark matter, gas, stars, and black holes, providing detailed information on each component's spatial distribution, temperature, and kinematics. However, simulating large cosmological volumes requires compromises in capturing the detailed physics of key processes occurring below their resolution limits, such as star formation and black hole feedback. As a result, these simulations rely on phenomenological ``subgrid recipes” that contain adjustable parameters calibrated to reproduce a given set of observations. Additionally, the computational expense of hydrodynamical simulations also limits their ability to explore variations in subgrid models across large volumes.

In contrast, SAMs offer a computationally efficient alternative by treating galaxies and halos as unresolved entities and relying solely on ``subgrid” recipes to simulate their formation and evolution. Built on merger trees extracted from dark matter-only N-body simulations or constructed using semi-analytic approaches based on the extended Press-Schechter formalism \citep{bond_excursion_1991, bower_evolution_1991, lacey_merger_1993, somerville_how_1999} SAMs create a physics-based link between the hierarchical growth of dark matter halos and observed galaxy populations by employing simplified analytical models of key physical processes relevant to galaxy evolution \citep{white_galaxy_1991, kauffmann_formation_1993, somerville_semi-analytic_1999}. In a SAM, a galaxy is typically represented as a collection of mass reservoirs, such as the stellar disk, stellar bulge, cold gas disk, hot halo gas, and ejected material. As a result, SAMs essentially function as ``flow models,” solving ordinary differential equations to track the flow of baryons between each component, with the flows governed by the physical prescriptions included in the model. The specific processes included in a given SAM—and the level of detail used to model them—lie at the discretion of the model’s creator. This flexibility has led to a range of implementations, from comprehensive frameworks like the Santa Cruz SAM \citep{somerville_semi-analytic_1999,somerville_explanation_2008}, the Munich model \textsc{L-Galaxies} \citep{henriques_galaxy_2015, henriques_l-galaxies_2020}, SAGE \citep{croton_many_2006}, DARK SAGE \citep{stevens_dark_2024}, GALFORM \citep{cole_recipe_1994,cole_hierarchical_2000,granato_infrared_2000,benson_what_2003,bower_breaking_2006,gonzalez-perez_how_2014,lagos_dynamical_2013, lacey_unified_2016}, and SHARK \citep{lagos_shark_2018}, to modular platforms that support user-defined prescriptions and calibration strategies \citep[e.g.,][]{benson_galacticus_2012}. It also allows SAMs to be tailored to specific scientific questions and computational resources, making them a valuable tool for understanding the ``big picture’’ of the complexities shaping galaxy evolution.

Despite their different approaches, both SAMs and large-volume hydrodynamical simulations rely on phenomenological prescriptions to model fundamental processes in the baryon cycle. This has led to the practice of ``tuning” or calibrating these models to reproduce a selection of observations, introducing additional uncertainty when comparing and interpreting results. Early comparisons of SAMs and hydrodynamical simulations \citep[e.g.,][]{benson_comparison_2001, helly_comparison_2003, yoshida_gas_2002, saro_gas_2010, stringer_analytic_2010, hirschmann_galaxy_2012} found general agreement in predicting properties such as the overall distribution of galaxies and their cold gas mass fractions, but differences in quantities like the efficiency of gas cooling, star formation histories, and merger histories. More detailed comparisons \citep[e.g.,][]{stringer_analytic_2010, monaco_semi-analytic_2014, guo_galaxies_2016, mitchell_comparing_2018, cote_validating_2018} similarly found that SAMs can approximate key galaxy properties such as stellar mass functions and specific star formation rates to a reasonable degree of accuracy when compared to hydrodynamical simulations. However, they also highlighted significant differences in gas cooling, star formation efficiencies, galaxy sizes, and the evolution of the star formation rate density. These results suggest that discrepancies between the two methods may arise from specific parameterizations and models used, rather than inherent limitations of each approach. 

Recognizing SAMs and hydrodynamical simulations as complementary, rather than competing, techniques, can significantly enhance their predictive power. For example, the detailed physical insights from hydrodynamical simulations can help to refine the subgrid models in SAMs, as demonstrated by \cite{pandya_first_2020}, who compared the FIRE-2 hydrodynamical simulations \citep{hopkins_fire-2_2018} with the Santa Cruz SAM (SC SAM) for halos with masses between $10^{10}$ and $10^{12}$ M$_\odot$. They found substantial agreement between FIRE-2 and the SC SAM for stellar and cold gas masses, but significant disagreements in the properties of the hot halo gas — an observable not widely available for Milky Way and lower mass halos to calibrate or validate SAMs. By adjusting the SAM to account for the suppression of halo gas accretion via stellar winds, particularly in dwarf galaxies, \cite{pandya_first_2020} reproduced the reduced halo gas accretion efficiencies of the FIRE-2 complementary galaxies remarkably well. 

Similarly, \cite{cote_validating_2018} refined \textsc{GAMMA}, a SAM aimed at understanding the chemical evolution of low-mass galaxies and the origins of metal-poor stars, by calibrating it against the observable properties of the most massive galaxy in the \cite{wise_birth_2011} high-redshift hydrodynamical simulation. They found that implementing a non-uniform mixing model was key to reproducing the observable quantities in the hydrodynamical simulation. Lastly, in a more extensive effort to incorporate insights from hydrodynamical simulations into semi-analytic models, \cite{pandya_unified_2023} used FIRE-2's predictions for mass and energy flows in a given set of halos to calibrate their {\sc Sapphire} SAM, which emphasizes the role of the CGM and its energy content in governing galaxy evolution. These calibrations allowed their model to successfully emulate the detailed baryon cycle found in FIRE-2 halos.

Of course, SAMs can also enhance the predictive power of hydrodynamical simulations. For instance, the IllustrisTNG model for galactic-scale, star-formation-driven kinetic winds incorporates several refinements over its predecessor in Illustris, as described in detail in Section \ref{subsec:tng}. Among these, the TNG model introduced a scaling of wind velocity with the local dark matter velocity dispersion $\sigma_\text{DM}$ and a redshift-dependent factor, inspired by the \cite{henriques_simulations_2013} SAM. This change enabled TNG to better match observed stellar mass and luminosity functions across redshift, demonstrating the bidirectional synergy between SAMs and hydrodynamical simulations.

In this work, we focus on the former direction--using hydrodynamical simulations to inform semi-analytic models--by comparing the predictions of IllustrisTNG (hereafter TNG) with the Santa Cruz SAM, both of which have demonstrated success in reproducing a wide range of observations. Previous work has examined differences between these two models at $z=0$ \citep{gabrielpillai_galaxy_2022}, but has largely focused on comparing their predictions. Here, we instead ask a different question: can the galaxy population produced by a state-of-the-art hydrodynamical simulation be reproduced within a semi-analytic framework? To address this, we treat TNG as a self-consistent reference model and calibrate the SAM to emulate its baryon cycling behavior, with a specific focus on replicating TNG's treatment of stellar feedback. While TNG broadly reproduces many observed galaxy properties, it also exhibits known tensions, such as in the mass-metallicity relation, where the slope at $z=0$ is shallower than observations \citep{torrey_evolution_2019}. As a result, the calibration implemented here is designed to improve agreement with the galaxy population modeled in TNG rather than with observations, although the latter may occur as a byproduct.

To avoid the added complexity and modeling uncertainties associated with AGN feedback, we limit our analysis to lower-mass halos, where stellar-driven processes dominate the baryon cycle. By calibrating the SAM with measurements of galaxy- and halo-scale inflow and outflow rates for a subset of TNG galaxies extracted by \cite{oren_cosmic_2025}, we aim to emulate the underlying baryon cycle modeled in TNG, resulting in the creation of a new “TNG SAM.” This endeavor serves four primary goals: (1) to simplify and distill TNG’s complex feedback and baryon cycling behaviors into a more interpretable semi-analytic framework, (2) to facilitate direct comparisons between SAMs and hydrodynamical simulations, clarifying their respective strengths and limitations, (3) to develop a framework that enables efficient exploration of alternative subgrid physics models that are otherwise too computationally expensive to vary directly within hydrodynamical simulations, and (4) to ultimately extend the predictive power of SAMs to volumes beyond the reach of hydrodynamical simulations.

\renewcommand{\arraystretch}{0.85}
\begin{deluxetable*}{lccc}
\tabletypesize{\scriptsize}
\tablecaption{Approaches to modeling galaxy formation in TNG100, the Santa Cruz SAM, and the TNG SAM.}
\label{tab:tng_sam_comparison}
\tablewidth{0pt}
\tablehead{
\colhead{} &
\colhead{TNG100} &
\colhead{SC SAM} &
\colhead{TNG SAM} 
}
\startdata
\hline
& & & \\
\textit{Box Size} &
$75\,h^{-1}$ Mpc &
$75\,h^{-1}$ Mpc &
$75\,h^{-1}$ Mpc \\
& & & \\
\hline
& & & \\
\textit{Mass Resolution} &
$m_{\rm DM} = 5.06 \times 10^6\,M_\odot/h$ &
mass$_{\rm root\ min} = 8.85 \times 10^8\,M_\odot$ &
mass$_{\rm root\ min} = 8.85 \times 10^8\,M_\odot$ \\
& & & \\
\hline
& & & \\
\textit{Halo Gas Accretion} &
\nodata &
Static halo cooling &
$f_{\rm in}^{\rm CGM} \propto f(M_{\rm halo}, z)$ \\
& & \citep{white_galaxy_1991} & \\
& & & \\
\hline
& & & \\
\textit{Hot Halo Gas Re-incorporation} &
\nodata &
$\chi_{\rm re\text{-}infall} = 0.1$ &
Bundled into $f_{\rm in}^{\rm CGM}$ \\
& & & \\
\hline
& & & \\
\textit{Hot Halo Gas Cooling} &
\nodata &
$r_{\rm cool} < r_{\rm vir}$: cold mode &
$t_{\rm cool} \propto f(M_{\rm halo}, z)$ \\
 & &  $r_{\rm cool} > r_{\rm vir}$: hot mode  & \\
& & & \\
\hline
& & & \\
\textit{Star Formation} &
Kennicutt--Schmidt;  &
H$_2$-based SF; gas partitioning &
$\tau_* \propto f(M_{\rm halo}, z)$, \\
 &
density threshold; &
\citep{gnedin_environmental_2011};  &
$\dot m_* = M_{\rm cold}/\tau_*(M_{\rm halo}, z)$ \\
 & $\Sigma_{\rm SFR} \propto \rho^{1.5}$; & $\tau_{*,0}=1.0$ Gyr&  \\
  &  $\tau_{*,0}=2.1$ Gyr & &  \\
  & & & \\
\hline
& & & \\
\textit{Chemical Enrichment} &
Delayed recycling;&
Instantaneous recycling;  &
Instantaneous recycling; \\
  & AGB, SNII, SNIa yields;& $y=1.2$;& $y=2.0$;\\
    & tabulated enrichment& $f_{\rm recycle}=0.43$& $f_{\rm recycle}=0.43$\\
& & & \\
\hline
& & & \\
\textit{Stellar Feedback} &
Outflows from star-forming particles; &
Galaxy-scale outflows: &
Galaxy-scale outflows: \\
& Injection velocity $v_w \propto [350$ km s$^{-1}$, & SNe ejection efficiency $\epsilon_{\rm SN}=1.7$,& $\eta_{\rm ISM} \propto f(v_w,e_w,M_{\rm halo},z)$;\\
& $f(\sigma_{\rm DM},H(z))]; $ &  $V_{\rm eject}=110$ km/s, &  Halo-scale outflows \\
& Wind energy $e_w \propto f(Z, E_{\rm SNII})$; & mass deposited in ejected reservoir& \citep{henriques_galaxy_2015}:\\
& Wind mass loading $\eta_w \propto f(v_w, e_w)$ & & $\eta_{\rm CGM} \propto f(\eta_{\rm ISM},M_{\rm halo},z)$ \\
& & & \\
\hline
& & & \\
\textit{Metal Circulation} &
Wind metal loading $\gamma_w=0.4$ &
$\frac{\dot{M}_{\rm Z}^{\rm in/out}}
{\dot{M}_{\rm gas}^{\rm in/out} \cdot Z_{\rm gas}} = 1$ & $ \frac{\dot{M}_{\rm Z}^{\rm in/out}}
{\dot{M}_{\rm gas}^{\rm in/out} \cdot Z_{\rm gas}} \propto f(M_{\rm halo},z)$ \\
& & & \\
\hline
& & & \\
\textit{Calibrations} &
cosmic SFR density over redshift;& stellar mass vs. halo mass; & TNG100: \fin, \tcool, \sfe, \\
& galaxy stellar mass vs radius; &  stellar mass function; & \etaism, \etahalo, \zetaism, \\
& stellar mass vs. halo mass; & stellar mass vs. metallicity & \zetahalo \\
& BH mass versus stellar bulge mass; & cold gas fraction vs stellar mass& \\
& hot gas fraction in galaxy clusters & for disc-dominated galaxies; & \\
& & BH mass versus stellar bulge mass& \\
\hline
& & & \\
\textit{References} &
\citet{pillepich_simulating_2018}  &
\cite{ somerville_semi-analytic_2008, somerville_star_2015} &
This work \\
& \citet{springel_cosmological_2003}& & \\
& & & \\
\enddata
\end{deluxetable*}
\renewcommand{\arraystretch}{1.0}

This paper is organized as follows: In Section \ref{sec:models}, we provide an overview of the IllustrisTNG hydrodynamical simulations and the Santa Cruz SAM, while Section \ref{sec:comp_tng_sam} describes the methods used to compare their respective outputs. Section \ref{sec:creating_tng_sam} outlines the modifications made to the SC SAM’s galaxy formation framework that underpin the development of the new “TNG SAM.” The results of these modifications, namely the performance of the TNG SAM in reproducing key global and flow properties of the galaxies modeled in TNG, are presented in Section \ref{sec:tng_sam_results}. In Section \ref{sec:discussion}, we analyze how each modification contributes to the TNG SAM’s success, and also discuss the model's current limitations and potential areas for further refinement. Finally, Section \ref{sec:summary} provides a concise summary of our results and their implications for developing the next generation of SAMs.

\section{Model Descriptions}
\label{sec:models}

\hspace{1em} In this section, we describe the two models used: the IllustrisTNG hydrodynamical simulations and the Santa-Cruz semi-analytic model. We summarize their key parameters in Table \ref{tab:tng_sam_comparison} and outline the methods used to compare the outputs of the Santa Cruz SAM with those of TNG.

\subsection{The IllustrisTNG Simulations}
\label{subsec:tng}

\hspace{1em} The IllustrisTNG simulations are a suite of graveto-magnetohydrodynamical simulations that model the physical processes governing the formation and evolution of galaxies across different cosmological volumes \citep{springel_first_2018, weinberger_simulating_2017, pillepich_first_2018, nelson_first_2018}. The suite includes nine simulations, with cubic volumes of roughly 50, 100, and 300 Mpc side length, each run at three different resolutions (1 being the highest and 3 the lowest; \citealt{marinacci_first_2018, naiman_first_2018, springel_first_2018, nelson_first_2019, pillepich_first_2019}). Each simulation has a companion dark matter-only (DMO) and full-physics (FP) run. The cosmological parameters used in each simulation are taken from \cite{planck_collaboration_planck_2016}, with matter density $\Omega_{M,0} = 0.3089$, baryon density $\Omega_{b,0} = 0.0486$, dark energy density $\Omega_{\Lambda,0} = 0.6911$, Hubble constant $H_0 = 67.74$ km s$^{-1}$ Mpc$^{-1}$, power spectrum normalization factor $\sigma_8= 0.8159$, and spectral index $n_s = 0.9667$.

TNG builds upon the original Illustris model \citep{vogelsberger_model_2013, torrey_model_2014}, which uses the AREPO code \citep{springel_e_2010} to solve coupled equations of self-gravity and magnetohydrodynamics \citep{pakmor_magnetohydrodynamics_2011, pakmor_simulations_2013}. To achieve improved consistency with observations, TNG refined the original Illustris models for active galactic nuclei (AGN) feedback, galactic winds, and magnetic fields \citep{weinberger_simulating_2017, pillepich_simulating_2018}. 

Given that this paper focuses on low mass halos $M_{\text{halo}} < 10^{12} M_\odot$, TNG's treatment of star formation and stellar feedback, particularly the treatment of supernova-driven galactic winds \citep{pillepich_simulating_2018}, is most relevant for our study. In both the TNG and Illustris models, stars form in dense gas regions that meet a threshold for star formation, following a subgrid implementation of the Kennicutt-Schmidt law \citep{kennicutt_global_1998}. Specifically, in the TNG model, gas above a critical density enters a two-phase ISM regime, where cold, star-forming clouds coexist with a hot, ionized medium. In this regime, the local star formation rate density scales as $\dot{\rho}_* \propto \rho^{1.5}$, where $\rho$ is the cold gas volume density \citep{springel_cosmological_2003, diemer_modeling_2018}. 

Building on the original Illustris simulation \citep{vogelsberger_model_2013}, TNG models supernova-driven galactic winds that are launched isotropically from star-forming gas cells in the dense ISM. The winds are modeled as collisionless particles, with the kinetic energy release rate determined by a mass-loading function that depends on the gas-phase metallicity and the dark matter velocity dispersion $\sigma_{DM}$. This velocity dispersion is calculated using a weighted kernel over the nearest 64 DM particles, and the wind velocity is further modulated by $\sigma_{DM}$ and redshift. These outflows are powerful enough to eject mass from the ISM \citep{nelson_first_2019} and, as we discuss in Section \ref{sec:creating_tng_sam}, can push gas entirely out of the halo.

Several physical processes in TNG—such as star formation, stellar and AGN feedback, black hole seeding, and accretion—are implemented using sub-grid recipes calibrated to align with various observations, similar to the methodology used in SAMs. However, the specific observations and the precision of calibration differ. For TNG, calibration targets include the cosmic star formation rate density as a function of redshift, stellar mass functions, the relationship between black hole mass and stellar mass, the hot gas fraction in galaxy clusters, and the galaxy stellar mass versus radius relation \citep{pillepich_simulating_2018}. Additionally, the distribution of galaxy optical colors in stellar mass bins was used to refine AGN feedback parameters, as detailed in \cite{nelson_first_2018}. While these calibration choices overlap with those used for the Santa Cruz SAM, described in the section below, there are differences in the specific observations and calibration techniques used. 

In this paper, we use both the hydrodynamical (full-physics) and dark matter only (N-body) outputs of the intermediate-sized box ($L_{\rm box} = 75$ Mpc h$^{-1}$) at its highest resolution, TNG100-1. We adopt TNG100 rather than TNG50 ($L_{\rm box} = 35$ Mpc h$^{-1}$) or TNG300 ($L_{\rm box} = 205$ Mpc h$^{-1}$) as a compromise between resolution and volume. While TNG50 offers higher resolution, its smaller volume limits statistical sampling, whereas TNG300 provides larger volumes at the expense of resolving the baryon cycling processes central to this work. Additionally, the fiducial TNG galaxy formation model was calibrated primarily on TNG100, making it the most internally self-consistent choice for our emulation-based approach. We defer a systematic exploration of resolution and volume effects across the TNG suite to future work (see Section \ref{subsubsec:future}).

\subsection{The Santa Cruz semi-analytic model}
\label{subsec:scsam}

\hspace{1em} The Santa Cruz SAM for galaxy formation, first introduced in \cite{somerville_semi-analytic_1999}, and subsequently updated in \cite{somerville_semi-analytic_2008, somerville_galaxy_2012}, \cite{porter_understanding_2014}, \cite{popping_evolution_2014}, and \cite{somerville_star_2015}, traces the evolution of baryons within galaxies by partitioning them into four distinct reservoirs: the cold gas in the galaxy disk, the hot gas associated with the dark matter halo, the ejected gas reservoir, and the stellar component, the latter of which is further divided into disk and bulge populations. To model the mass and energy transfer between these reservoirs, the Santa Cruz SAM incorporates physically and observationally motivated prescriptions for a variety of baryonic processes. These include gas cooling, star formation, stellar feedback, gas recycling, chemical enrichment, and the growth of supermassive black holes and their associated feedback. 

In this work, we use the latest version of the Santa Cruz SAM published in \cite{somerville_star_2015} and recently used in  \cite{gabrielpillai_galaxy_2022} and \cite{yung_semi-analytic_2023}. Since we limit our focus to central galaxies with $M_{\rm halo} < 10^{12}$\msun, we turn all AGN feedback ``off.” Below, we briefly summarize the key processes modeled in the SAM, and refer the reader to \cite{somerville_semi-analytic_2008} and \cite{somerville_star_2015} for details. \\

\subsubsection{Gas Cooling and Accretion } 
\label{subsubsec:scsam_cooling}

\hspace{1em} For any given halo, the SAM estimates the dark matter accretion rate, $\dot{m}_{\rm DM}$, for each halo by using the virial mass provided by the halo merger tree and calculating its rate of change between consecutive snapshots. Prior to reionization, the SAM assumes that gas accretion $\dot{m}_{\rm acc}$ closely follows the accretion of dark matter, scaled by the universal baryon fraction $f_b$ as:
\begin{equation}
\dot{m}_{\rm acc} = f_{\rm b} \dot{m}_{\rm DM},
\end{equation}
where $f_{\rm b} = 0.1573$ \citep{planck_collaboration_planck_2016}. 

Following reionization, the photoionizing background suppresses gas accretion into halos, with the fraction of baryons able to collapse determined by $f_{\rm coll}$, a function of halo mass and redshift \citep{okamoto_mass_2008}. The SAM also accounts for the recycling of gas previously expelled by stellar feedback, described in detail in Section \ref{subsubsec:scsam_sf}. The rate of gas re-accretion is given by:
\begin{equation}
\dot{M}_{\rm CGM, in, recycled} = \chi_{\rm re-infall} \Big(\frac{M_{\rm ejected}}{t_{\rm dyn}} \Big),
\end{equation}
\noindent where $M_{\rm ejected}$ represents the total mass of gas in the ejected reservoir, and $\chi_{\rm re-infall}$ specifies the fraction of this gas that can return to the hot halo at each time step. Typically, in the previously published SC SAMs, ${\chi}_{\rm re-infall} = 0.1$, meaning that the ejected gas is recycled over $\sim 10$ dynamical times. The dynamical time is defined as $t_{\rm dyn} \equiv R_{\rm vir}/V_{\rm vir}$, where $V_{\rm vir}$ is the circular velocity of the halo at the virial radius. 

When haloes merge, the SAM assigns the galaxy with the most massive progenitor as the central galaxy, while galaxies associated with less massive progenitors become satellites.  Satellite galaxies orbit within the host halo and may eventually merge with the central galaxy due to orbital decay driven by dynamical friction, with merger timescales estimated following \citet{boylan-kolchin_resolving_2009}. The stripping and potential disruption of satellites are treated using the prescriptions described in \citet{somerville_semi-analytic_2008}. Once a halo becomes a subhalo, its circumgalactic gas is assumed to be transferred to the hot gas reservoir of the host halo.

Following the framework of \cite{white_galaxy_1991}, the SAM assumes that the CGM is uniformly at the virial temperature of the halo at each time step. The radiative cooling time, which determines how quickly the gas loses thermal energy via radiation, is calculated using the cooling function $\Lambda(T_{\rm vir}, Z_{\rm h})$ \citep{sutherland_cooling_1993}, where $T_{\rm vir}$ is the virial temperature and $Z_{\rm h}$ is the metallicity of the hot halo gas. The gas density profile is modeled as a singular isothermal sphere, and the cooled gas mass within the radius $r_{\rm cool}$—the radius within which all gas can cool within the cooling time $t_{\rm cool}$—determines the ISM mass accretion rate.

If $r_{\rm cool}$ is smaller than the virial radius $r_{\rm vir}$, the SAM applies a standard cooling flow model, with $t_{\rm cool}$ calculated following \cite{somerville_semi-analytic_2008} under the assumption of an isothermal and isobaric gas density profile. When $r_{\rm cool}$ exceeds $r_{\rm vir}$, the cooling rate matches the gas accretion rate into the halo, representing ``cold/fast/filamentary” mode accretion. In these cases, the SAM ignores radiative cooling predictions and sets the ISM accretion rate equal to the halo gas accretion rate. Variations in this cooling model, such as adjustments to the cooling time definition or changes to the gas density profile, can impact the ISM accretion rate by a factor of 2-3 \citep{somerville_explanation_2008}.

\subsubsection{Star Formation and Stellar Feedback} 
\label{subsubsec:scsam_sf}

In the most recently used Santa Cruz SAM, gas accreted into the ISM is partitioned into \ion{H}{1}, H$_2$, \ion{H}{2}, and metals, with their respective mass surface densities tracked in radial disk annuli \citep{popping_evolution_2014, somerville_star_2015}. The default prescription for the star formation rate (SFR) surface density is based on the molecular hydrogen gas phase alone, accounting for a higher conversion efficiency above a critical $H_2$ surface density \citep{bigiel_star_2008, narayanan_general_2012}:
\begin{equation}
\dot{\Sigma}_{\rm SFR} = A_{\rm SF} \left(\frac{\Sigma_{H_2}}{10 M_\odot \rm pc^{-2}}\right) \left(1 + \frac{\Sigma_{H_2}}{\Sigma_{\rm H_2, crit}}\right)^{N_{\rm SF}},
\end{equation}
\noindent where $A_{SF}$, $N_{SF}$ and $\Sigma_{\rm H_2, crit}$ are free parameters. The molecular gas surface density $\Sigma_{H_2}$ is estimated using the metallicity-dependent partitioning scheme of \citet{gnedin_environmental_2011}.

Star formation proceeds through two distinct modes: a quiescent disk mode operating in isolated galaxies, and a merger-driven
starburst mode triggered during galaxy mergers \citep{somerville_star_2015}. The disk mode follows the molecular gas–based prescription described above, while merger-driven bursts temporarily increase the star formation efficiency on short timescales. The efficiency and duration of these bursts depend on the merger mass ratio and the gas fractions of the progenitor galaxies, with more equal-mass and gas-poor mergers producing stronger responses. This behavior is calibrated using hydrodynamical simulations of galaxy mergers \citep{robertson_fundamental_2006, hopkins_how_2009}. In addition to enhancing star formation, mergers redistribute angular momentum within galaxies and contribute to the growth of spheroidal stellar components. More major and gas-poor mergers are more effective at destroying disks and building spheroids, while gas-rich mergers tend to preserve disk structure.

Stellar feedback ejects cold gas from the ISM, with the ejection rate modeled as a power law:
\begin{equation}
\dot{m}_{\rm eject} = \epsilon_{\rm SN} \left(\frac{200 \rm km/s}{V_{\rm disk}}\right)^{\alpha_{\rm rh}} \dot{m}_*,
\end{equation}
\noindent where the free parameters $\epsilon_{\rm SN}$ and $\alpha_{\rm rh}$ modulate the efficiency and dependence on the disk's circular velocity $V_{\rm disk}$, and $\dot{m}_*$ is the star formation rate. $V_{\rm disk}$ is approximated as the circular velocity of the uncontracted halo at twice the Navarro-Frenk-White (NFW) scale radius $r_s$ \citep{navarro_structure_1996}. All ejected gas is either expelled from the halo entirely or deposited in the ``ejected” reservoir, depending on whether the halo's circular velocity falls below a critical threshold. A fraction of this gas is allowed to return to the hot halo on dynamical timescales, as described in Section \ref{subsubsec:scsam_cooling}.

\subsubsection{Metal Production} 
\label{subsubsec:scsam_metal}

\hspace{1em} The Santa Cruz SAM models metal production using the instantaneous recycling approximation, a commonly employed approach in SAMs. When a mass of stars, $dm_*$, forms, it generates a corresponding mass of metals, $dM_Z = y \cdot dm_*$, which are instantly mixed with the cold gas in the disk. The yield, $y$, is typically calibrated to match observational data, such as the normalization of the stellar metallicity–mass relation from \citealt{gallazzi_ages_2005}, as shown in \citealt{somerville_star_2015}. 

The metallicity of the cold gas \zcold\ evolves over time, with newly formed stars inheriting this metallicity. Supernova feedback drives both gas and metals out of the disk, transferring them to the hot halo or ejecting them into an external reservoir at rates proportional to the mass of outflowing gas. These ejected metals are not lost permanently; they can be re-accreted into the halo, contributing to ongoing chemical enrichment over time.  

\subsubsection{Observational Calibration} 
\label{subsubsec:scsam_calib}

\hspace{1em} The Santa Cruz SAM is calibrated by adjusting free parameters to match key observations at $z \sim 0$, such as the stellar-to-halo mass relation, stellar mass function, stellar mass-metallicity relation, cold gas fraction versus stellar mass for disk-dominated galaxies, and the black hole mass vs. bulge mass relation. Observational constraints for these calibrations are sourced from \cite{rodriguez-puebla_constraining_2017}, \cite{bernardi_massive_2013}, \cite{gallazzi_ages_2005}, \cite{peeples_budget_2014}, \cite{calette_hi-_2018}, and \cite{mcconnell_revisiting_2013}. Additional cross-checks involve the cold gas phase mass-metallicity relation and the $H_2$ mass function, using constraints from \cite{obreschkow_simulation_2009}, \cite{keres_co_2003}, \cite{anderson_lessigreaterwiselessigreater_2014}, \cite{zahid_chemical_2013}, and \cite{boselli_cold_2014}. The calibration considers the sensitivity of gas fraction, stellar metallicity, and star formation efficiency, fine-tuning the parameters to balance these observational constraints while adjusting for the effects of AGN and SNe feedback to match both high-mass and low-mass galaxy populations. For a more detailed description of the observations used for calibration, refer to Appendix B in \cite{yung_semi-analytic_2023}. 

\subsubsection{Modifications to the Santa Cruz SAM} 
\label{subsubsec:scsam_modifications}

\hspace{1em} The version of the Santa Cruz SAM presented in this work revises how the SAM computes the rate of gas condensation from the CGM into the ISM. When \rcool $>$ \rvir, the original Santa Cruz SAM disregards the radiative cooling prediction and equates the ISM accretion rate to the halo gas accretion rate. This approach, however, results in unrealistically low hot halo gas masses for dwarf galaxies when compared to predictions from hydrodynamical simulations \citep{pandya_first_2020}. To rectify this issue, we follow \cite{pandya_semi-analytic_2021} and adopt a modified approach inspired by \cite{guo_dwarf_2011}: when $r_{\rm cool} > r_{\rm vir}$, the rate of cooling is limited solely by the freefall time of the hot halo gas, such that:
\begin{equation}
\dot{M}^{\rm in}_{\rm ISM} = \begin{cases} 
\frac{M_{\rm CGM}}{t_{\rm dyn}} \frac{r_{\rm cool}}{r_{\rm vir}} & \text{when } r_{\rm cool} < r_{\rm vir} \\ 
\frac{M_{\rm CGM}}{t_{\rm dyn}} & \text{when } r_{\rm cool} \geq r_{\rm vir}. 
\end{cases}
\end{equation}

While this modification no longer predicts extremely low CGM masses in dwarf galaxies (See Figure \ref{fig:new_csam_v_asam_halo}), it may instead overestimate CGM masses by underestimating the total gas inflow into the galaxy. Additionally, when \rcool $>$ \rvir, our new version of the model assumes a hot accretion mode from the entire mass within \rvir. However, this approach does not take into account the contribution from cold accretion in streams or filaments from the IGM. Thus, the model represents an upper bound on the actual accretion rate.

\section{Comparing TNG and the SC SAM}
\label{sec:comp_tng_sam}
\subsection{Matching Galaxies Between TNG FP and SC SAM Run on TNG-DMO}
\label{subsec:match_tng_sam}

\hspace{1em} The identification of halos and construction of merger trees forms the backbone of all SAMs. Thus, to facilitate a direct comparison between the Santa Cruz SAM and TNG, we extract merger trees from a dark matter only (DMO) simulation run with the same initial conditions as the Full Physics (FP) TNG simulation. The SC SAM employs the \textsc{ROCKSTAR} halo finder and \textsc{consistent trees}, a gravitationally consistent merger tree algorithm developed by \cite{behroozi_rockstar_2012}.

TNG, however, utilizes a different halo finder based on the friends-of-friends (FoF) algorithm \citep{davis_evolution_1985} to identify ``groups,” and the \textsc{SUBFIND} algorithm \citep{springel_populating_2001} to identify substructure. These merger trees are not compatible with the Santa Cruz SAM.

To create bijective matches between the two sets of halo catalogs, \cite{gabrielpillai_galaxy_2022} ran the \textsc{ROCKSTAR} halo finder and \textsc{consistent trees} on TNG100-1-DM. They then matched halos identified by \textsc{ROCKSTAR} and \textsc{SUBFIND} catalogs with SubLink, a software tool developed by \cite{rodriguez-gomez_merger_2015} that uses a merit function for matching subhalos. This produced successful bijective matches between the \textsc{ROCKSTAR} and FoF/\textsc{SUBFIND} halos for central galaxies, achieving a 99\% match rate at \zzero. The fraction of bijective matches for non-central subhalos, however, is often substantially lower, with a match rate from 50–70\%. As a result, we limit our analysis to central galaxies only. 

Although  $99$\% of central galaxies have bijective matches, a direct comparison between the halo mass of the central galaxies between the two catalogs reveals slight offsets up to 20\% for low mass halos (see Figure 2 in \citealt{gabrielpillai_galaxy_2022}). These discrepancies are likely primarily influenced by baryonic physics rather than differences in halo-finding algorithms. For a more detailed description of the halo finder, merger tree algorithms, and method used to create these bijective matches, we refer the reader to Section 3 of \cite{gabrielpillai_galaxy_2022}. 

\begin{figure}
    \includegraphics[width=\linewidth]{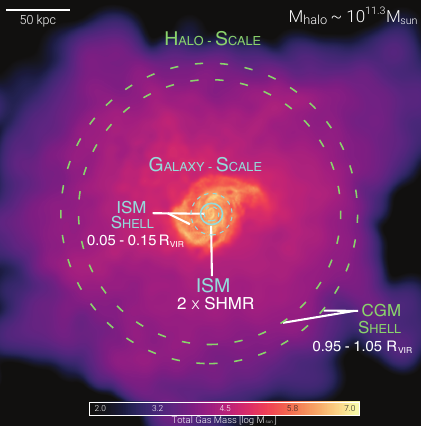}
    \caption{A sample \mhalo $\sim 10^{11.3} M_\odot$ central subhalo (ID: 609911) from TNG100 is shown to illustrate how TNG’s outputs are compared to the SAM. The map is color-coded by total gas mass, with brighter regions indicating higher mass concentrations. In TNG, we define the interstellar medium (ISM) as all material within twice the stellar half-mass radius (solid blue circle), and the circumgalactic medium (CGM) as all material within the central galaxy's virial radius (green circle). To analyze baryon flows on galaxy and halo scales, we measure inflow and outflow rates across the dashed shells at $0.05$–$0.15$\rvir\ (galaxy scale) and $0.95$–$1.05$\rvir\ (halo scale).}
    \label{fig:tngsam_definitions}
\end{figure}

\subsection{Selecting Comparable Galaxy and Halo Properties}
\label{subsec:galaxy_selection}
\hspace{1em} We define two physical scales for comparing TNG and the SAM:

\begin{enumerate}
\item the galaxy-scale (ISM), and 
\item the halo-scale (CGM),
\end{enumerate}
as illustrated in Figure  \ref{fig:tngsam_definitions}. Properties at the ``galaxy scale” describe the stellar mass (\mstar), cold gas mass (\mcold), their respective metallicities (\zstar\ and \zcold), and the star formation rate (SFR) of the central galaxy within the halo. Properties at the ``halo scale” include the virial mass (\mhalo),  the mass of the hot halo (\mhot), and the metallicity of the gas in the hot halo (\zhot). Virial masses and radii are computed using the spherical overdensity definition of \citet{bryan_statistical_1998}, where \mhalo\ is the total mass enclosed within \rvir\ such that the mean density of dark matter haloes is $\Delta_{\rm c}$ times the critical density at a given redshift. 

Although the SAM naturally segments and outputs all quantities defined above, mapping these quantities onto TNG requires careful consideration. For the galaxy's cold gas, we compute \mcold\ and \zcold\ in two complementary ways. For comparison with the SC SAM, we define \mcold\ as the sum of the combined neutral atomic hydrogen (\ion{H}{1}) and molecular hydrogen (H$_2$) components to account for differences in treatment between the SC SAM and TNG’s star formation models. The SC SAM partitions gas into neutral, ionized, and molecular components following \citet{gnedin_environmental_2011}. Since TNG does not explicitly resolve “cold” gas, we measure \ion{H}{1}$+$H$_2$ using post-processed catalogs from \citet{diemer_modeling_2018, diemer_atomic_2019} and, as in \cite{gabrielpillai_galaxy_2022}, adopt the \citet{gnedin_environmental_2011} volumetric method to ensure consistency with the SC SAM’s gas-partitioning treatment. For comparison with the TNG SAM (described in Section \ref{sec:creating_tng_sam}), we instead measure \mcold\ and \zcold\ using all gas particles within twice the stellar half-mass radius.

All remaining galaxy- and halo-scale properties are defined consistently when comparing the SC SAM and the TNG SAM to TNG. Although the ISM mass is defined as all gas within twice the stellar half-mass radius, we do not confine stellar properties to this aperture. We define \mstar\ as the total mass of bound star particles, the SFR as the total star formation rate of star-forming gas cells bound to the subhalo, and the stellar metallicity \zstar\ from the metallicities of all bound star particles, assuming that stars form from the ISM and may subsequently migrate throughout the subhalo. For halo-scale properties like \mhot\ and \zhot, we sum all gas in the host group minus the contribution from gas lying within twice the stellar half-mass radius of satellite group members. The corresponding field names used in both the SAM and TNG outputs can be found in the Appendix's Table \ref{tab:tng_sam_def}.

To enable reliable comparison between TNG and the SAM, we restrict the bulk of our analysis to well-resolved TNG halos. Following the general guidelines of \citet{gabrielpillai_galaxy_2022}, we require halos to contain at least 100 particles to be considered well-resolved. We also require that the root halo be at least 100 times more massive than the smallest resolvable progenitor to accurately capture its merger history. For TNG100-1 DMO, with a particle mass resolution of $8.9 \times 10^6 M_\odot$, this results in a conservative minimum root halo mass of $8.9 \times 10^8 M_\odot$. Since the SAM’s resolution is determined by the N-body simulation used to extract merger trees, we adopt this minimum root halo mass as our resolution threshold.

For the hydrodynamical TNG100-1 simulation, each gas or star particle has a mass of $1.4 \times 10^6$ \msun. Since we are primarily interested in tracking the gas flows between galaxies, we restrict our sample to galaxies with at least $\sim 100$ stellar particles ($M_{\rm star} \geq 1.4 \times 10^8$ \msun), and 100 gas particles ($M_{\rm gas} \geq 1.4 \times 10^8$ \msun). The gas particle requirement naturally sets a lower limit on the SFR. Following \cite{donnari_star_2019}, we adopt a minimum SFR of \(1.0 \times 10^{-3} \, M_\odot \, \text{yr}^{-1}\). These criteria yield a cleaner, albeit more restrictive, sample that is biased toward systems with well-resolved gas reservoirs and ongoing star formation (discussed further in Section \ref{subsubsec:resolution}). Applying these cuts results in a final sample of $19, 650$ galaxies with halo masses ranging from $\sim 4 \times 10^{10} - 10^{12} M_\odot$ at $z=0$. 

\subsubsection{Flow Cycle Measurements in TNG}
\label{subsubsec:tng_flow}

\hspace{1em} To establish a common language for describing the baryon cycle between TNG and the SC SAM, we draw on the analysis recently carried out by \citealt{oren_cosmic_2025} (hereafter \citetalias{oren_cosmic_2025}). \citetalias{oren_cosmic_2025} extracted measurements of gas mass, metal, and energy inflows and outflows at galaxy and halo scales for a randomly selected subset of $\sim 9,522$ central galaxies in TNG100-1 across 10 redshifts ranging from $z = 0$ to $z = 10$. At each redshift, \citetalias{oren_cosmic_2025} selected halos in bins of 0.3 dex in virial mass, ranging from $M_{\rm vir} = 10^{10} M_{\odot}$ to the most massive halo at that redshift.

Using an Eulerian approach, \citetalias{oren_cosmic_2025} calculated instantaneous gas flows by defining volumetric shells of thickness $\Delta r \sim 0.05$ \rvir\ from $0.05$\rvir\ to $1.5$\rvir. The mass and velocity of the gas particles crossing these boundaries were tracked between consecutive simulation outputs, directly measuring the inflow and outflow rates. For further details on the methodology used to derive these measurements, we refer the reader to \citetalias{oren_cosmic_2025}.

To minimize the impact of resolution effects on our results, we divided the \citetalias{oren_cosmic_2025} “flow sample” into “resolved” and “unresolved” categories using the same resolution criteria applied to the full TNG100 sample. While these resolution criteria were originally developed to assess the reliability of global galaxy properties (such as \mstar\ and SFR), we apply them here as a practical filter to exclude galaxies that are likely to be poorly resolved overall. The resolved subsample comprises $\sim 400$ galaxies per redshift, with the remaining galaxies classified as unresolved. For each subsample, we define the halo boundary from $0.95 - 1.05$ \rvir, and the galaxy boundary from $0.05-0.15$ \rvir, both illustrated as dashed circles in Figure \ref{fig:tngsam_definitions}. We volumetrically summed the inflow and outflow rates within each boundary to obtain the gas and metal flow rates shown in Sections \ref{sec:creating_tng_sam} and \ref{sec:tng_sam_results}. 

Both the resolved and unresolved samples are used in this work to ensure a comprehensive analysis across all halo masses. While we primarily rely on the resolved sample for calibrating the TNG SAM, we also make use of the unresolved halos to inform our extrapolation to lower masses, where many key phases of early galaxy evolution occur and resolution limitations become more severe (see Section \ref{subsubsec:resolution}). 

\subsection{Baseline Comparison Between TNG and the SC SAM}
\label{subsec:compare_tng_sam}

 With bijective matches between central halos in TNG and the SC SAM established in Section \ref{subsec:match_tng_sam}, we now assess how well the SC SAM reproduces the galaxy–halo relationships found in the full-physics TNG model. \cite{gabrielpillai_galaxy_2022} first performed this statistical comparison between the SC SAM and TNG at $z=0$ for a subset of key global properties, namely stellar mass, cold gas mass, star formation rate, and hot gas mass. Building upon their initial work, here we extend the comparison to cover a wider redshift range from $z=0$–$6$ and a broader set of global quantities, now including the metallicities of the stars, cold gas, and hot halo gas.

\subsubsection{Galaxy-Scale}
\label{subsec:compare_tng_sam_ism}

\begin{figure*}
    \includegraphics[width=\linewidth]{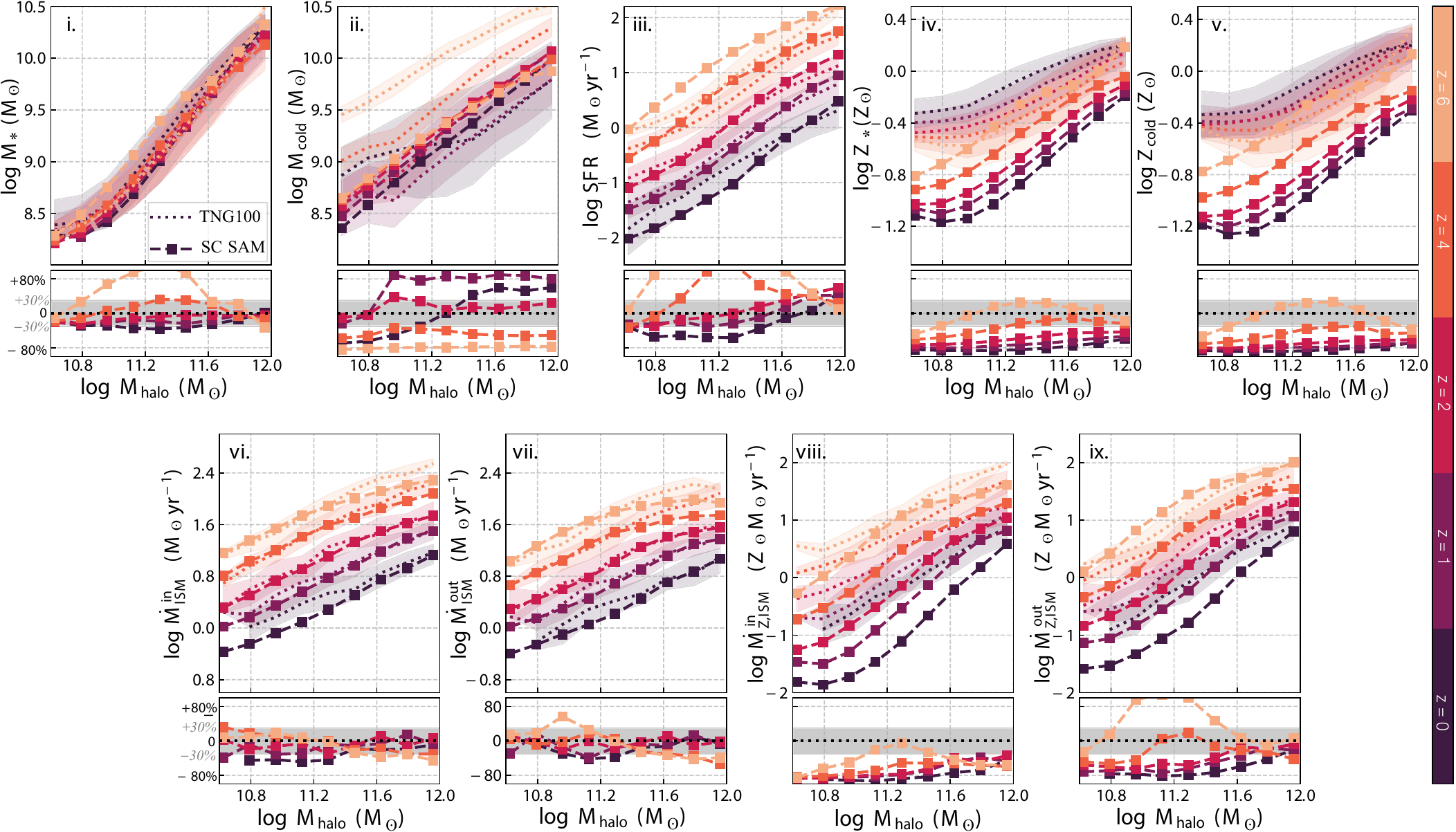}
\caption{Comparison of global galaxy-scale properties (panels $i.$–$iv.$) and flow rates (panels $v.$–$ix.$) between TNG100 (dotted lines) and the Santa Cruz SAM (dashed squares). Each panel shows median trends as a function of halo mass, color‐coded by a representative subset of redshifts ($z=0,1,2,4,6$): darker colors correspond to lower redshift and lighter colors to higher redshift, ranging from dark purple to yellow (see color bar to the right). This redshift color scheme is used consistently across all figures unless specified otherwise. Global-property medians are measured from the full resolved TNG100 sample, while flow-rate medians are drawn from the \citetalias{oren_cosmic_2025} subsample. Shaded regions show the 16–84th percentiles for TNG100. Unless otherwise noted, lightly shaded regions in all subsequent figures represent the same TNG100 percentile range. Below each main panel, the residual panel shows the fractional difference of the SC SAM relative to TNG100, with a horizontal grey band highlighting $\pm30\%$ agreement. Positive residuals indicate that the SC SAM predicts larger values than TNG, while negative residuals indicate smaller values. The SC SAM mostly matches TNG's predictions within $\pm30\%$ for the stellar mass (\mstar), and the star formation rate (SFR) at low redshift, but deviates substantially outside this range for the cold gas mass (\mcold) and the metallicities of the stars (\zstar) and cold gas (\zcold). Gas and metal inflow/outflow rates also vary, with metal flow differences often exceeding $80\%$ across redshift.}
    \label{fig:scsam_gal}
\end{figure*}

In the top row of Figure \ref{fig:scsam_gal}, we compare TNG100 (dotted lines) and the SC SAM (dashed squares) for the evolution of five galaxy-scale properties as a function of halo mass: stellar mass, cold gas mass, star formation rate, stellar metallicity and cold gas metallicity. Median trends are shown across multiple redshifts from $z=0$ to $z=6$, with residual panels beneath each relation illustrating the fractional difference between the SC SAM and TNG.

Starting with the stellar mass - halo mass relation (panel \textit{i.}), the SC SAM and TNG predictions generally agree within $\sim30\%$ across the halo mass and redshift range shown. The SC SAM tends to predict up to $\sim 40\%$ lower stellar masses than TNG at low redshift ($z<4$) and more than $\sim 100\%$ higher masses at high redshift ($z=6$). For the cold gas mass (panel \textit{ii}), the agreement between the SC SAM and TNG100 is considerably weaker. Across halo mass and redshift, the SC SAM typically differs from TNG by more than $\pm30\%$, with deviations mostly spanning $\pm80\%$. At low redshift ($z<4$), the SC SAM tends to underpredict $M_{\rm cold}$ by up to $\sim100\%$, whereas at high redshift ($z\ge4$) it increasingly overpredicts the cold gas content, again by as much as $\sim100\%$. A similar trend with redshift appears for the star formation rate (panel \textit{iii}): the SC SAM predicts up to $\sim 40\%$ lower rates of star formation than TNG at low redshift ($z<4$), and rates exceeding $\sim 100\%$ at high redshift ($z=6$).

Although the SAM and TNG's stellar and star formation rate predictions broadly track one another slope-wise across redshift, the divergence between the two models becomes much clearer for the metal populations. At $z<4$, the SC SAM predicts stellar and gas-phase metallicities that are nearly 100\% lower than those in TNG (panels \textit{iv.}–\textit{v.}). When $z\sim4$, where the SC SAM yields begins to predict higher stellar masses, the two models agree better, but at $z=6$, the SC SAM predicts higher metallicities than TNG, reflecting the SC SAM's significantly larger stellar masses.

\begin{figure*}
    \includegraphics[width=\linewidth]{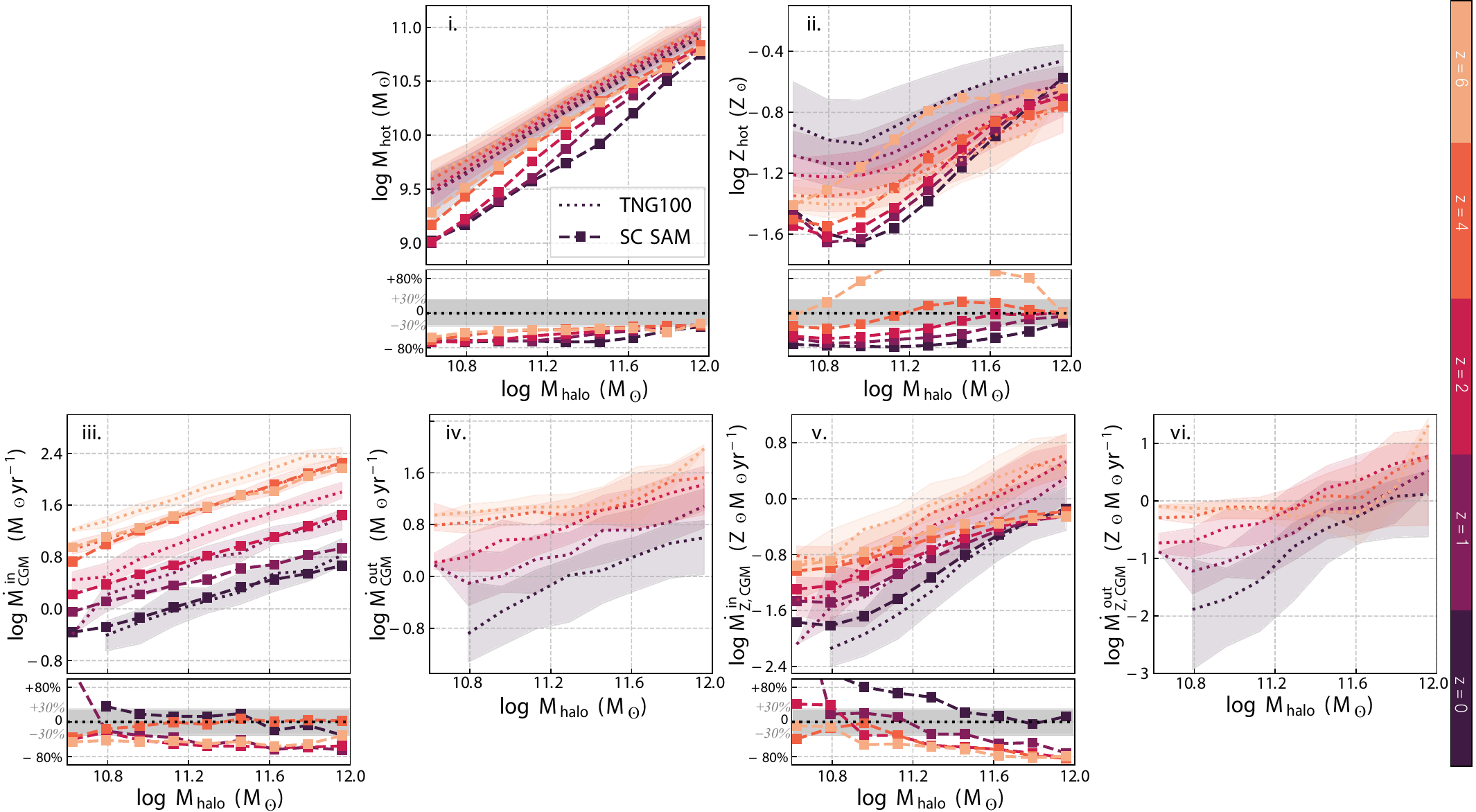}
\caption{As in Figure \ref{fig:scsam_gal}, we compare median relations from TNG100 (dotted lines) and the SC SAM (dashed squares) over redshift. Here we show halo-scale properties: \textit{i.} hot gas mass ($M_{\rm hot}$), \textit{ii.} hot gas metallicity ($Z_{\rm hot}$), and the corresponding gas and metal flow rates (\textit{iii.} - \textit{vi.}). Across redshift, the SC SAM tends to significantly underpredict both $M_{\rm hot}$ and $Z_{\rm hot}$, with differences reaching $\sim 80\%$. Gas and metal inflow rates (panels $iii.$ and $v.$) vary widely, remaining within $\sim 30\%$ of TNG at some redshifts, and with many deviations often exceeding $\sim 80\%$ (e.g., $z=1, 2, 6$ for the gas). Notably, the SC SAM does not include any explicit halo outflow channel for gas or metals (panels \textit{iv} and \textit{vi}), whereas in TNG the halo outflow rates are comparable in magnitude to the inflow rates.}
    \label{fig:scsam_halo}
\end{figure*}

In the bottom row of Figure \ref{fig:scsam_gal}, we compare galaxy-scale baryon flow rates measured in the \citetalias{oren_cosmic_2025} TNG subsample (dotted lines) with those predicted by the SC SAM (dashed squares). Considering first the inflow of gas from the CGM into the ISM (\mdotinism, panel \textit{vi.}), the SC SAM generally predicts rates within $\sim30\%$ of TNG, though at $z<4$ it tends to predict lower inflow rates, with discrepancies reaching up to $\sim60\%$. For the outflow of gas from the ISM driven by stellar feedback, the SC SAM and TNG generally agree within 30\%, with deviations reaching up to 70\% at high redshift ($z\geq4$) most closely in higher-mass halos ($\log$ \mhalo$ > 11$).

In panels $viii.$ and $ix.$, we compare the rates of metal inflow into the ISM \zmdotinism\ and metal outflow from the ISM \zmdotoutism\ between the SC SAM and TNG. As expected from the differences in the global metal distributions, the underlying metal flow rates also differ: the SC SAM predicts rates mostly $\sim80\%$ lower than TNG across redshift. The agreement only improves at higher redshift ($z\ge4$) for the metal outflow rate, where the SAM's predictions either agree with TNG's within 30\% or exceed them by more than 100\%. 

\subsubsection{Halo-Scale}
\label{subsec:compare_tng_sam_cgm}

In the top row of Figure \ref{fig:scsam_halo}, we compare the evolution of the total hot gas mass and its metallicity across halo mass and redshift. The SC SAM systematically predicts lower values of \mhot\ and \zhot\ than TNG by up to 80\% across most redshifts. The exception occurs in \zhot\ when $z=6$, where the SC SAM exceeds TNG’s predictions by up to $\sim100\%$, likely reflecting the larger stellar masses the SAM predicts at high redshift (see Figure \ref{fig:scsam_gal} panel $i.$).

Although the SC SAM predicts that galaxies have a more depleted CGM than TNG, the disagreement is noticeably smaller than reported in \citet{gabrielpillai_galaxy_2022} (see Appendix Figure \ref{fig:new_csam_v_asam_halo}). This improvement stems from the updated cooling model adopted here: rather than assuming that gas cools into the ISM at the full accretion rate once $r_{\rm cool} > r_{\rm vir}$, the current SC SAM limits the cooling rate by the freefall time of the hot halo gas, motivated by the behavior of dwarf galaxies in hydrodynamical simulations \citep{pandya_first_2020} (see Section \ref{subsubsec:scsam_modifications}).

To understand how different baryon cycling drives the large discrepancies in halo-scale properties, the lower panels of Figure \ref{fig:scsam_halo} compare the rates of gas and metal inflow and outflow. In panels $iii.$ and $iv.$, the SC SAM reproduces TNG’s halo gas inflow rates \mdotincgm to within $\sim30\%$ across most masses and redshifts, though it predicts lower values for them by up to $\sim80\%$ for higher-mass halos at $z\sim1,2, 6$. In contrast, the SC SAM includes no explicit channel for gas to leave the halo, whereas TNG predicts non-negligible halo outflows — typically only a factor of $\sim2$–$3$ times lower than the inflow rates.

A similar pattern holds for the metals. The SC SAM generally predicts either lower or higher metal inflow rates (\zmdotincgm, panel $v.$) into the CGM at higher halo masses by up to $\sim100\%$. Meanwhile, TNG predicts substantial metal outflows from the CGM (\zmdotoutcgm, panel $vi.$), often comparable to (and in some cases larger than) the metal inflow rates. 

The comparisons across the galaxy and halo scale show that although the SC SAM and TNG often arrive at broadly similar endpoints for properties that can be calibrated against observations, such as stellar mass, cold gas mass, and SFR trends, both models reach those outcomes through very different patterns of baryon cycling. At the scale of the galaxy, the SC SAM occasionally matches TNG’s stellar masses and star formation rates within tens of percent, yet differs more strongly in how material is exchanged within the ISM, with the largest discrepancies in ISM metal flows. These differences are further amplified at the scale of the halo, where the SC SAM lacks a channel for mass or metals to leave the CGM, whereas TNG continuously cycles both in and out, with the rate of outflowing gas/metals from the halo comparable to that of the inflows. As a result, the SC SAM ends up with a more depleted and less enriched CGM, even while producing global galaxy properties that look broadly similar to TNG’s. 

\begin{figure*}
    \includegraphics[width=\linewidth]{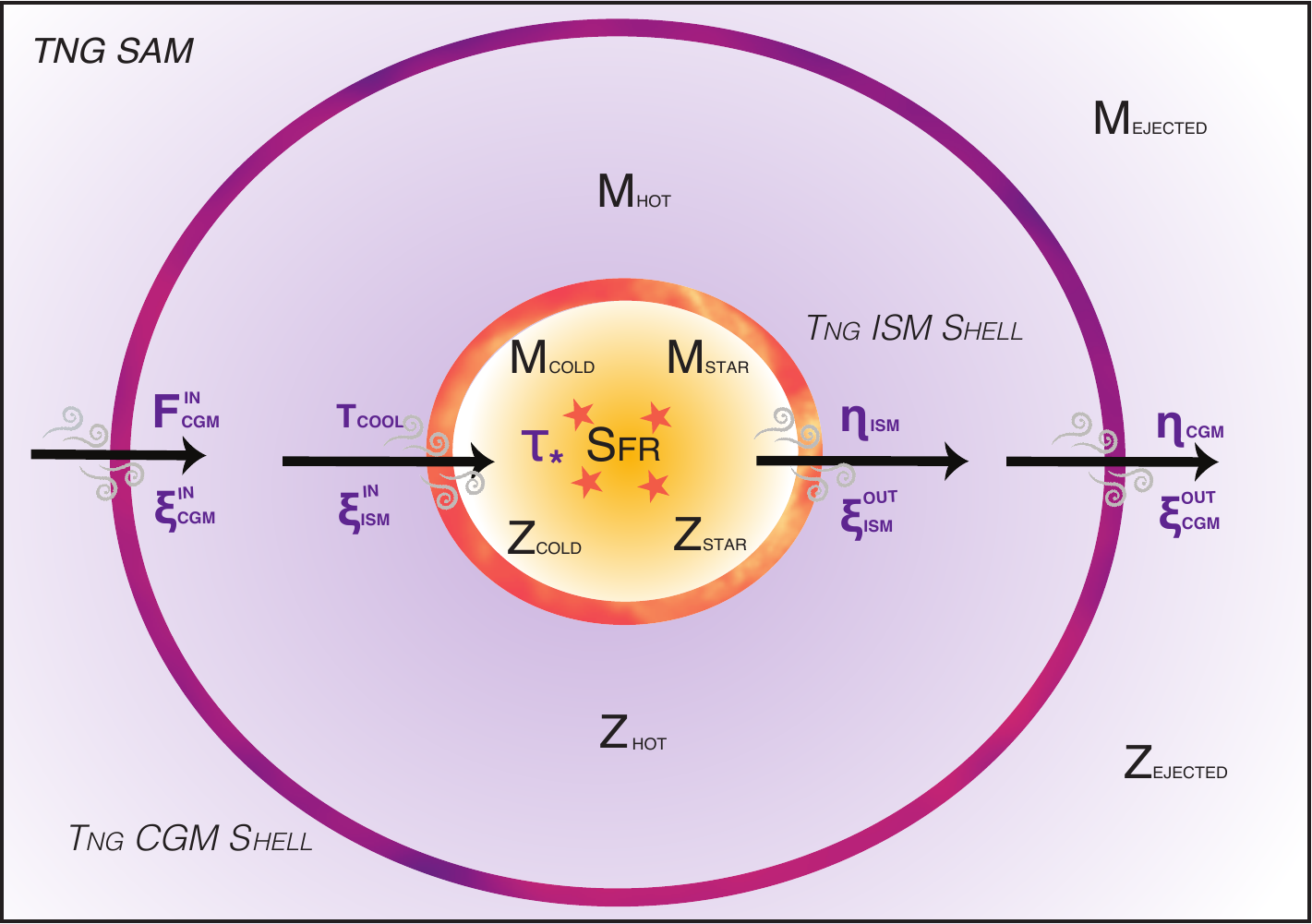}
    \centering
    \caption{Illustration of the TNG SAM, calibrated to reproduce the baryon cycle of central galaxies in the TNG100 simulation. The diagram shows the main baryonic reservoirs—cold gas ($M_{\rm{cold}}$), stars ($M_{\rm star}$), hot halo gas ($M_{\rm{hot}}$), and ejected gas ($M_{\rm{ejected}}$)—and the flows of gas and metals between them. Arrows represent baryon flow directions, with governing parameters labeled in purple (e.g., \tcool\ regulates cooling from the hot halo to the ISM, while \etaism\ and \etahalo\ control gas outflows). Metallicity enrichment is tracked for each reservoir as $Z_{\rm{cold}}$, $Z_{\rm{star}}$, $Z_{\rm{hot}}$, and $Z_{\rm{ejected}}$. Each parameter was calibrated using inflow and outflow rates measured at the ISM and CGM boundaries defined in Figure \ref{fig:tngsam_definitions}. }
    \label{fig:tngsam}
\end{figure*}

Below, we present the modifications made to the SC-SAM to better emulate the underlying gas and metal cycling pathways that inform TNG's predictions. 

\section{Translating TNG's Detailed Physics into the SC SAM's Galaxy Formation Model}
\label{sec:creating_tng_sam}
\begin{figure*}
    \includegraphics[width=\linewidth]{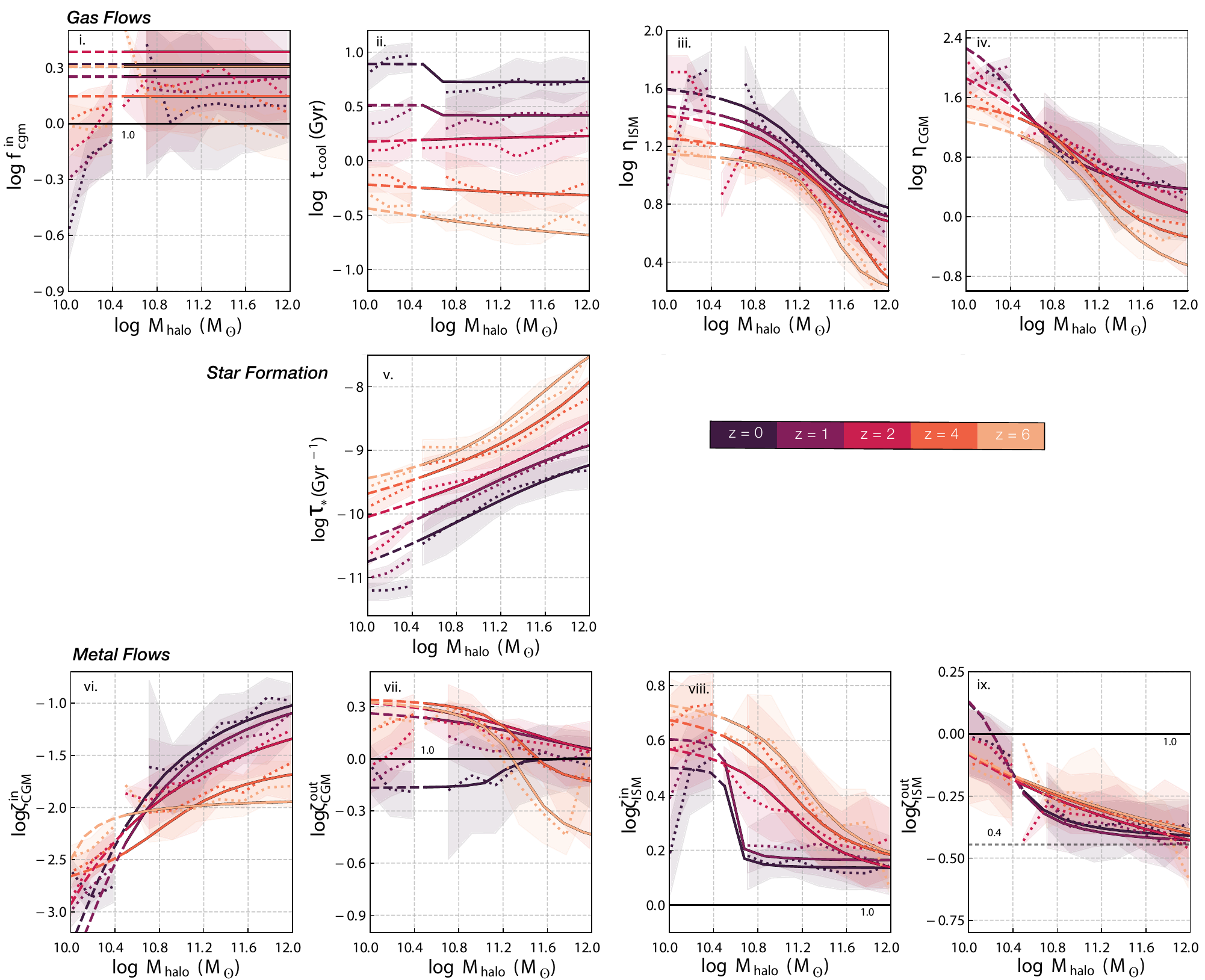}%
    \caption{Overview of the scaling relations measured from TNG100 and the subset of TNG100 galaxies analyzed by \citetalias{oren_cosmic_2025}, organized by relations regulating gas flows, star formation, and metal flows. Each panel shows the distribution of TNG100 values as a function of halo mass and over redshift as dotted lines colored by redshift. Solid lines show analytic fits based on the well-resolved sample (100-400 galaxies per redshift for \citetalias{oren_cosmic_2025} sample); dashed lines show fits based on the total sample ($>1000$ galaxies per redshift).  The parameters shown are: (\textit{i.}) CGM inflow efficiency (\fin), (\textit{ii.}) Time for gas to cool from the CGM - ISM (\tcool), (\textit{iii - iv.}) ISM and CGM outflow mass-loading factors (\etaism, \etahalo), (\textit{v.}) star formation efficiency ($\tau_*$), (\textit{vi - vii.}) CGM/ISM metal inflow enrichment factors ($\zeta^{\rm in}_{\rm CGM}$, $\zeta^{\rm in}_{\rm ISM}$), and (\textit{viii–ix.}) ISM/CGM outflow metal outflow enrichment factors ($\zeta^{\rm out}_{\rm ISM}$, $\zeta^{\rm out}_{\rm CGM}$). The analytic functions (accurate to within $\sim30\%$) used in the TNG SAM are provided in Table \ref{tab:app_tuned_params}. }
    \label{fig:tng_tuned_params}
\end{figure*}

\hspace{1em} To translate the complex physics of a hydrodynamical simulation like TNG into the simplified framework of the SC SAM, we model how gas and metals move between galaxies and their halos over time, as summarized in Figure \ref{fig:tngsam}. We derive these prescriptions using gas and metal inflow and outflow measurements from the \citetalias{oren_cosmic_2025} subsample of  TNG galaxies, which comprises $\sim 100-1000$ galaxies per redshift (see Section \ref{subsubsec:tng_flow}). Treating this subset as representative of the entire TNG central galaxy population, we derive halo mass- and redshift-dependent scaling relations that describe gas recycling, cooling, star formation, mass loading, and metal enrichment. For star formation, we leverage the full sample of $\sim 20,000$ TNG galaxies. These relations, shown in Figure \ref{fig:tng_tuned_params}, form the physical basis for calibrating the SAM to emulate TNG's baryon cycling processes.

To implement these scaling relations in the SAM, we adopt a two-pronged approach that leverages both the ``resolved” and ``unresolved” populations of the \citetalias{oren_cosmic_2025} sample. For halos within the well-resolved mass range of $10.5 < \log M_{\rm halo} < 12$, we directly apply the median scaling relations derived from the resolved sample, as shown by the solid lines in Figure \ref{fig:tng_tuned_params}. When halos fall below this mass range, we extrapolate the scaling relations to lower-mass halos (\(\log M_{\rm halo} < 10.5\)) by analyzing trends in the complete \citetalias{oren_cosmic_2025} sample, which includes both resolved and unresolved populations down to $\log M_{\rm halo} \sim 10$. These relations, shown as dashed lines in Figure \ref{fig:tng_tuned_params}, are applied to all halos with $\log M_{\rm halo} < 10.5$.  In practice, we use the medians at $\log M_{\rm halo}=10.4$ and $\log M_{\rm halo}=10.6$ as anchoring points to construct a smooth transition at $\log M_{\rm halo}\approx10.5$. 

Although a fully continuous transition between the resolved and unresolved regimes would be ideal, for some quantities (e.g., the star formation efficiency (panel \textit{v.})), the behavior inferred from the unresolved population does not smoothly align with the trends seen in the resolved halos. In these cases, we prioritize continuity with the resolved relations, where the flow measurements are most reliable, rather than forcing the low-mass extrapolation to exactly reproduce the behavior observed in the unresolved sample. Overall, this approach preserves the accuracy of the SAM in the regime where the calibration is best constrained, while still enabling it to model baryon cycling in lower-mass halos. Of course, extrapolating the relations to the low-mass regime naturally introduces additional uncertainty, which we discuss in greater detail in Section \ref{subsec:limits}. 

The calibrated SAM fits analytic functions to all scaling relations across halo mass and time, allowing the relations to be evaluated at intermediate halo masses and redshifts. While the physical processes governing gas flows are not expected to fundamentally change with redshift, TNG shows that their behavior evolves over time, likely driven by complex and still poorly understood underlying mechanisms. Each fitted relation reproduces the TNG medians to within $\sim30\%$ accuracy and is listed in Table \ref{tab:app_tuned_params}. Rather than directly imposing median flow rates from TNG, this approach enables the SAM to compute baryon flows self-consistently from a compact set of physically motivated parameters. We refer to this newly calibrated model as the `TNG SAM.'

\subsection{Halo Accretion and Cooling}

\subsubsection{IGM - CGM Accretion and Recycling}
\label{subsubsec:tng_igm_cgm}

In the traditional SAM framework, the rate at which gas accretes into the CGM is tightly coupled to the dark matter accretion rate, as detailed in Section \ref{subsubsec:scsam_cooling}. In the TNG SAM, we introduce the parameter \fin\ to regulate total gas inflow into the CGM:  
\begin{equation}
    f_{\rm in, CGM} \equiv \dot{M}_{\rm in, CGM}/(f_b \, \dot{M}_{\rm halo}),
\label{eqn:fin}
\end{equation}
\noindent where \fin\ measures the efficiency of gas accretion relative to the expected baryonic inflow, without distinguishing between pristine and recycled gas. Here, $\dot{M}_{\rm in, CGM}$ is the gas inflow rate into the CGM, $f_b$ is the cosmic baryon fraction ($f_b = 0.1573$), and $\dot{M}_{\rm halo}$ is the net mass accretion rate of the halo in TNG. 

We adopt the \textit{net} accretion rate, rather than the total inflow rate to remain consistent with how the SAM calculates $\dot{M}_{\rm halo}$. The SAM estimates the halo growth rate by finite differencing the total halo mass between consecutive timesteps, capturing the net change in halo mass from both smooth accretion and mergers. Additionally, since the SAM is run on the TNG DMO merger trees, whose halo masses differ from TNG FP's by up to 20\% in low mass halos (see Sections \ref{subsec:match_tng_sam} \& \ref{subsubsec:fp_vs_dm}), we rescale the analytic relation for \fin so that the resulting $\dot{M}^{\rm in}_{\rm CGM}$ in the SAM matches TNG within $\pm30\%$.

When \fin $ = 1$, the inflow rate matches the expected baryon-to-dark matter ratio. Values of \fin\ different from unity reflect departures from simple cosmological accretion. In particular, \fin\ $<1$ indicates suppressed gas accretion relative to halo growth, consistent with the effects of ``preventive" feedback that limits baryon inflow (see \cite{somerville_physical_2015} for a review). Conversely, \fin\ $>1$ indicates that the net inflow includes contributions beyond first-time accretion, such as the re-accretion of previously ejected gas and gas delivered by merging satellites. Figure \ref{fig:tng_tuned_params}, panel $i.$, shows that in TNG, \fin\ fluctuates around unity across halo mass and redshift, with most fluctuations above 1 when $\log$ \mhalo $>10.6$, and below 1 at the lower end of the halo mass range. We further explore the implications of this behavior in Section \ref{subsubsec:agree_fin}. We note that the halos in our analysis are well above the mass where we expect the metagalactic photoionizing background to reduce baryonic accretion into halos \citep[e.g.,][]{okamoto_mass_2008}. 

\subsubsection{CGM-ISM Cooling}
\label{subsubsec:tng_cgm_ism}

At the scale of the galaxy, the SC SAM, like most conventional SAMs, relies on the cooling radius \rcool\ to determine the rate at which gas cools into the ISM, as described in Section \ref{subsubsec:scsam_cooling}. In the traditional model, the cooling time \tcool—the time required for gas to radiate away its thermal energy—is estimated using
\begin{equation}
t_{\text{cool}} (r) \equiv \frac{3}{2} \frac{\mu m_p kT}{\rho_g(r)\Lambda(T, Z_h)},
\label{eqn:tcool_standard}
\end{equation}
where $\mu m_p$ is the mean molecular mass, $T$ is the virial temperature $T_{\rm vir} = 35.9 (V_{\rm vir}/(\rm km/s))^2 \  K$, $\rho_g(r)$ is the radial density profile of the gas, $\Lambda (T, Z_h)$ is the temperature and metallicity dependent cooling function \citep{sutherland_cooling_1993}, and $Z_h$ is the metallicity of the hot halo gas \citep{somerville_semi-analytic_2008}.

In hydrodynamical simulations, this equation is often approximated as the time required to radiate away the thermal energy $ t_{\rm cool} \sim E_{\rm th} / \Lambda n^2$, where $E_{\rm th}$ is the thermal energy of the gas \citep[e.g., Equations 21 and 27 in][]{pandya_unified_2023}. In this framework, cooling follows a ``cold” mode (\rcool $>$ \rvir) when the gas cools efficiently and falls directly into the galaxy, with the cooling time scaling with the dynamical time. Conversely, in ``hot” mode accretion (\rcool $<$ \rvir), the gas is shock-heated, leading to significantly longer cooling times.

Since the TNG SAM does not track the thermal energy of the hot halo gas, we approximate the cooling time using
\begin{equation}
t_{\rm cool} \equiv \frac{M_{\rm CGM}}{\dot{M}_{\rm cool, CGM-ISM}},
\label{eqn:tcool_cgm_ism}
\end{equation}
where $\dot{M}_{\rm cool, CGM-ISM}$ describes the rate that gas cools from the hot halo into the ISM. Rather than tracking the time for gas to radiate away its thermal energy, as in the standard definition of \tcool, this formulation instead measures how quickly gas transitions from the CGM to the ISM.

Figure \ref{fig:tng_tuned_params} panel $ii.$, shows the resulting cooling times as a function of halo mass and redshift. Across all halo masses, the cooling time exceeds the dynamical time, indicating that the traditional “cold” versus “hot” accretion dichotomy does not describe cooling in TNG (discussed further in Section \ref{subsubsec:agree_tcool}). In the new TNG SAM, we use the cooling times measured in TNG to determine the rate at which gas accretes onto the galaxy via Equation \ref{eqn:tcool_cgm_ism}.

\subsection{Star Formation}

TNG models star formation using the subgrid two-phase ISM model of \cite{springel_cosmological_2003}, where an effective equation of state is adopted for the ISM. In this model, dense, cold clouds are embedded in a hot tenuous medium, with the two phases in rough pressure equilibrium. For gas volume densities greater than a critical value $\rho > \rho_{\rm th}$, star formation is assumed to take place at the rate:
\begin{equation}
\frac{d\rho_*}{dt} = \rho_c/t_*,
\end{equation}
where $\rho_c$ is the density of gas in the cold phase. 

The gas consumption time is then given by $t_* = t_0^* \ (\rho/\rho_{\rm th})^{1/2}$. The parameters $t_0^*$ and $\rho_{\rm th}$ are tuned to reproduce the global observed Kennicutt-Schmidt (KS) relation, given by \citet{kennicutt_global_1998}:
\begin{equation}
\begin{split}
\Sigma_{\rm SFR} &= (2.5 \pm 0.7)\times10^{-4}
\left(\frac{\Sigma_{\rm gas}}{M_\odot\,{\rm pc}^{-2}}\right)^{1.4 \pm 0.15} \\
&\quad M_\odot\,{\rm yr}^{-1}\,{\rm kpc}^{-2}
\end{split}
\label{eqn:ks}
\end{equation}

In the TNG SAM, we do not implement the KS relation, even though it is available as an option in the SC SAM's framework.  This choice is motivated by the fundamental mismatch in how galaxy sizes are defined and interpreted in the SAM versus in TNG. In the SC SAM, stars are assumed to reside in a rotationally supported exponential disk, and this disk scale length sets the gas surface density in the KS relation. In contrast, in TNG, the stellar component comprises dispersion-dominated spheroids, and the reported galaxy sizes (e.g., the stellar half mass radii) reflect the full three-dimensional stellar distribution rather than a rotationally supported disk.  

\begin{figure*}
    \includegraphics[width=\linewidth]{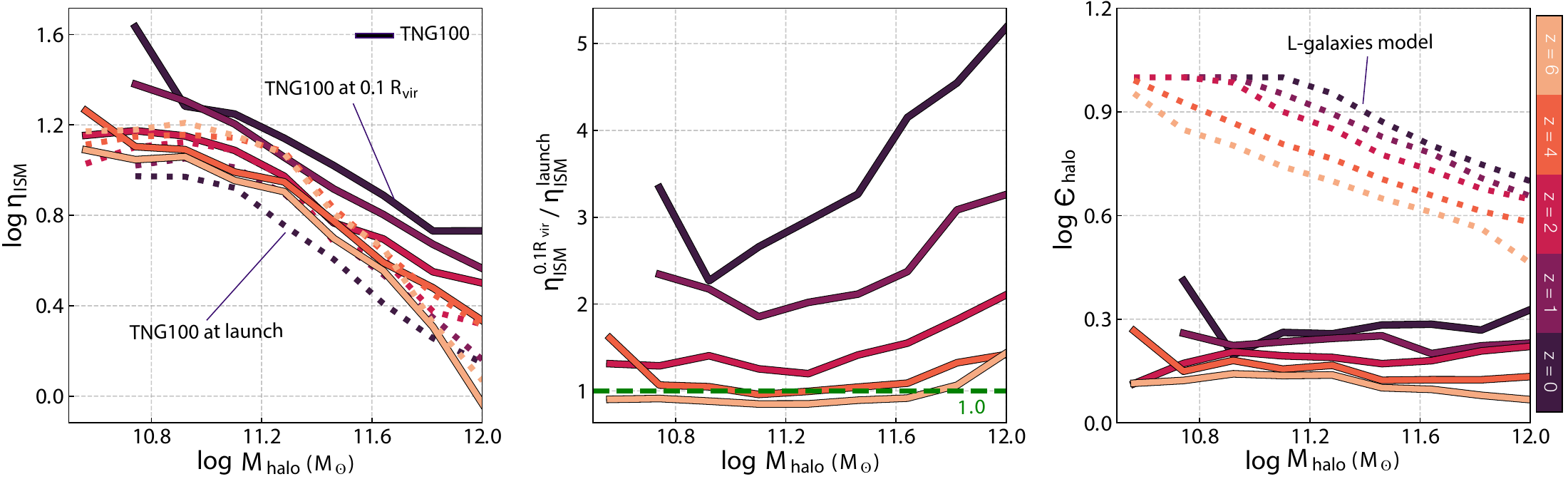}
    \caption{Functions that govern the TNG SAM's stellar feedback-driven outflows at the scale of the galaxy (left and middle panels) and halo (right panel) across redshift. The left panel shows that the mass loading factor \etaism\ (solid lines), defined as the ratio of the outflow rate to the star formation rate, almost always exceeds the launch value from star-forming gas cells in TNG (dotted lines), suggesting additional mass entrainment as winds reach the galaxy’s radius. To account for this, the TNG SAM scales the launch value by the ratio of galaxy-scale to launch-scale values, which vary with halo mass and redshift, as shown in the middle panel. At the halo scale, the outflow efficiency is captured by \etahalo, defined analogously as the outflow rate through the halo boundary divided by the star formation rate, and is modeled using the dimensionless parameter $\epsilon_{\rm halo}$. The values of $\epsilon_{\rm halo}$ adopted in the L-galaxies SAM (dotted lines) are much higher than those directly measured in TNG (solid lines) using Equation \ref{eqn:eps_halo}, which we used as inputs in the TNG SAM.}
    \label{fig:tngsam_stellar_feedback}
\end{figure*}

As a result, we model star formation directly through the star formation efficiency \sfe, defined as

\begin{equation}
\tau_* = \frac{\dot{m}_*}{M_{\rm cold}}.
\label{eqn:sfe}
\end{equation}

\noindent We measure \sfe\ as a function of halo mass and redshift using the full TNG100 galaxy sample (see Figure \ref{fig:tng_tuned_params} panel \textit{v}.), and calibrate the star formation rate with:

\begin{equation}
\dot{m}_* = \tau_*(M_{\rm halo}, z) \cdot M_{\rm cold}.
\label{eqn:sfe_sfr}
\end{equation}

\subsection{Stellar Feedback}
\label{subsec:tngsam_feedback}

\subsubsection{Galactic Winds and Galaxy-scale Outflows}
\label{subsec:tngsam_feedback_galaxy}

To emulate TNG’s treatment of stellar feedback, the TNG SAM incorporates the galactic wind model from \citealt{pillepich_first_2018}. Inspired by the results of semi-analytic models, the TNG model launches wind particles with an initial speed that scales with the local dark matter velocity dispersion  $\sigma_{\rm DM}$. It also includes a redshift-dependent factor and a minimum wind velocity, such that the wind velocity $v_w$ scales as:
\begin{align}
    v_w &=  {\rm max} \Big[ \kappa_w \sigma_{\rm DM} \Big(\frac{H_0}{H(z)}\Big)^{1/3}, v_{w,{\rm min}}\Big]
\label{eqn:wind_velocity}
\end{align}
\noindent where $\kappa_w = 7.4$ and $v_{w,{\rm min}} = 350$ \kms\ in the fiducial TNG model. This form ensures a physically plausible scaling of wind velocities with redshift and halo mass, while preventing unphysically large mass loading factors in low-mass halos. 

The TNG model further refines wind generation by linking the wind mass loading factor $\eta_w$ to the specific energy available for wind generation $e_w$. This energy is primarily determined by the energy released from Type II supernovae, with a fraction $\tau_w$ being thermal. The wind energy itself is a function of the metallicity of the star-forming gas, leading to a metallicity-dependent galactic wind model. 

The wind mass loading factor $\eta_w$, defined as the ratio of the mass outflow rate $\dot{m}_{\rm ISM}^{\rm out}$ to the star formation rate $\dot{m}_*$, quantifies the efficiency of these winds in removing gas from the ISM, and is related to $v_{\rm w}$ and the wind energy, $e_w$, as:
\begin{equation}
    \eta_w = \frac{2}{{v_w}^2}\, e_w(1-\tau_w)
\end{equation}
\noindent where $e_w$ is given by:
\begin{multline}
    e_w = \bar{e}_w \left[ f_{w,Z} + \frac{1-f_{w,Z}}{1+(Z/Z_{\rm w, ref})^{\gamma_{w,Z}}} \right]\\ \times N_{\rm SNII} E_{\rm SNII}\, 10^{51} {\rm erg\, M^{-1}_\odot}.
\end{multline}
Here, $\bar{e}_w$ is the average energy per unit stellar mass formed, $f_{w,Z}$ modulates the metallicity dependence, $Z$ represents the gas metallicity, $Z_{\rm w, ref}$ is a reference metallicity, $\gamma_{w,Z}$ controls the strength of the metallicity dependence, $N_{\rm SNII}$ denotes the number of Type II supernovae, and $E_{\rm SNII,51}$ is the energy released per supernova in units of $10^{51}$ erg.

In the TNG model, $\eta_w$ is implemented at the scale of the launched wind particle. In the TNG SAM, we instead apply $\eta_w$ at the boundary of the ISM, which we denote as \etaism. The left panel of Figure \ref{fig:tngsam_stellar_feedback} shows that the value of \etaism\ at the scale of the galaxy (solid lines) can differ significantly from the value at launch (dotted lines). When $z<6$, the galaxy-scale outflow rates are systematically higher than the launch rates, indicating that winds entrain additional material as they propagate outward. At $z=6$, the opposite trend is seen: the galaxy-scale outflow rates are lower than the launch rates, suggesting that winds stall, likely due to the higher density and pressure of the CGM. To capture this behavior in the SAM, we model the ratio between $\eta_w$ at the launch scale and $\eta_{\rm ISM}$ at the galaxy scale in TNG, shown in the middle panel of Figure \ref{fig:tngsam_stellar_feedback}.

In the TNG SAM, the material ejected from the ISM is always deposited in the CGM. Unlike the fiducial SC SAM, which allows $\eta_w$ to grow to extremely large values in low-mass halos (exceeding $10^3$ at $10^8 \, M_\odot$), the TNG-calibrated scaling relations yield values that remain near $\sim100$ in this regime, without the need for an explicit cap.

\subsubsection{Halo-scale Outflows}
\label{subsec:tngsam_feedback_halo}

In TNG, galactic winds can not only expel gas from the ISM but also entrain additional material from the CGM, ejecting it into the IGM.  In the SC SAM, however, stellar feedback ejects gas directly from the ISM into a separate ejected reservoir. The ejected gas returns to the CGM indirectly when a portion of it is re-accreted into the hot halo at later times. In the new TNG SAM, we explicitly account for the CGM-IGM outflow channel with the parameter \etahalo, defined as the ratio of the mass outflow rate $\dot{m}_{\rm CGM}^{\rm out}$ to $\dot{m}*$. Complementing the existing \etaism\ at the galaxy scale, \etahalo\ quantifies the efficiency with which stellar feedback ejects gas from the halo.

The TNG SAM model for \etahalo\ is inspired by the halo-scale outflow prescription in the L-galaxies SAM \citep[][H15]{henriques_galaxy_2015}. In the L-galaxies framework, excess supernova energy first reheats gas in the ISM and ejects it into the CGM. If enough energy remains, it can further expel hot halo gas into the IGM, with:
\begin{equation}
    \frac{1}{2} \Delta m_{\rm eject} V_{\rm vir}^2 = \Delta E_{\rm SN} - \Delta E_{\rm reheat},
\end{equation}
where $\Delta E_{\rm reheat} = \frac{1}{2} \Delta m_{\rm reheat} V_{\rm vir}^2$.

Assuming the quantity $\Delta m_{\rm eject}/\Delta m_*$ is equivalent to our $\eta_{\rm halo} \equiv \dot{m}_{\rm CGM-IGM}/\dot{m}_*$, we derive the expression for \etahalo\ as:
\begin{equation}
\eta_{\rm halo} = \frac{2 \epsilon_{\rm halo} V_{\rm SN}^2}{V_{\rm vir}^2} - \eta_{\rm ISM}
\label{eqn:eps_halo}
\end{equation}
where $V_{\rm SN}=630$ km/s is constant and $\epsilon_{\rm halo}$ is a dimensionless efficiency parameter capped at unity and given by: 
\begin{equation}
    \epsilon_{\rm halo} = \epsilon_{\rm halo, 0} \left[0.5 + \left(\frac{V_{\rm max}}{V_{\rm eject}} \right)^{-\beta_2} \right].
\end{equation}

In \citetalias{henriques_galaxy_2015}, the best-fit values for $\epsilon_{\rm halo}$ are $\epsilon_{\rm halo, 0} = 0.62, \beta_2 = 0.80,$ and $V_{\rm eject} = 100$ km/s. However, as shown in the right panel of Figure \ref{fig:tngsam_stellar_feedback}, applying these values in TNG (dotted lines) results in values for $\epsilon_{\rm halo}$ significantly higher than what is measured in TNG (solid lines), overestimating the efficiency by up to a factor of 10. To account for this, the TNG SAM instead parameterizes $\epsilon_{\rm halo}$ based on its actual dependence on halo mass and redshift in TNG (solid lines).

In the TNG SAM, gas ejected from the halo does not re-accrete. Instead, the \fin\ parameter accounts for a fraction of this gas returning to the halo. However, the reaccretion process is not explicitly modeled, as the distinction between pristine and recycled gas cannot be resolved with the Eulerian method used by \citetalias{oren_cosmic_2025}.

\subsection{Metal Enrichment and Circulation}

In TNG, metals play a dual role: they trace star formation and chemical enrichment while also regulating galactic winds and cooling rates. The TNG model explicitly tracks the gradual release of metals from asymptotic giant branch stars, core-collapse supernovae, and Type Ia supernovae, allowing the simulation to follow their redistribution within the ISM and CGM. In contrast, the SC SAM employs a simplified, instantaneous approach to metal production and circulation, described in Section \ref{subsubsec:scsam_metal}.

To better replicate TNG’s metal flows in the SAM, we introduce metal-enrichment factors that quantify the fraction of metals transported into and out of reservoirs relative to the total gas flow. Following \citet{peeples_constraints_2011}, we define the metal enrichment factor for outflows as:
\begin{equation}
\zeta^{\rm out}_{\rm gas} = \frac{\dot{M}_{\rm Z}^{\rm out}}
{\dot{M}_{\rm gas}^{\rm out} \cdot Z_{\rm gas}},
\label{eqn:zeta_out}
\end{equation}
where $\dot{M}_{\rm Z}^{\rm out}$ is the rate at which metals are outflowing from the ISM/CGM, $\dot{M}^{\rm out}_{\rm gas}$ is the rate at which all gas is outflowing from the ISM/CGM, and $Z_{\rm gas}$ describes the metallicity of the gas reservoir from which the outflows originated. 

Traditionally, the SC SAM, along with most SAMs (e.g., GALFORM, SAGE, and \textsc{L-GALAXIES}), assumes $ \zeta^{\rm out}_{\rm gas} = 1$, such that outflowing gas has the same metallicity as its originating reservoir.  Some more recent models have implemented different assumptions; for example, the latest version of \textsc{DARK SAGE} adopts metal-poor outflows based on an energy-based argument, in which stellar feedback imparts comparable specific energy to all particles, leading to under-enriched winds at launch. The TNG simulation, by directly tracking the metal content of outflows, allows these assumptions to be tested. Panels $vii.$ and $ix.$ in Figure \ref{fig:tng_tuned_params} show that outflow metallicities in TNG do not follow a simple scaling with the originating reservoir's metallicity. At the scale of the galaxy, $\zeta^{\rm out}_{\rm ISM}$ consistently falls below unity, indicating that the winds eject fewer metals relative to the average metallicity of the ISM. This result is not too surprising, for the fiducial TNG model sets the wind metal loading factor $\gamma_w = 0.4$ \citep{vogelsberger_model_2013}, leading to under-enriched winds compared to the average metallicity of the ISM. However, as the winds reach the scale of the galaxy, they entrain additional material, increasing their metal content such that the effective metal loading exceeds the launch value, as shown by the grey dashed line in panel \textit{ix.}.

At the scale of the halo, for $z=0$, $\zeta^{\rm out}_{\rm CGM}$ often deviates from unity, mostly lying above 1 for halos with $\log$ \mhalo $< 11.2$. This indicates that winds leaving the halo are more metal-enriched than the mean metallicity of the hot halo gas. In contrast, for more massive halos, and across all halo masses at $z=0$, $\zeta^{\rm out}_{\rm CGM}$ remains below unity, indicating that gas leaving the halo is less metal-enriched than the average metallicity of the hot halo.

In addition to calibrating metal enrichment factors for gas outflows, we also created inflow enrichment factors to capture the metal mass transported by gas accreted into the ISM and CGM from the CGM and IGM, respectively. For the ISM, we define $\zeta^{\rm in}_{ISM}$ as:
\begin{equation}
\zeta^{\rm in}_{ISM} = \frac{\dot{M}^{\rm in}_{\rm Z,ISM}}{\dot{M}^{\rm in}_{ISM} \cdot Z_{\rm CGM}},
\end{equation}
where $\dot{M}_{\rm Z, ISM}^{\rm in}$ is the metal inflow rate into the ISM, $\dot{M}_{\rm ISM}^{\rm in}$ is the total gas inflow rate, and $Z_{\rm CGM}$ is the metallicity of the CGM. For the CGM, since the SAM does not track the metallicity of the IGM, we define $\zeta^{\rm in}_{CGM}$ as the fraction of metals entering the CGM compared to the total rate of gas flowing into the CGM:
\begin{equation}
\zeta^{\rm in}_{CGM} = \frac{\dot{M}^{\rm in}_{\rm Z, CGM}}{\dot{M}^{\rm in}_{\rm CGM}}.
\end{equation}

Notably, Figure \ref{fig:tng_tuned_params} panel \textit{viii.} shows that the fraction of metals accreting into the ISM in TNG consistently exceeds unity across redshift. This indicates that the gas returning to the ISM is more metal-rich than the average CGM, suggesting efficient cooling of metal-enriched gas. At the scale of the halo, panel $vi.$ shows a non-negligible inflow of metals into the CGM given the non-zero values of $\zeta^{\rm in}_{CGM}$. This indicates that some fraction of the metals ejected by stellar feedback can escape the halo and subsequently be re-accreted, emphasizing the role of metal recycling across both galactic and halo scales.

To further improve the SAM’s agreement with TNG for the metal populations, we increased the stellar yield $y$ from the fiducial SC SAM value of $1.2\,Z_\odot$ to $1.5\,Z_\odot$. While TNG does not assume a fixed yield, \citealt{torrey_evolution_2019} estimate an effective global yield of $y_{\rm global} \approx 0.05$, implying that roughly 5\% of the stellar mass formed is returned to the ISM as metals. In comparison, our adopted yield of $1.5 Z_\odot$, assuming $Z_\odot = 0.02$, corresponds to a metal production efficiency of $\sim 3\%$ by mass. Although this is somewhat lower than TNG’s global value, the comparison is not truly one-to-one since TNG tracks delayed enrichment and mixing, whereas the SAM assumes a fixed, instantaneous yield deposited directly into the ISM at the time of star formation.

\section{Reproducing TNG100's Results Over 12 Gyr}
\label{sec:tng_sam_results}

As discussed in Section \ref{sec:intro} and observed in Section \ref{subsec:compare_tng_sam}, hydrodynamical simulations and SAMs can yield similar global galaxy properties even though they model gas and metal flows in fundamentally different ways. The primary goal of the TNG SAM is to use the calibrations presented in Section \ref{sec:creating_tng_sam} to not only match TNG’s global properties but also replicate its underlying gas and metal flow cycles. Here, we evaluate how well the TNG SAM reproduces both the large-scale gas and metal flows and the global galaxy and halo properties in TNG. 

\begin{figure*}
    \includegraphics[width=\linewidth]{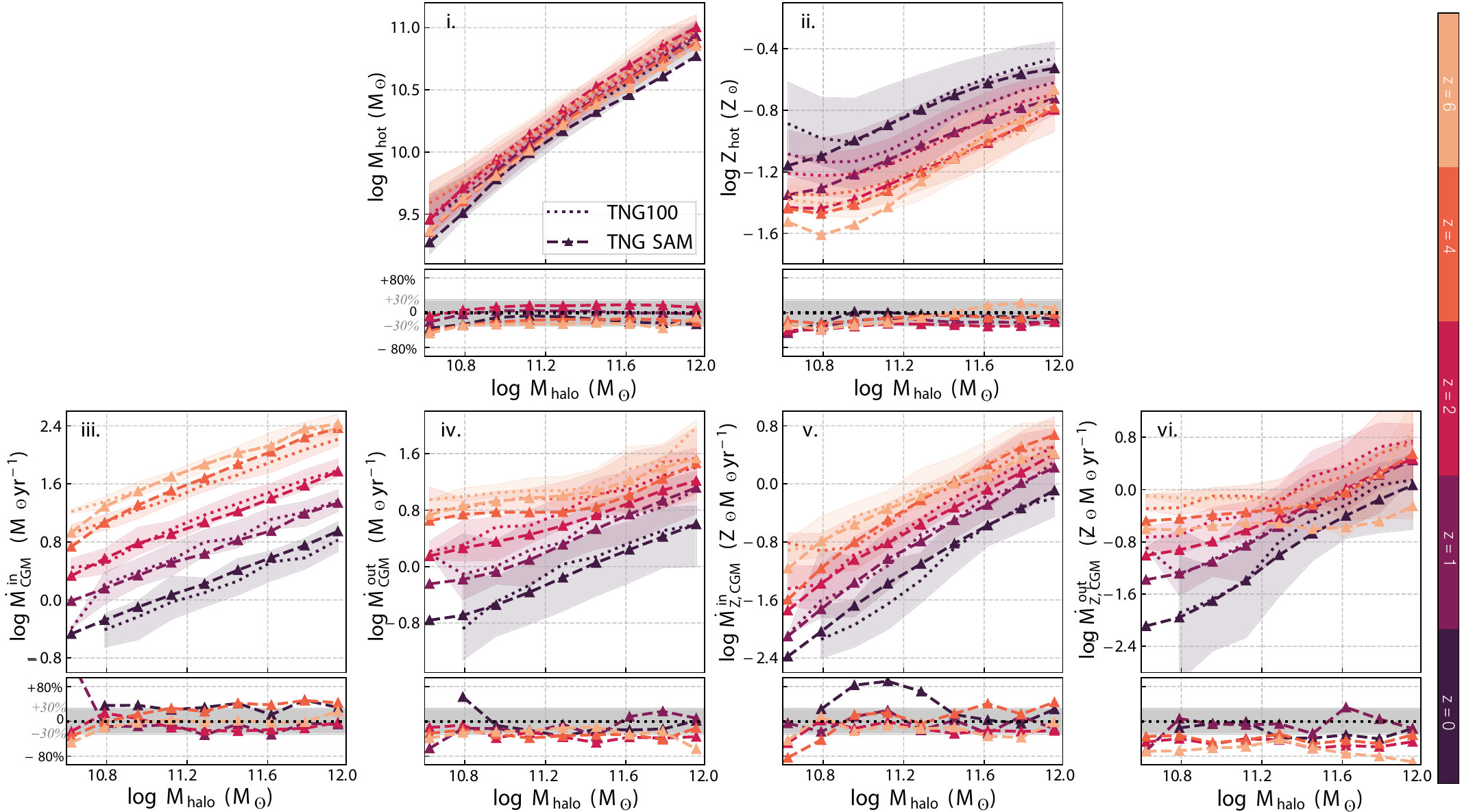}
\caption{Comparison of halo-scale quantities (panels \textit{i–ii.}) and the corresponding gas and metal flow rates (panels \textit{iii–vi.}) between TNG100 (dotted lines) and the newly calibrated TNG SAM (dashed triangles). The TNG SAM reproduces the hot gas mass and hot gas metallicity within $\pm30\%$ across most halo masses and redshifts. Gas and metal inflow rates are similarly well-matched, typically within $\pm30\%$ of TNG. A major improvement over the SC SAM is that the TNG SAM now captures the halo outflow channels with comparable accuracy, reproducing both gas and metal outflows to mostly within $\sim30\%$. }
    \label{fig:tngsam_halo}
\end{figure*}

\subsection{The Galaxy Scale}
\label{subsec:compare_tngsam_galaxy}

\begin{figure*}
    \includegraphics[width=\linewidth]{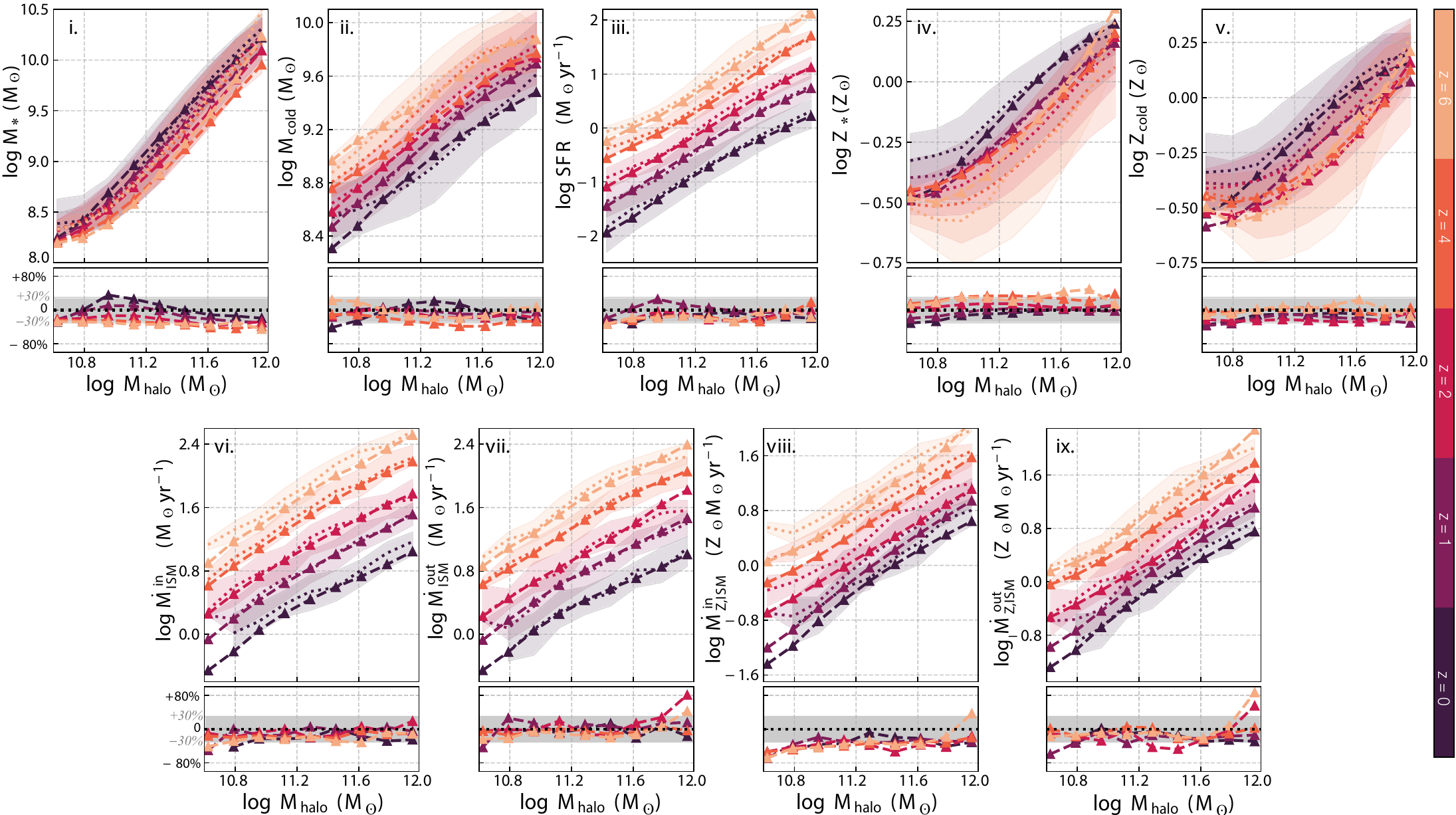}
\caption{Comparison of galaxy-scale properties (panels \textit{i–v.}) and flow rates (panels\textit{ vi–ix.}) between TNG100 (dotted lines) and the newly calibrated TNG SAM (dashed triangles). As in Figure \ref{fig:scsam_gal}, medians are shown as a function of halo mass, and redshift is indicated by the same color scheme used throughout the paper. Overall, the TNG SAM reproduces both the baryon flow rates and global properties of TNG with substantially improved accuracy relative to the fiducial SC SAM, particularly for the metal quantities. Across most halo masses and redshifts, the TNG SAM matches TNG within $\pm 30\%$ for stellar mass, cold gas mass, and star formation rate, metals, and both gas and metal inflow/outflow rates.}
    \label{fig:tngsam_gal}
\end{figure*}

In Figure \ref{fig:tngsam_gal}, the top row (panels \textit{i–v.}) compares the stellar mass, cold gas mass, star formation rate, and the metallicities of stars and cold gas as a function of halo mass between TNG100 (dotted lines) and the newly calibrated TNG SAM (dashed triangles). 
The TNG SAM generally reproduces the stellar mass–halo mass relation (panel \textit{i.}) to within $\sim 20-30\%$ over most halo masses and redshifts. At intermediate halo masses ($ \log$  \mhalo $\sim  11$), there is a modest excess at \zzero, where the TNG SAM predicts stellar masses up to $\sim 40\%$ higher than TNG. This offset correlates with the slightly elevated cold gas mass and SFR at the same halo mass and redshift (both still within $\sim 20-30\%$ of TNG's predictions), indicating that the SAM is converting a somewhat larger available gas reservoir into stars over time. 

The fact that the TNG SAM's ISM inflow and outflow rates are now reproduced to within $\sim 30\%$ across halo mass and redshift (panels \textit{vi–vii.}) is directly connected to the improvements in $M_{\rm cold}$ and SFR. Still, \mdotinism tends to lie near the lower end of this agreement at $z=6$. This behavior is consistent with the modest deficit in hot gas mass at this redshift (Figure \ref{fig:tngsam_halo}, top left panel), which reduces the rate at which gas can cool onto the ISM. Excluding behavior at the edges of the halo mass range (i.e., $\log$ \mhalo $\sim 10.6$ or $\log$ \mhalo $\sim 12.0$), the SAM still generally balances gas inflow into the ISM and gas removal via outflows in a manner consistent with TNG, allowing the cold gas reservoir to remain well regulated over time.

Among the global properties, the metals provide the most direct evidence that reproducing a hydrodynamical simulation’s baryon cycle leads to improved agreement with its predictions. Panels \textit{iv–v.} show that the TNG SAM matches TNG's stellar and cold gas metallicities within $\sim30\%$ across halo mass and redshift, with deviations up to $\sim80\%$ limited to the stellar metallicity at $z=6$ in high mass halos. This stands in stark contrast to the SC SAM, which predicted ISM metallicities a factor of 2 times lower. Specifically, at $z=0$, the resulting increase in metallicities brings the mass–metallicity relation into better agreement with observations than the fiducial SC SAM, demonstrating that calibrating to TNG can improve observational agreement, even though it is not the model's primary objective. The improvement stems from the fact that the TNG SAM now reproduces metal inflow and outflow rates to within $\sim30\%$ of TNG (panels \textit{viii.} and \textit{ix.}), with larger deviations confined to the edges of the halo mass range, whereas the SC SAM differed by factors of $\sim2$.

\subsection{The Halo Scale}
\label{subsec:compare_tngsam_halo}

In Figure \ref{fig:tngsam_halo}, the top row (panels \textit{i} \& \textit{ii.}) compares the total hot gas mass and hot gas metallicity between TNG100 (dotted lines) and the TNG SAM (dashed triangles) as a function of halo mass and redshift. The TNG SAM reproduces both quantities to within $\sim 30\%$ across most halo masses and redshifts, with small deviations up to $\sim 40\%$ mostly near the low and high ends of the mass distribution. This marks a substantial improvement over the SC SAM (Figure \ref{fig:scsam_halo}), which predicted CGM gas masses and metallicities $\sim 80\%$ lower than TNG.

Panels \textit{iii–vi.} show the corresponding baryon flows into and out of the halo. Gas inflow rates (\textit{iii.}) are typically reproduced to within $\sim 30\%$ across mass and redshift, with some larger deviations (up to $\sim 80\%$) appearing at the edges of the resolved mass range around $z=1$ and $z=6$. Metal inflow rates (\textit{v.}) show similar overall agreement, although differences of up to $\sim 80\%$ appear at $z=4$, mostly confined to the edges of the mass range shown. These wider deviations, and the broader spread around the 30\% band, likely arise because \mdotincgm\ itself already exhibits significant scatter. In the SAM, the halo inflow rate is obtained by finite differencing halo masses from the DMO simulation, which differ from those in the hydrodynamical TNG run. Discussed further in Section \ref{subsubsec:fp_vs_dm}, these halo mass differences reach $\sim 20\%$ for low mass halos and increase toward higher redshift, directly impacting accretion rate estimates. The scatter in \mdotincgm\ is then propagated and amplified when computing the metal inflow rate, since \zmdotincgm\ is determined via the enrichment factor $\zeta_{\rm CGM}^{\rm in}$. Despite this, the TNG SAM still captures the overall scaling of CGM gas and metal inflows with halo mass and redshift.

\begin{figure*}
    \includegraphics[width=\linewidth]{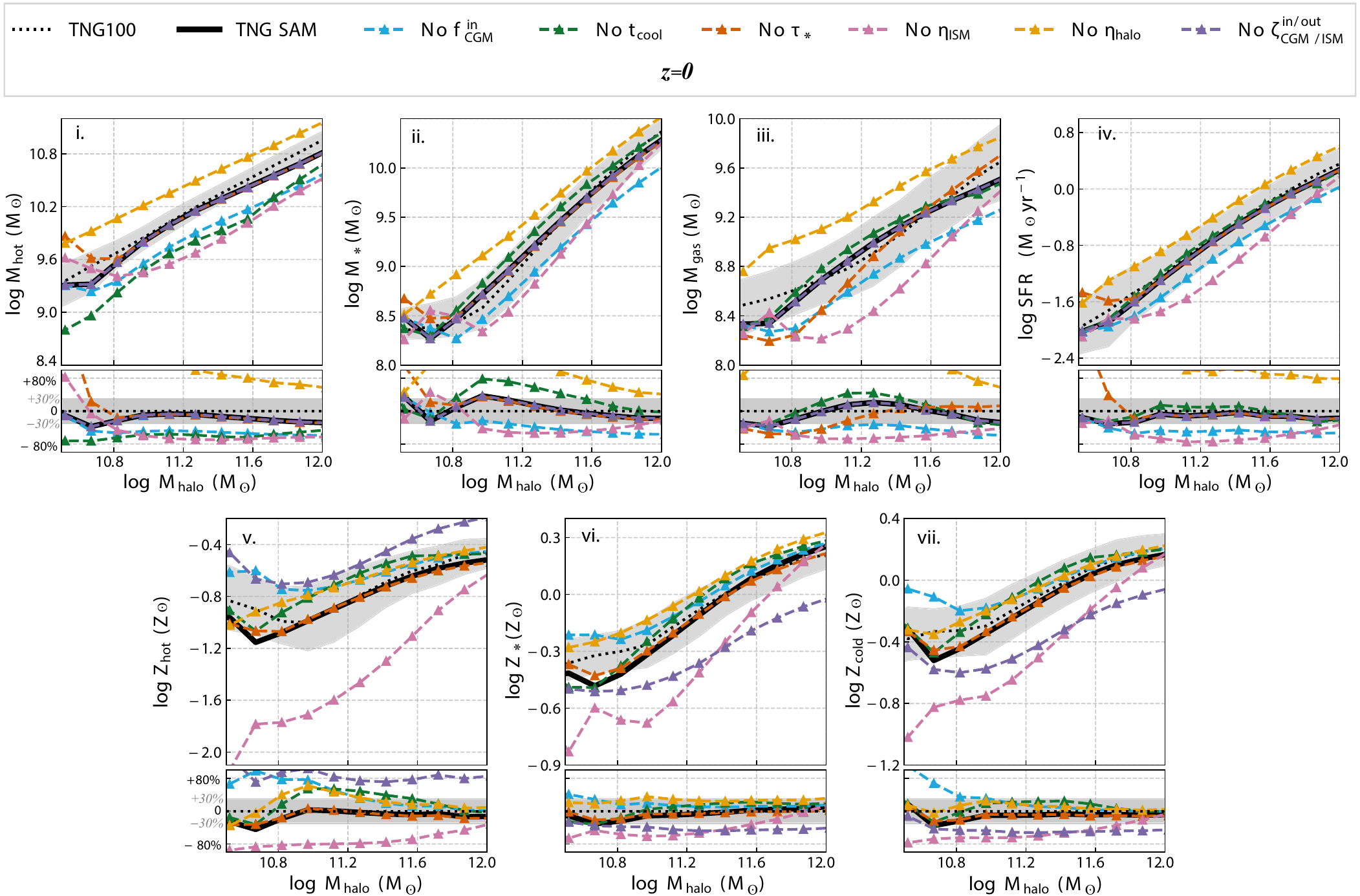}
    \caption{To show the importance of each modification made to the TNG SAM, we illustrate how reverting individual model components to their original SC SAM formulations impacts the TNG SAM’s ability to reproduce TNG’s global results at the galaxy and halo scales at \zzero. The solid black line represents the 50th percentile of the fully calibrated TNG SAM, while the shaded gray region marks $\pm 30\%$ of the TNG100 population, with dotted lines indicating the upper and lower boundaries. Solid lines with circles represent the TNG SAM’s performance when specific calibrations are reverted, with each line colored according to the removed new model component.}
    \label{fig:tngsam_variations}
\end{figure*}

Another stark improvement over the SC SAM appears in the halo outflow channels. The TNG SAM predicts both gas and metal outflow rates to mostly within $\sim 30\%$ of TNG; by contrast, the SC SAM contained no explicit mechanism for material to leave the halo (Figure \ref{fig:scsam_halo}). The TNG SAM's gas outflow rates show only mild deviations (occasionally up to $\sim 80\%$ near the lower mass edge of the halo mass range) from TNG. The metal outflow rates show a slightly wider spread near the higher mass end of the halo mass range, with deviations reaching up to $\sim80\%$ at $z=6$. This is not unexpected, since \zhot\ also deviates more strongly from TNG at $z=6$, and directly sets the mass of metals carried by halo-scale outflows.

Although the TNG SAM reproduces halo inflows and outflows less accurately than galaxy-scale flows, capturing the correct qualitative and quantitative trends still has important physical consequences. In the SC SAM, the absence of halo-scale outflows, combined with reduced inflow rates and lower ISM metallicities, produced a significantly under-enriched, less massive hot halo. In contrast, the TNG SAM allows material to both enter and exit the halo at rates comparable to TNG, and also regulates the amount of hot gas that cools into the ISM (panel \textit{vi.} in Figure \ref{fig:tngsam_gal}). Together, these effects produce a CGM that better retains both gas and metal mass (although \mhot\ remains slightly low at $z=6$), bringing the halo-scale properties into substantially closer agreement with TNG.

\section{Discussion}
\label{sec:discussion}
\subsection{Why do the SAM and TNG Agree?}
\label{subsec:tng_sam_agreement}

The TNG SAM’s ability to reproduce the median galaxy- and halo-scale properties of $\sim 20,000$ TNG galaxies mostly within $\sim 30\%$ highlights the effectiveness of recalibrating SAMs using insights from hydrodynamical simulations. Below, we examine the specific physical modifications that facilitated this agreement for the galaxy- and halo-scale properties compared. Figures \ref{fig:tngsam_variations} and \ref{fig:tngsam_variations_mdot} summarize the effect of each modification, showing the TNG SAM’s performance before and after these adjustments at $z=0$ and their impact on each of the compared properties.

\begin{figure*}
    \includegraphics[width=\linewidth]{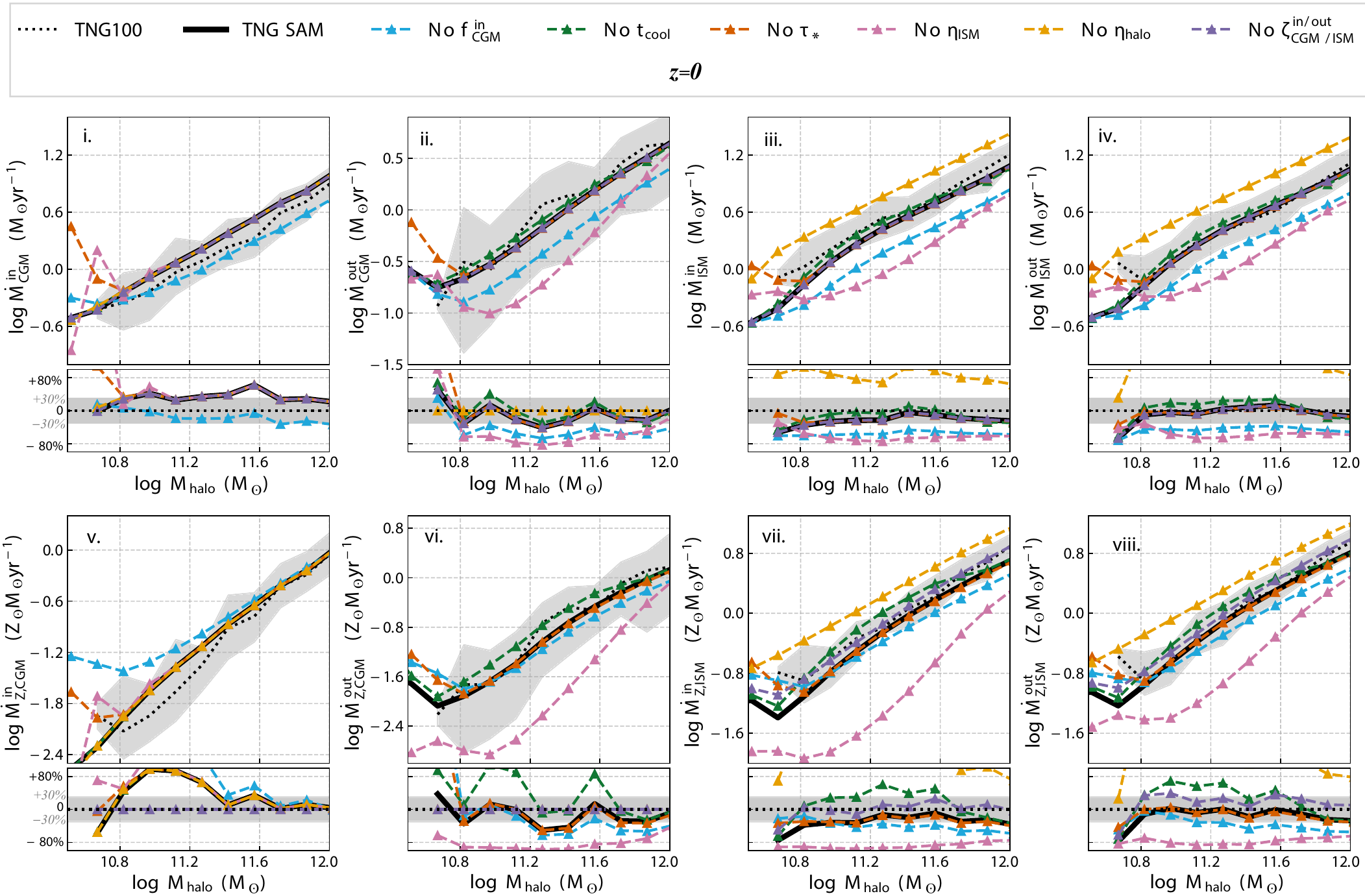}
    \caption{Continuation of Figure \ref{fig:tngsam_variations}. Visualization of how individual calibration adjustments to the TNG SAM impact the flows of gas and metals into and out of galaxies and their surrounding hot halos.}
    \label{fig:tngsam_variations_mdot}
\end{figure*}

\subsubsection{Efficient Halo (Re-)accretion: Strong Gas Recycling and Weak Preventive Feedback}
\label{subsubsec:agree_fin}

The efficiency of gas accretion into halos, regulated by \fin, plays a significant role in the TNG SAM’s ability to replicate TNG’s baryon cycle. As shown in Figure \ref{fig:tng_tuned_params}, \fin\ often exceeds unity when $\log$ \mhalo $> 10.6$, particularly at lower redshifts, indicating that gas inflow frequently meets or surpasses the expected baryon-to-dark matter ratio. In contrast, \fin\ falls below unity in some other hydrodynamical simulations like FIRE \citep{pandya_first_2020, pandya_semi-analytic_2021} and EAGLE \citep{mitchell_galactic_2020, mitchell_how_2022, wright_impact_2020}, reflecting stronger “preventive” feedback that heats the IGM and limits gas accretion onto halos. TNG’s elevated \fin\ suggests that ejected gas re-enters halos relatively rapidly at the halo mass range explored, mitigating the impact of any preventive feedback that may be present on the halo scale. We note, however, that there is evidence for preventive feedback at both lower halo masses ($\log$ \mhalo $< 10.6$) and higher halo masses ($\log$ \mhalo $> 12$), which lie outside the primary mass range considered in this work. We refer the reader to Section 5.2 of \citetalias{oren_cosmic_2025} for further discussion.

With \fin\ remaining high across all well-resolved halo masses, the question becomes: what's driving such efficient gas recycling in TNG? For low-mass halos ($10.5 < \log$ \mhalo $< 11$), stellar feedback plays the primary role in ejecting the gas that gets recycled, as shown in \citetalias{oren_cosmic_2025}. At higher masses, AGN feedback becomes increasingly important, such that the population is split roughly evenly between SN– and AGN–dominated systems, though stellar feedback remains slightly dominant. Despite this transition, \fin\ remains consistently elevated, indicating that both stellar and AGN feedback are effective at driving gas beyond the galaxy and into the CGM, but neither fully unbinds the gas from the larger potential surrounding the halo. This result is consistent with recent work arguing that much of the gas classified as ``ejected” in galaxy formation models remains gravitationally bound and therefore readily recyclable, rather than escaping permanently into the IGM \citep[e.g.,][]{voit_bound_2025}. As a result, a substantial fraction of ejected material returns to the halo over time.

In the TNG SAM, the function \fin\ replaces the model for the return of ejected gas, governed by the parameter $f_{\rm return}= 0.1$ used in the SC SAM. Most SAMs also rely on static values for gas recycling (e.g., 0.64 in GALFORM, 1.0 in \textsc{L-galaxies}), although the parameterization of the re-accretion time in terms of halo mass and dynamical time varies (see e.g., \citetalias{henriques_galaxy_2015}). However, static return fractions clearly oversimplify the dynamic accretion and recycling processes seen in hydrodynamical simulations like TNG.

When \fin\ is not included, the TNG SAM underpredicts the CGM inflow rate relative to TNG, typically lying near the lower edge of the $\sim30\%$ agreement band at $z=0$ (Figure \ref{fig:tngsam_variations_mdot}, panel~\textit{i.}) and deviating more strongly at higher redshift (not shown). This behavior likely stems from how inflows are defined and measured. The SC SAM’s $\dot{M}_{\rm CGM}^{\rm in}$ output includes only smooth accretion and the re-accretion of previously ejected gas -- it does not account for gas delivered to the CGM by merging satellites, whose hot or ejected gas reservoirs are stripped and incorporated into the central halo. In contrast, the TNG flow sample measures gas inflow directly across a thin spherical shell, capturing smooth accretion, re-accretion, and merger-driven contributions alike. As a result, the SC SAM should underpredict the CGM inflow rate relative to TNG.

While the reduction in \mdotincgm\ is modest at $z=0$ without \fin, it has significant cumulative consequences. The reduced inflow rate limits the buildup of hot gas mass by $\sim50\%$ (Figure \ref{fig:tngsam_variations}, panel \textit{ii.}), which in turn suppresses cooling into the ISM, lowering both the mass of the cold gas reservoir and the star formation rate (Figure \ref{fig:tngsam_variations}, panels \textit{iii. \& iv.}). This also propagates through the baryon cycle: reduced hot-halo mass leads to diminished cold-gas accretion (Figure \ref{fig:tngsam_variations_mdot} panel \textit{iii.}), which produces less star formation and thus lower stellar mass growth. 

Despite the effectiveness of \fin\ in the SAM, uncertainties remain regarding the precise nature of gas recycling in TNG. TNG's mesh-based approach makes it difficult to track whether accreted gas is pristine or recycled, unlike particle-based simulations like EAGLE, where gas flows can be explicitly traced. In EAGLE, for instance, \citet{mitchell_how_2022} found that the halo recycling efficiency increases monotonically with halo mass and redshift. Looking ahead, the next generation of SAMs would greatly benefit from incorporating more flexible and physically motivated recycling models that are guided by the explicit tracking of gas flows observed in hydrodynamical simulations. While the \fin\ parameter offers a useful step in replacing the ad hoc static recycling fractions used in most SAMs, further refinement—such as directly modeling the recycling efficiency as it evolves with halo mass and redshift could also be helpful. For example, based on fits to multiple hydrodynamical simulations, the SAGE SAM \citep{croton_semi-analytic_2016} dynamically adjusts the reincorporation rate so that gas is returned to the halo more efficiently in massive halos.

\subsubsection{Revised Cooling Model: Limitations of the Cold-Mode vs. Hot-Mode Dichotomy}
\label{subsubsec:agree_tcool}
\begin{figure*}
    \includegraphics[width=\linewidth]{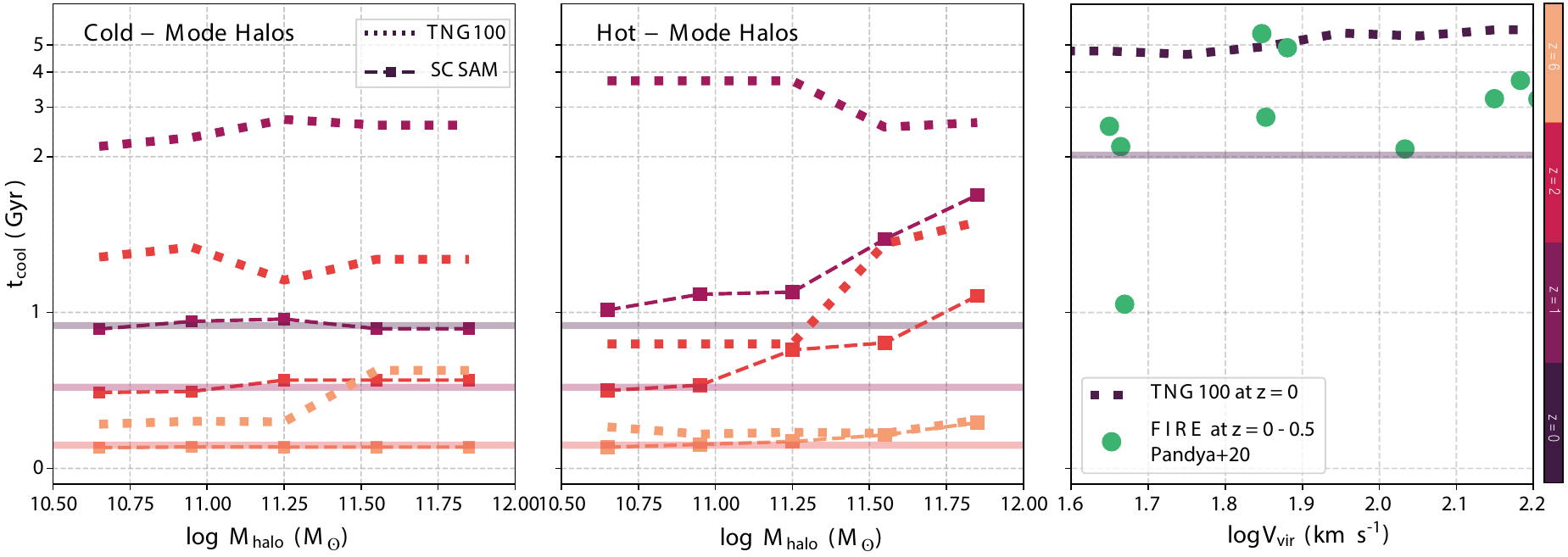}
    \caption{Cooling times (\tcool) for TNG (dotted lines) and the SC SAM (dashed squares). The left and middle panels show the time evolution of \tcool\ for cold-mode ($r_{\rm cool} > $ \rvir) and hot-mode ($r_{\rm cool} < $ \rvir) halos, respectively, at $z=1, 2, 6$. The solid lines represent the dynamical time at the given redshift. In TNG, \tcool\ generally remains above the dynamical time for both modes, highlighting the limitations of the traditional binary accretion classification in SAMs. The right panel compares \tcool\ as a function of virial velocity (\vvir) in the median TNG100 population at $z=0$, and individual FIRE halos (blue-green circles, $z=0-0.5$; Figure 4.22 in \citealt{pandya_semi-analytic_2021}). Despite differing methodologies, both TNG and FIRE predict \tcool\ values generally longer than the dynamical time (solid line).}
    \label{fig:cold_v_hot_mode}
\end{figure*}
Another significant improvement to the TNG SAM came from revising the model for how rapidly the CGM cools and accretes into the central galaxy. The overwhelming majority of SAMs published in the literature (e.g., Galacticus (\citealt{benson_galacticus_2012}), Shark (\citealt{lagos_shark_2018}), GAEA (\citealt{hirschmann_galaxy_2016}), Morgana (\citealt{monaco_morgana_2007})) classify gas accretion as either ``cold-mode” or ``hot-mode” based on the ratio of the cooling radius to the virial radius. In these models, radiative cooling is governed by well-established processes such as collisional excitation, ionization, recombination, and bremsstrahlung, with the assumption that the hot gas is shock-heated to the virial temperature of the host halo and that chemical abundances are well-mixed throughout the gas. As cooling depends on the gas density and metallicity, denser gas at the center of the halo is assumed to cool faster than outer gas, resulting in an inside-out cooling pattern. While this simplified approach captures the basic mechanisms of gas cooling, it has been shown to lead to systematic discrepancies when compared to cosmological hydrodynamical simulations, with SAMs systematically underpredicting gas accretion rates onto low-mass halos and overpredicting them for massive halos \citep{lu_algorithms_2011}.

In TNG, we find the traditional cooling model has the opposite trend for low-mass halos. The left panels of Figure \ref{fig:cold_v_hot_mode} show that cooling times in TNG often exceed the expected timescales for both cold- and hot-mode accretion.  To distinguish between cold and hot accretion modes, we compare the cooling radius \rcool — defined as the radius within which the gas cooling time equals the halo dynamical time — to the halo virial radius \rvir. If \rcool $\geq$ \rvir, the halo is classified as cold-mode dominated, implying gas can cool rapidly without shock-heating \citep{white_galaxy_1991}. Conversely, if \rcool $<$ \rvir, the halo is considered hot-mode dominated, with gas shock-heating near the virial radius and cooling inefficiently thereafter. Even in halos where the cooling radius exceeds the virial radius (the cold-mode regime), the cooling time frequently surpasses the dynamical time (dashed line). If cooling were dominated by these rapid, cold-mode flows, we would expect significantly shorter cooling times and higher accretion rates into the ISM than those found in TNG. 

However, the TNG model incorporates additional local factors—including gas density, temperature, element-based metal line cooling, and radiation from the UV background and nearby AGN—that regulate radiative cooling rates. As a result, CGM gas cools more gradually, more closely resembling the delayed cooling expected from hot-mode accretion rather than a distinct cold-mode process. Instead of two separate cooling channels, cooling in TNG seems to follow a single mode.  Although identifying the exact refinements in TNG’s radiative cooling model that lead to disagreement with SAMs is beyond the scope of this paper, such an investigation could provide valuable insights for refining SAMs in the future.

Interestingly, this behavior is seen in other hydrodynamical simulations besides TNG. The right panel of Figure \ref{fig:cold_v_hot_mode} compares cooling times in TNG and \fire, where each yellow circle represents an individual \fire\ halo taken from Figure 4.22 of \citet{pandya_semi-analytic_2021}. Both simulations yield cooling times generally longer than those expected under the classic cold-mode accretion model, even in halos where the cooling radius exceeds the virial radius. This similarity persists despite the two simulations modeling feedback very differently: FIRE uses high-resolution, explicit feedback to capture bursty, localized stellar-driven outflows for a limited number of halos, while TNG applies sub-grid prescriptions that distribute feedback energy over larger volumes across thousands of halos.  This result suggests that the traditional cooling model used in SAMs may miss key aspects of gas dynamics in realistic galaxy environments.

To address this shortcoming, we revised the TNG SAM's cooling model to align with the cooling times in TNG.  This adjustment significantly improves the SAM’s predictions of halo-scale properties compared to the traditional cooling model, as shown in panel $i.$ of Figure \ref{fig:tngsam_variations}. In the traditional model, shorter cooling times lead to faster depletion of the hot gas reservoir, reducing the total hot gas mass by $\sim 60\%$ and shifting more gas into the ISM. While the instantaneous inflow rate into the ISM appears roughly equivalent between the two models (Figure \ref{fig:tngsam_variations_mdot} panel $iii.$), this reflects the lower available hot gas mass rather than the true long-term cooling behavior. The excessive gas delivery into the ISM in the traditional model leads to an overly massive cold gas reservoir, which fuels higher star formation, yielding stellar masses that are systematically larger than those in TNG. Because more stars are formed, the metal populations across the ISM and halo also increase.

While the TNG SAM's revision preserves a more realistic hot halo, better aligning with TNG’s predictions, it is important to note that our definition of \tcool\ serves as an empirical depletion timescale for CGM-to-ISM gas transfer (see Section \ref{subsubsec:tng_igm_cgm}), rather than a physically motivated radiative cooling model. It captures the net rate at which hot gas transitions into the ISM, bundling the effects of radiative cooling, feedback, and other local processes into a single effective timescale, rather than isolating the individual contributions from each process. 

Moving forward, refining cooling models to incorporate empirical constraints and better capture radiative processes will likely require not only explicitly tracking mass flows but also modeling energy flows within SAMs. \citet{pandya_unified_2023}, \citet{carr_regulation_2023}, and \citet{voit_equilibrium_2024, voit_equilibrium_2024-1}  demonstrated that accounting for the transfer of feedback energy to the CGM using energy-tracking ordinary differential equations naturally slows cooling rates, producing timescales consistent with those observed in FIRE-2 halos. Given that both TNG and FIRE-2 halos exhibit slow cooling times, such an energy-based framework may also achieve cooling times that better align with TNG’s results.

\subsubsection{Cooling and Star Formation Efficiency Regulate ISM Agreement}
\label{subsubsec:agree_sf}
Despite the fundamental difference in spatial scales at which star formation recipes are applied in TNG—locally at the scale of $10^6$ \msun\ gas cells—and in the TNG SAM—globally across the entire galaxy—the TNG SAM's calibration to the full TNG100 sample's star formation efficiency successfully reproduces global ISM properties very well. The cold gas mass, stellar mass and star formation rate all closely follow TNG's trends within 30\%, with the cold gas mass and star formation rate performing better than the stellar mass. Panels \textit{ii–iv.} of Figure \ref{fig:tngsam_variations} show that the fiducial SAM's \cite{bigiel_star_2008}-inspired Kennicutt–Schmidt star formation recipe leads to good agreement with TNG's star formation rate, cold gas mass and stellar mass at higher halo masses ($\log$ \mhalo $> 11$). At lower halo masses, however, the agreement in the cold gas mass and star formation rate degrades, with deviations reaching up to $\sim 80\%$. At face value, the TNG SAM's agreement isn't particularly unexpected given that the SAM’s star formation rate is set directly by the star formation efficiency. However, this agreement does not imply that any globally applied star-formation law will perform equally well.

Although not shown, we find that even if the TNG SAM applies a global Kennicutt-Schmidt star formation recipe calibrated to match the global star formation rate density vs gas density in TNG galaxies, the SAM still struggles to reproduce the correct star formation rate, stellar mass, and cold gas mass within 30\% across cosmic time. This mismatch is mostly driven by the discrepancy between disk sizes predicted by the SAM and those produced by TNG. Most SAMs, including the SC SAM, determine disk sizes through angular momentum conservation under the assumption of an exponential disk. While this method achieves reasonably good agreement with observed radial disk sizes as a function of stellar mass up to $z \sim 2$ in the SC SAM \citep{somerville_explanation_2008}, it does not align with observations as well as TNG's predictions \citep{genel_size_2018}. Thus, improving disk size models is a necessary prerequisite for reliably applying surface-density–based star-formation laws in SAMs.

Given the current uncertainties in predicting disk sizes in SAMs, tuning the star formation efficiency provides a more robust and flexible way to regulate star formation in the TNG SAM. However, this does not mean that the good agreement between the TNG SAM and TNG is purely a consequence of calibrating $\tau_*$. The TNG SAM succeeds because it gets both components of the baryon cycle broadly right: (i) realistic cooling times deliver the correct amount of gas to the ISM (Section \ref{subsubsec:agree_tcool}), and (ii) $\tau_*$ regulates how efficiently that gas forms stars. When cooling proceeds too efficiently (as in the traditional model), the extra cold gas fed to the ISM leads to more star formation (Figure \ref{fig:tngsam_variations} panel ii.) even though $\tau_*$ is tuned.

\subsubsection{ISM and Halo-Scale Outflows Modulate the Stellar Mass and the Mass of the Hot Halo}
\label{subsubsec:agree_outflows}

While galaxy-scale outflows are well-established as important regulators of galaxy evolution in numerical simulations, recent studies have underscored the importance of outflows at the halo scale as well \citep{wright_baryon_2024, pandya_characterizing_2021, mitchell_galactic_2020}. Consistent with these findings, we observe that both galaxy and halo-scale outflows are essential for accurately modeling the gas content in the TNG SAM.

Incorporating TNG’s metallicity-dependent wind model into the TNG SAM significantly improved agreement with TNG’s outflow rates. Without this modification, the SC SAM’s prescription underestimates ISM outflow rates up to $\sim80\%$ across the halo masses explored (Figure \ref{fig:tngsam_variations_mdot}, panel $iv.$). Because this gas is immediately ejected rather than cycled through the CGM, the SC SAM fails to build up a more massive CGM reservoir (Figure \ref{fig:tngsam_variations}, panel $i.$). As a result, the SC SAM predicts lower inflow rates into the ISM (Figure \ref{fig:tngsam_variations_mdot}, panel $iii.$), suppressed cold gas masses, reduced star formation, and ultimately lower stellar masses across the resolved mass range (Figure \ref{fig:tngsam_variations}, panel $ii-iv.$). In contrast, the TNG SAM’s scaling relations matches the ISM outflow rate within 30\%, yielding values that closely track those found in TNG.

Although not obvious in Figure \ref{fig:tngsam_variations} panel $ii.$, we also find that correctly extrapolating $\eta_{\rm ISM}$ at lower halo masses plays an important role in predicting the stellar mass. If $\eta_{\rm ISM}$ is too low in lower-mass halos, the stellar mass exceeds TNG's predictions by nearly 40\% because more cold gas remains available for star formation. The difference relative to the SC SAM is not immediately obvious because the SC SAM expels even more gas from low-mass halos, with $\eta_{\rm ISM}$ exceeding $1000$ in $10^8 \ M_\odot$ halos at \zzero. Whether such values are realistic remains a subject of debate, but in practice the TNG-calibrated scaling relations keep $\eta_{\rm ISM}$ near $\sim100$ in low-mass halos, rather than driving it to the high levels applied in the SC SAM.

A critical update to the TNG SAM’s feedback model was the explicit inclusion of halo-scale outflows.  Many traditional SAMs \citep[e.g.,][]{croton_many_2006, croton_semi-analytic_2016, lacey_unified_2016, hirschmann_galaxy_2016}, including the SC SAM, assume that gas expelled from the ISM is transferred directly to an ejected reservoir/IGM without first passing through the hot halo. Some more recent SAMs \citep[e.g.,][]{henriques_galaxy_2015, benson_galacticus_2012, lagos_shark_2018} explicitly model gas cycling through the CGM before ejection. Similarly, in the TNG SAM, gas first leaves the ISM and enters the CGM, where it can later be expelled into the ejected reservoir/IGM, forming the CGM-to-IGM outflow channel. This two-step process provides a more realistic depiction of how gas is cycled and expelled, as opposed to the simplified ISM-to-ejected reservoir flow used in the SC SAM. 

When halo-scale outflows are neglected, the hot halo grows far too massive and enriched: \mhot\ and \zhot\ exceed TNG by factors of $2-3$ and 40\%, respectively (Figure \ref{fig:tngsam_variations}, panels \textit{i \& v.}). This excess mass then cools too efficiently into the ISM, driving cold gas masses, SFRs, and stellar masses to overshoot TNG's predictions by factors of $2-3$ (panels \textit{ii–iv.} in Figure \ref{fig:tngsam_variations}).

\subsubsection{Metal Cycling Efficiencies Improve Metallicity Predictions}
\label{subsubsec:agree_zetas}
Incorporating metallicity-weighted mass-loading factors greatly enhanced the TNG SAM’s ability to replicate the evolution of the metallicities found in TNG100. If we instead retain the SC SAM's original assumptions, namely that the proportion of metals flowing into and out of the ISM is unity (i.e., \zetaism$ = 1$), and that there is no explicit prescription for the proportionality of metals entering or leaving the CGM (i.e., \zetahalo $=0$), the TNG SAM underpredicts the cold gas and stellar metallicities by up to $\sim 80\%$, while overpredicting the hot gas metallicity by more than $\sim 80\%$, as shown in Figure \ref{fig:tngsam_variations} panels \textit{v-vii}. The excess in $Z_{\rm hot}$ arises because metals expelled from the ISM into the CGM are retained in the halo due to the absence of a halo-scale metal outflow channel, leading to a buildup of metals in the hot halo. We note that this comparison is not strictly one-to-one with the SC SAM. In the SC SAM, metals entering the CGM are implicitly tied to the return fraction of gas ejected from the halo, whereas setting \zetahalo$=0$ in the TNG SAM removes the halo-scale metal inflow and outflow channels without reinstating the associated reaccretion pathway.

Most SAMs in the literature use instantaneous recycling models for metal production, although some have incorporated more sophisticated multi-element galactic chemical evolution models that relax the instantaneous recycling approximation \citep[e.g.,][]{arrigoni_galactic_2010, yates_relation_2012, kobayashi_simulations_2007, hirschmann_galaxy_2012}. However, nearly all SAMs, including the Santa Cruz SAM, also assume that metals are exchanged between baryonic reservoirs in direct proportion to the exchanged gas mass, without accounting for metal-enhanced or metal-depleted inflows and outflows. 
A few SAMs, such as \textsc{SAG} \citep{collacchioni_semi-analytic_2018} and GALFORM, have experimented with incorporating metal enrichment factors. However, these models found little impact on reproducing the observed stellar mass-metallicity relation, perhaps because their supernova feedback models are not tied to the metallicity of the gas. \citetalias{oren_cosmic_2025} and our work show that both inflows and outflows can be either metal-enhanced or metal-depleted relative to the reservoir of origin, and these metal enrichment factors can have a complex dependence on various halo properties. Incorporating these metal enrichment factors into future SAMs will be critical for accurately modeling how metals cycle through galaxies and their halos. 

\begin{figure*}

    \includegraphics[width=\linewidth]{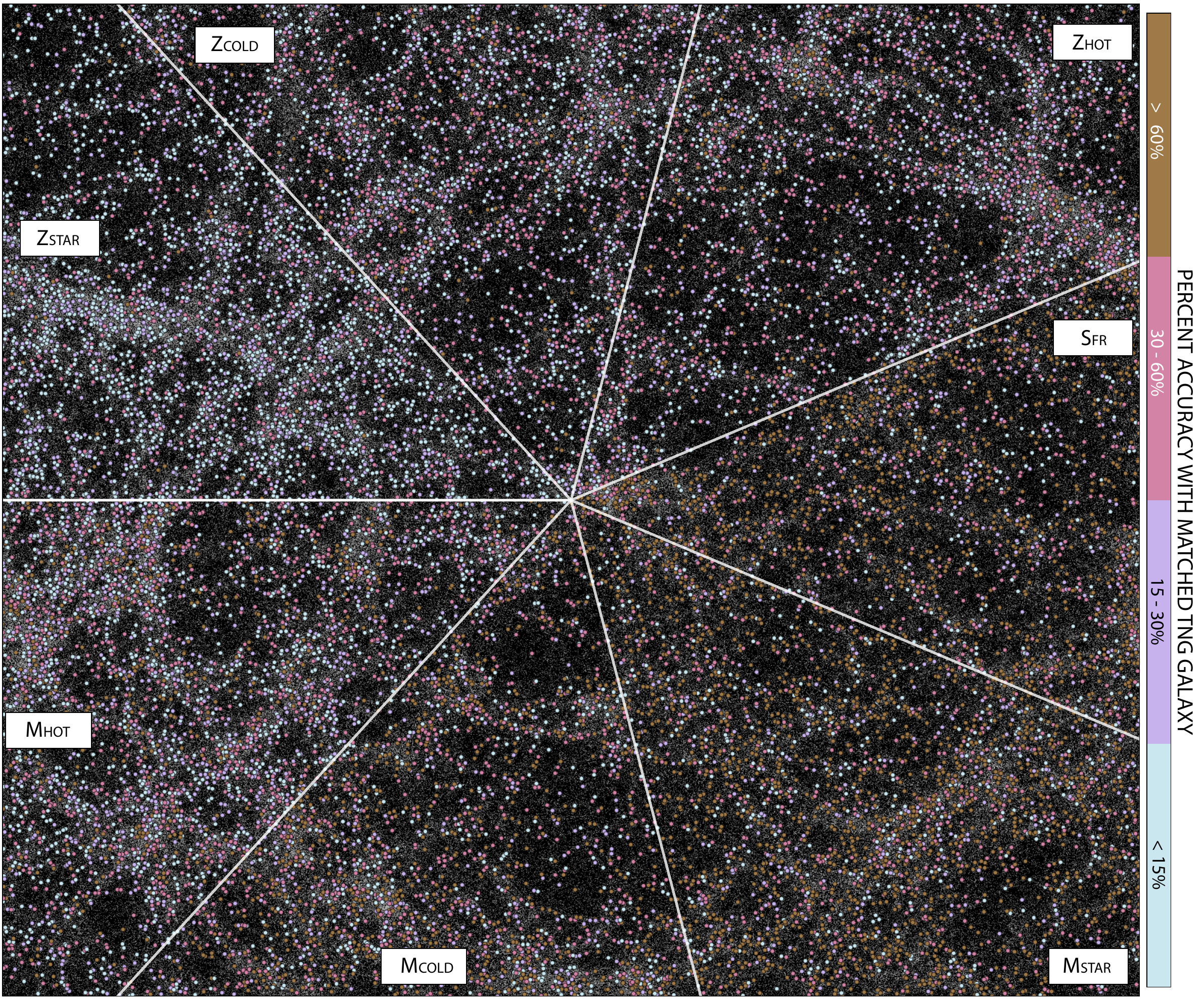}
    \caption{Spatial distribution of matched halos between TNG100 and the TNG SAM at \zzero, plotted in physical space (x vs. y position). All TNG100 halos with $10.6 < \log M_{\rm halo} < 12$ are shown in white. The map is divided into seven sectors corresponding to a specific galaxy property: hot gas mass, cold gas mass, stellar mass, star formation rate, and metallicities of hot gas, cold gas, and stars. Each matched galaxy is colored by the percent difference between the TNG SAM and TNG100 predictions for the given property in each sector. Blue and purple points represent agreement within 30\%, while pink and brown points highlight discrepancies greater than 30\%. While the majority of galaxies agree within 30\% for the metal populations and mass of the hot gas, notable variance is observed for galaxy-scale properties like \mstar\ and \mcold, illustrating the challenges of calibrating to median trends alone.}
    \label{fig:tngsam_sg}
\end{figure*}

\subsection{Limitations and Future Directions}
\label{subsec:limits}

Although Section \ref{sec:tng_sam_results} shows that the median galaxy population produced by the TNG SAM agrees quite well with TNG given the physical updates made, there are still several limitations to our approach, discussed below. 

\subsubsection{Modeling Individual Galaxies with Median-Based Calibrations}
\label{subsec:tng_sam_single_gal}

The TNG SAM is calibrated using median relations that reflect the overall behavior of galaxies within TNG rather than tailoring the model to fit individual galaxies or halos. To evaluate how well the TNG SAM's median-based calibrations replicate the detailed growth histories of individual TNG galaxies, we directly matched subhalos between TNG and the SAM to perform an object-by-object comparison. Although the TNG SAM’s median galaxy population is mostly within 30\% of TNG's medians, Figure \ref{fig:tngsam_sg} shows that the spread increases significantly when comparing individual galaxies. The TNG SAM still demonstrates generally good agreement for several key properties, namely \mhot\ and all metal populations, in which the majority of all matched galaxies agree within 30\% of TNG. The agreement is less consistent for \mstar, \mcold\ and the star formation rate. Here, the TNG SAM predicts each property within 30\% accuracy for only 30-40\% of galaxies.

The disagreement among the gas properties at the galaxy scale likely stems from mismatches in how scatter arises in the star formation rate. In TNG, galaxy–to–galaxy variation in SFR and cold–gas mass reflects both local physical processes and gas cycling in the ISM. In the TNG SAM, the global star–formation efficiency regulates the mean gas consumption rate successfully, but does not reproduce the full spread in individual systems. In addition, the cold gas mass used to form stars in the SAM corresponds to gas within twice the stellar half mass radius, whereas in TNG star formation can proceed from gas that was, or is, located outside this aperture. As a result, the SAM may assign slightly too much or too little cold gas to star formation in individual galaxies, leading to deviations in scatter even when median trends agree well.

Despite this limitation, the results are encouraging. Nearly half of the 20,000 matched galaxies agree with TNG's predictions within 30\%, even though the SAM was primarily calibrated using medians derived from a much smaller resolved sample of just \(\sim 400\) galaxies. This level of agreement is particularly promising given the computational simplicity of the SAM compared to the detailed hydrodynamical simulation. Extending the work presented here will provide a foundation for developing the next generation of flexible, computationally efficient, yet physically detailed SAMs needed to model galaxy formation across wider cosmological volumes.

\subsubsection{TNG100 DMO vs Hydrodynamical Simulations}
\label{subsubsec:fp_vs_dm}

A significant limitation to our approach arises from running the TNG SAM on the TNG100-1-DMO simulation while comparing results to the hydrodynamical TNG100-1 simulations. As noted in \citet{gabrielpillai_galaxy_2022}, the halo masses in the TNG100 DMO and TNG100 FP simulations differ by up to 20\% for low halo masses at $z=0$. This discrepancy stems from the inclusion of baryonic processes in the hydrodynamical simulation, and varies across redshift. 

This variation has significant implications for the TNG SAM, particularly in how the model calculates the rate of gas accretion. The SAM derives the rate of gas accretion by finite differencing the halo masses reported by the DMO simulation. The mismatch between halo masses in the N-body and hydrodynamical simulations affects how well we can model parameters like \fin, which governs the growth of the hot halo, which in turn regulates the build up of cold gas and stars in the ISM. Due to this discrepancy, directly tuning \fin\ in the SAM to match the hydrodynamical simulation was not feasible. Instead, we inferred the value of \fin\ by aligning the rate of gas entering the CGM with TNG, which performed remarkably well (see Section \ref{subsubsec:agree_fin}).

\subsubsection{Sample Size and Resolution}
\label{subsubsec:resolution}

The TNG SAM's ability to reproduce TNG100's results is further constrained by the halo mass and temporal resolution of the calibration sample.  We consider halos well-resolved if they contain more than 100 star particles ($\log M_{\rm halo} \gtrsim 10.6$); however, the baryon flow measurements used to calibrate TNG SAM are extracted from radial shells 0.1\rvir\ thick within each halo. For lower-mass halos, particularly those around $10^{10}\ M_{\odot}$, these shells often contain far fewer resolution elements, producing noisy estimates and contributing to discrepancies in the SAM’s predictions at the low-mass end, where deviations are most pronounced. Moreover, the $10^{10}$ \msun\ halos modeled at \zzero\ evolved from progenitors with halo masses of $\sim 10^8$ \msun\ at \zsix, which contain poorly resolved stellar and gas components. 

These resolution limits precluded us from tracking the baryon cycle's time evolution in a select group of galaxies selected at \zzero. Instead, we had to rely on aggregate measurements at each redshift, which introduces additional uncertainties for low-mass systems. To account for unresolved galaxies below the well-resolved mass range, we extrapolated their behavior using baryon flow trends observed for halos in the range $10 < \log M_{\rm halo} < 10.5$. While this approach introduces uncertainty—since it is unclear whether these trends hold for $\log M_{\rm halo} < 10$—it plays a critical role in achieving the TNG SAM’s overall 30\% agreement with TNG for higher-mass halos. This was particularly important for \mstar, which exceeded TNG's distribution by 40\% in halos with $\log M_{\rm halo} < 11$ if \etaism\ was too low for $\log M_{\rm halo} < 10.5$ halos. However, this extrapolation also introduces unquantifiable uncertainty, raising questions about the robustness of the TNG SAM's overall 30\% agreement with TNG.

\begin{figure*}
    \centering
    \includegraphics[width=\linewidth]{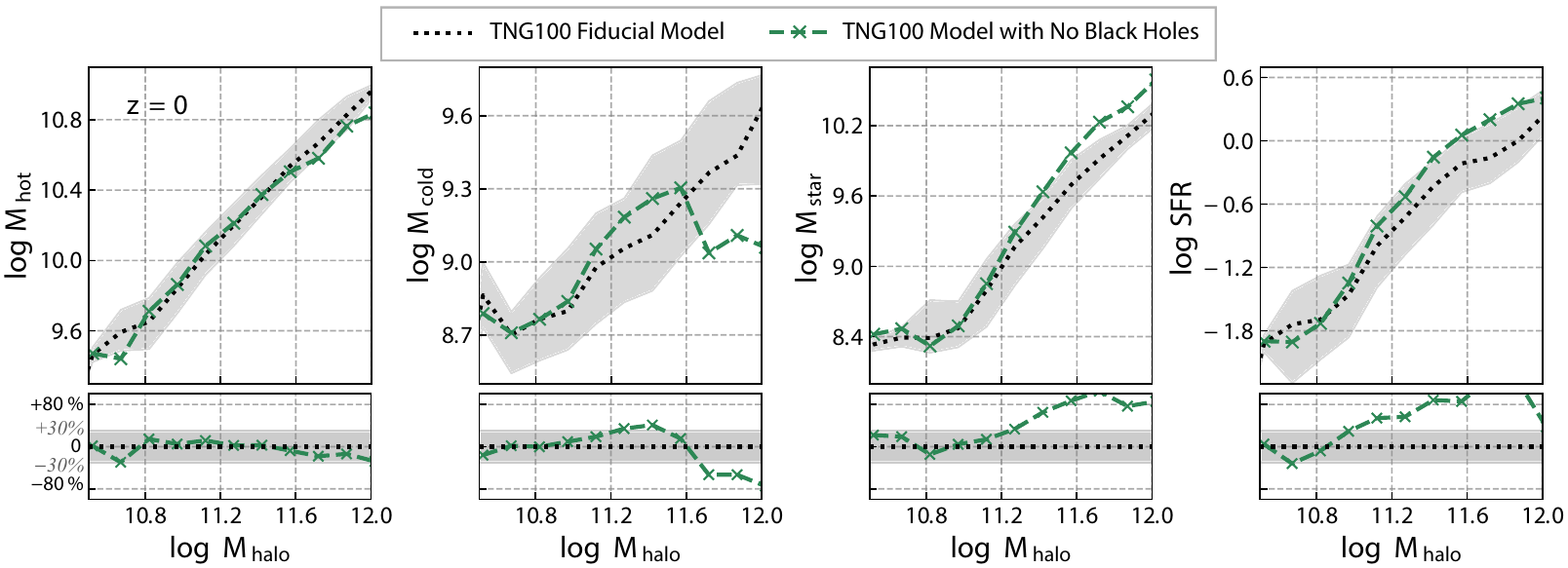}
    \caption{Comparison of \mhot, \mcold, \mstar and SFR between the fiducial TNG100 model implemented in a $36.9^3$ comoving Mpc$^3$ volume (black dotted lines) and an otherwise identical variant that excludes black holes (green dashed crosses). At lower halo masses ($\log$ \mhalo $< 11.2$), the two models generally agree within $\pm 30\%$. At higher halo masses ($\log$ \mhalo $\gtrsim 11.2$), galaxy-scale properties differ substantially by up to a factor 2, coincident with the regime where black hole feedback becomes increasingly important in the fiducial TNG model.}
    \label{fig:tngnobh}
\end{figure*}

These resolution requirements also introduce a selection bias in the calibration sample. By construction, the TNG galaxies used to constrain the baryon cycle are limited to systems with well-resolved stellar and gas components, and are therefore biased toward more gas-rich, actively star-forming systems. Consequently, the sample does not represent the full galaxy population, preventing robust interpretations of population-level statistics like the quenched galaxy fraction or cosmic star formation rate density. Applying this approach to the higher-resolution TNG50 (our future goal; see Section \ref{subsubsec:future}) will reduce this bias by incorporating a broader range of galaxies, including lower-mass and weakly star-forming systems.

Finally, the limited size of the calibration sample used to calibrate the TNG SAM naturally raises two questions: (1) could the TNG SAM’s accuracy improve with a larger calibration set, and (2) how many galaxies are required to maintain reliable predictions? Although the model already reaches $\sim 30\%$ accuracy using only $\sim 400$ resolved TNG galaxies, expanding the calibration sample is an obvious path toward reducing noise, particularly at the low-mass end where the measurements are most uncertain. Since the extraction of baryon flow quantities is a one-time task and not computationally prohibitive, scaling to larger samples is entirely feasible and a worthwhile direction for future work, which we discuss further in Section \ref{subsubsec:future}.

Interestingly, we find that the TNG SAM maintains agreement with TNG at the galaxy scale to within $\sim40\%$ even when calibrated using only 100 randomly selected galaxies per redshift. Across redshift, aggregate galaxy- and halo-scale properties are generally reproduced at this level, although the cold gas mass shows the weakest agreement, with deviations reaching $\sim80\%$ in some regimes (see Figures \ref{fig:app_tngsam_100gal} and \ref{fig:app_tngsam100_halo}). The baryon flow rates are less tightly matched: gas and metal inflow and outflow rates are typically reproduced to within $\sim40\%$, while the rates of gas and metals leaving the halo show the largest discrepancies, with divergence up to $\sim80\%$. These results suggest that a larger sample size would likely improve the accuracy currently achieved by the TNG SAM. They also show that reproducing the correct evolutionary behavior of gas and metal cycling, rather than matching instantaneous flow rates as closely as possible, is sufficient to achieve good agreement with TNG’s global properties.

These results are particularly promising for future efforts, such as those by the SMAUG\footnote{\url{https://www.simonsfoundation.org/flatiron/center-for-computational-astrophysics/galaxy-formation/smaug/}} and Learning The Universe\footnote{\url{https://learning-the-universe.org/}} collaborations, to develop next-generation SAMs and hydrodynamical simulations using detailed physical insights from zoom-in simulations and observations. They demonstrate that the quality of the calibration sample—its ability to represent key physical processes—can be as impactful as computationally expensive increases in sample size. This is also promising for multi-wavelength observational surveys, which are often constrained by the difficulty in securing enough observing time across numerous telescope time allocation committees. For example, large surveys like CALIFA \citep{sanchez_califa_2012} comprise around 600 galaxies, while highly detailed surveys such as PHANGS feature $\sim 90$ galaxies \citep{lee_phangs-hst_2022}. The TNG SAM’s success with just 100 well-sampled galaxies per snapshot suggests that small but carefully selected observational datasets could also help to refine feedback prescriptions in numerical simulations. While observational data inevitably include greater uncertainties than simulations, high-quality, representative datasets can still aid in improving our understanding of the baryon cycle and galaxy evolution across cosmic time. 

\subsubsection{AGN Feedback In Low Mass Halos}
\label{subsec:tng_sam_agn}

Given the halo mass range explored in this work, we disable AGN feedback in the SAM. However, as discussed in Section \ref{subsubsec:agree_fin}, AGN feedback is not entirely absent in TNG at these masses. Although stellar feedback supplies most of the feedback energy at the low-mass end of the sample, the relative contribution from AGN increases with halo mass, leading to a regime in which stellar- and AGN-driven feedback contribute comparably, with stellar feedback remaining marginally dominant overall. This raises the question of whether some of the remaining discrepancies between the TNG SAM and TNG, particularly those confined to the $\pm30\%$ level, could be attributable to omitting explicitly modeling AGN feedback in the TNG SAM.

In addition to TNG100, the IllustrisTNG project includes a suite of smaller-volume simulations with identical initial conditions and resolution that vary individual components of the subgrid physics, enabling controlled, galaxy-by-galaxy comparisons of different variations to the fiducial model \citep{pillepich_first_2018}. Comparisons across different implementations of black hole feedback prescriptions have been carried out, focusing on more massive halos \citep[e.g.,][]{terrazas_relationship_2020}, but not on the lower-mass halo population studied in this work. 

To assess the impact of removing black hole feedback on TNG100's results, and by extension, our TNG SAM's results, Figure \ref{fig:tngnobh} compares the fiducial TNG model in the 36.9$^3$ Mpc$^3$ box to an otherwise identical run in which black holes are entirely removed. While \mhot\ remains consistent between the two models to within $\sim 30\%$, \mcold\ shows wider deviations up to $\sim 80\%$, with even larger differences in stellar mass and star formation rate. The differences mostly begin to exceed 30\% when $\log$ \mhalo $\gtrsim 11.2$, coincident with the halo mass at which energy from black hole feedback begins to make more significant contributions to total feedback energy.

The TNG SAM’s agreement with TNG at the $\sim30\%$ level likely arises because the influence of black hole feedback is implicitly absorbed into the calibrated scaling relations used to build the model (Section \ref{sec:creating_tng_sam}). This is most evident in Figure \ref{fig:tngsam_stellar_feedback}, where the ratio of the mass loading at the galaxy scale to that at the wind launch scale increases sharply when $\log$ \mhalo $\gtrsim 11.2$, plausibly reflecting the growing influence of AGN-driven feedback. This means that the TNG SAM's success does not depend on explicitly modeling all possible feedback channels, but rather on capturing the net balance of gas inflows, outflows, and recycling that regulates galaxy growth.

It is also important to note that when $\log$ \mhalo $< 11.2$, the two models still continue to differ at the 30\% level, even though black hole feedback is far more negligible in this regime. This sensitivity is consistent with the “butterfly effect” identified in cosmological simulations, where minute perturbations or modest changes to subgrid physics lead to $\sim5 - 25\%$ variations in individual galaxy properties over cosmic time \citep{genel_quantification_2019}. In this context, the residual differences between the TNG SAM and TNG, particularly when confined to the $\sim30\%$ level, are not unexpected. The observed agreement floor likely reflects a combination of factors from not explicitly modeling black hole feedback, the butterfly effect, and the limited size of the calibration sample.

\subsubsection{Future Directions}
\label{subsubsec:future}

This work set out to (1) directly compare the outputs between SAMs and hydrodynamical simulations, (2) treat the baryon cycling behavior of TNG into a more interpretable semi-analytic framework, (3) develop a framework for efficiently exploring alternative subgrid physics models in hydrodynamical simulations, and (4) ultimately extend the predictive power of SAMs to volumes beyond the reach of current hydrodynamical simulations. The TNG SAM's results clearly achieve the first two goals. However, further development will be required before the third and fourth goals can be fully realized.

A key next step is to test the TNG SAM across a wider range of the TNG suite. Although TNG100 was the more appropriate choice for this study given its use in the calibration of the fiducial TNG galaxy formation model and the known resolution-dependent discrepancies in TNG \citep{pillepich_simulating_2018}, expanding the analysis to galaxies in the TNG50 simulations, which resolves down to $\sim 10^8$ \msun will probe a broader population of stellar-feedback–dominated systems and clarify how reliable the model’s current low-mass extrapolations are. Applying the updated framework to the larger volumes of TNG300 and MillenniumTNG \citep{pakmor_millenniumtng_2023} will further enable us to assess convergence across the TNG suite and identify which aspects of the baryon cycle are most sensitive to resolution and simulation volume.

Another important direction is to refine how feedback is represented. The current TNG SAM captures the net impact of feedback through aggregate mass-flow scalings, but future work will aim to more explicitly separate the contributions from stellar and black hole feedback. While traditional SAMs, including this implementation, track mass flows, several recent studies have argued for explicitly following energy flows as well \citep{pandya_unified_2023, carr_regulation_2023, voit_equilibrium_2024, voit_equilibrium_2024-1}. A TNG SAM that incorporates an energy flow model that accounts for both energy sources (e.g., stellar and AGN feedback) and sinks (e.g., radiative cooling and turbulent dissipation) would provide a more physically complete description of how feedback regulates gas cycling, gas cooling, and star formation. \citetalias{oren_cosmic_2025} has already separated the energy contributions from stellar feedback and AGN, distinguishing between thermal and kinetic modes, demonstrating that this kind of implementation is feasible within the TNG framework. Embedding that structure within the SAM would enable direct comparisons to TNG runs in which stellar or black hole feedback prescriptions are altered, allowing us to isolate how each channel affects cooling, recycling, and galaxy growth. 

In the longer term, these refinements may allow feedback prescriptions explored in small-volume simulation variants to be implemented and tested within the SAM before being deployed in full-volume hydrodynamical runs. Furthermore, because the SAM can reproduce TNG’s galaxy populations and baryon cycling behavior when calibrated to only ~100 galaxies (see Appendix \ref{subapp:100gals}), it can be constrained by datasets that are inherently limited in size, such as high-resolution zoom simulations, which are computationally expensive, or detailed multiwavelength observational samples, which are time-intensive to assemble. Ultimately, this flexibility will enable efficient exploration of diverse galaxy-formation models across the vast survey volumes that will be mapped by the Vera C. Rubin Observatory Legacy Survey of Space and Time \citep{ivezic_lsst_2019} and the Nancy Grace Roman Space Telescope.\footnote{\url{https://roman.gsfc.nasa.gov/}}

\subsection{Comparison to Similar Work}
\label{disc:comparison_other_work}

Previous attempts to align the predictions of SAMs with hydrodynamical simulations have varied in methodology, scope, and success. Here, we compare our approach and results with those of \citet{stringer_analytic_2010}, \citet{neistein_hydrodynamical_2012} and \citet{mitchell_how_2022}, highlighting key differences and similarities, and the relative accuracy of each approach.

Early efforts to compare the results from SAMs and hydrodynamical simulations (e.g., \cite{helly_comparison_2003}, \cite{benson_comparison_2001}, and \cite{yoshida_gas_2002})  found that gas cooling rates from the hot halo into the ISM could be brought into good agreement when SAMs were modestly adjusted. For example, \citet{helly_comparison_2003} showed that reducing the core radius of the gas density profile and increasing cooling times in low-mass halos allowed semi-analytic models to match SPH simulations within 25–50\% across halo masses. \cite{stringer_analytic_2010} expanded this approach by attempting to reproduce the entire formation history of a single disk galaxy within the GASOLINE hydrodynamical simulation using a modified version of the GALFORM SAM. To achieve broad agreement with the simulation data, they adjusted parameters related to gas cooling, star formation, and feedback, demonstrating the potential of SAMs to reproduce the detailed evolution of individual galaxies. However, the focus on a single object left uncertainties about the general applicability of their findings.  

\cite{neistein_hydrodynamical_2012} (hereafter \citetalias{neistein_hydrodynamical_2012}) expanded on this to a wider range of galaxies by extracting efficiencies describing accretion, cooling, star formation, and feedback from the OverWhelmingly Large Simulations (OWLS) and applying them within the SAM presented in \cite{neistein_degeneracy_2010} to reproduce OWLS's results. In terms of approach and scope, \citetalias{neistein_hydrodynamical_2012}  is most similar to our work here with the TNG SAM. \citetalias{neistein_hydrodynamical_2012} also utilizes a set of efficiencies to describe the rate at which gas enters the halo, cools from the hot halo to the galaxy, forms stars, and is ejected by stellar feedback. For instance, the halo gas accretion efficiency parameter $f_a =  \dot{M}_{\rm CGM}^{\rm in} / \dot{M}_{\rm halo}$ in \citetalias{neistein_hydrodynamical_2012} is similar to the TNG SAM's \fin, except \fin\ adjusts for the baryon fraction and includes both first-time infall and gas recycling. Despite this nuance, both models similarly use their respective parameters to match the gas accretion rate found in their hydrodynamical simulations.

Overall, \citetalias{neistein_hydrodynamical_2012} used less detailed efficiencies to calibrate their SAM against OWLS, reflecting their primary focus on reproducing the hot gas mass, cold gas mass, stellar mass, and total galaxy mass. Their approach yielded very good agreement for these quantities, reporting a standard deviation ranging from 0.1 - 0.2 dex for each mass reservoir. However, SFR deviations were larger, with a standard deviation of 0.5 dex. The TNG SAM, despite aiming to reproduce a wider range of global galaxy and halo properties and extending to higher redshifts, achieves comparable success within 0.1-0.3 dex for almost all galaxy and halo scale properties explored, including the SFR.

Here, we note a broader issue that affects most comparisons between SAMs and hydrodynamical simulations: key galaxy- and halo-scale properties are not always defined in a consistent way. In the TNG SAM, we define \mstar\ as all stars bound to the subhalo, and \mcold\ within twice the stellar half-mass radius. Other works have adopted different choices. For example, \citetalias{neistein_hydrodynamical_2012} also define \mstar\ in their hydrodynamical comparison as the total stellar mass bound to the subhalo, but they define \mcold\ as the mass of all gas particles eligible to form stars and \mhot\ as all remaining gas, without clearly stating whether gas associated with satellite galaxies is also excluded. Such definitional choices directly affect derived properties, particularly \mhot, since what counts as “hot” depends directly on how “cold” is defined. This raises an important question: to what extent do reported levels of agreement in earlier works depend on their specific definitions, and how might their conclusions change under a definition consistent with ours (or vice versa)?

Despite these uncertainties, even comparisons with inconsistent definitions can provide valuable insights into the baryon cycle.  Recently, \citet[][hereafter MS22]{mitchell_how_2022} expanded on the \citetalias{neistein_hydrodynamical_2012} framework to emulate the results of the \textsc{EAGLE} simulations and investigate the parameters most critical to reproducing the stellar mass–halo mass (SHM) relation using a gas regulator model, which is a simplified version of a SAM. They found that EAGLE's SHM relation is primarily shaped by gas ejection via outflows for halos with \mvir $< 10^{12}$ \msun, with halo-scale preventative feedback and recycling of ejected gas playing secondary roles. They also found that the redshift evolution of the SHM relation is most sensitive to the efficiencies of first-time gas accretion and ejection by outflows, and is less sensitive to the efficiency of wind recycling, and of gas consumption by star formation. In the TNG SAM, we find similar trends, with Figure \ref{fig:tngsam_variations} panel \textit{ii.} showing that the stellar mass is most sensitive to \etahalo\ and \etaism.

\citetalias{mitchell_how_2022} also investigated how star formation and gas flows affect the relationship between halo mass and the masses of both the ISM and CGM. They found that the CGM mass is most sensitive to variations in gas inflows and outflows, particularly those occurring at the halo scale. This finding is echoed in our analysis of the TNG SAM, where we also observed a strong dependence of the CGM mass in TNG on halo-scale gas flows. \citetalias{mitchell_how_2022} also found that the ISM mass is more sensitive to halo-scale flows, whereas the stellar mass is more sensitive to galaxy-scale flows. Similarly, the TNG SAM shows that \etahalo\ significantly impacts both \mcold\ and \mstar, but with \mcold\ more impacted than \mstar. 

A significant difference between the TNG SAM and the above studies is its explicit tracking of metals across different mass reservoirs, including stars, cold gas, and hot gas, as well as the flows between them. It remains unclear whether the other approaches discussed above could replicate metal content as effectively as the TNG SAM. However, given our success using the metal enrichment factors, it is reasonable to assume that incorporating similar prescriptions in those models could yield comparable improvements.

\section{Summary and Conclusions}
\label{sec:summary}

In this paper, we introduced the TNG SAM, a modified version of the Santa Cruz semi-analytic model, designed to replicate the complex baryon cycle of galaxies in the IllustrisTNG cosmological hydrodynamical simulation. The TNG SAM bridges the detailed physical processes captured in hydrodynamical simulations with the computational efficiency of SAMs, offering a powerful tool to study the baryonic processes that shape galaxies.

Focusing on stellar feedback-dominated systems, we aimed to reproduce the baryon cycle in low- to intermediate-mass dwarf galaxies (\mhalo $\sim 10^{10} - 10^{11}$\msun) and Milky Way-mass galaxies (\mhalo $\sim 10^{12}$\msun) in TNG100. Using measurements of gas flows from a subset of $\sim 400$ central galaxies as a proxy for the larger TNG100 sample, we updated the SC SAM’s physical prescriptions for halo gas accretion, cooling, stellar feedback, and metal circulation. These updates, implemented as a function of halo mass and redshift, led to more accurate predictions of galaxy-scale properties such as stellar mass, cold gas content, and metallicity, as well as halo-scale properties like the overall baryon content. In doing so, the TNG SAM provides a framework for translating the complex behavior modeled in hydrodynamical simulations into a flexible semi-analytic form that can, in future work, be extended across the full TNG suite, used to test alternative feedback prescriptions explored in smaller simulation volumes, and eventually applied to larger cosmological volumes to generate predictions for upcoming wide-volume surveys.

Several key insights about the baryon cycle in TNG emerged from this work, with important implications to keep in mind when building the next generation of SAMs:

\begin{enumerate}

   \item \textit{The balance between gas recycling and preventative feedback varies across hydrodynamical simulations, requiring flexible (re-)accretion models in SAMs.} In TNG, the halo-scale gas inflow efficiency (\fin) frequently exceeds unity. This indicates that gas recycling, where previously ejected material re-enters the halo, plays a dominant role in setting the total inflow rate. This behavior contrasts with simulations like FIRE and EAGLE, which show lower \fin\ values due to stronger preventive feedback, defined here as processes that suppress gas accretion into halos. The majority of SAMs rely on fixed or static gas return fractions and therefore do not capture the dynamic recycling efficiencies seen in hydrodynamical simulations. For future SAMs, incorporating flexible (re-)accretion models that vary with halo mass and redshift such as the TNG SAM’s \fin\ will be critical for accurately predicting the baryon content of galaxies (Section \ref{subsubsec:agree_fin}, Figure \ref{fig:tngsam_variations}).
    
    \item \textit{The classic hot mode/cold mode cooling model used in nearly all SAMs does not provide a good description of cooling and halo gas accretion in TNG.} In TNG, cooling times often exceed dynamical timescales, a behavior that has also been observed in FIRE. This indicates that the simple cold-mode vs. hot-mode framework commonly used in SAMs does not adequately capture the cooling processes modeled in hydrodynamical simulations. The TNG SAM’s revised cooling model, calibrated to the cooling times in TNG, results in improved predictions of gas accretion onto galaxies and the mass of the hot halo. This highlights the need for SAMs to move beyond the traditional ``cold mode” vs. ``hot mode” dichotomy and adopt models that better reflect the full range of gas cooling timescales seen in hydrodynamical simulations (Section \ref{subsubsec:agree_tcool}, Figure \ref{fig:cold_v_hot_mode}).
    
    \item \textit{Directly calibrating the cooling time and star-formation efficiency yields more accurate global ISM predictions.} In the TNG SAM, the star formation rate is regulated by a star–formation efficiency calibrated from the full TNG100 sample, while gas condensation into the ISM is governed by cooling times matched to the TNG subsample. These two calibrations work jointly: realistic cooling times deliver the correct amount of gas to the ISM, and the calibrated star–formation efficiency regulates how that gas forms stars, producing stellar masses, cold gas masses, and star formation rates to mostly within $\sim$30\% of TNG's predictions across time. (Section \ref{subsubsec:agree_sf}, Figure \ref{fig:tngsam_variations} panels \textit{ii.–iv.}) Attempts to apply a global Kennicutt–Schmidt law resulted in large discrepancies because the fiducial SAM does not compute disk sizes in a manner comparable to TNG. In the absence of a reliable disk-size prescription for SAMs, calibration against the star formation efficiency offers a more stable route to reproducing TNG’s global ISM scaling relations.

    \item \textit{Modeling outflows at the galaxy- and halo-scale is essential for accurately predicting the gas content in galaxies.} While galaxy-scale outflows are a well-known driver of galaxy evolution, the TNG SAM highlights that halo-scale outflows are equally critical for predicting the distribution of baryons. Incorporating a two-step outflow model—where gas is first ejected from the ISM into the CGM, and then from the CGM into the IGM—provides a more realistic depiction of baryon cycling than traditional SAMs, which typically eject gas directly into an external reservoir, and improves the accuracy of stellar, cold gas, and hot gas mass predictions. (Section \ref{subsubsec:agree_outflows}, Figures \ref{fig:tngsam_variations} panels \textit{i- iv.} and \ref{fig:tngsam_variations_mdot} panels \textit{i. \& iv.}).
        
    \item \textit{Accurately modeling metal-enriched and metal-depleted flows between the ISM and CGM is essential for reproducing galaxy- and halo-scale metallicities.} The TNG SAM’s use of metallicity-weighted enrichment factors that capture how metal-enhanced or metal-depleted gas flows are relative to their source reservoirs significantly improves its ability to match TNG's predictions for the metallicity of the stars, cold gas and hot gas. Explicitly modeling these metal enrichment factors is therefore critical for accurately tracking the redistribution of metals between the ISM, CGM, and IGM. (Section \ref{subsubsec:agree_zetas}, Figures \ref{fig:tngsam_variations} panels \textit{v - vii.} and \ref{fig:tngsam_variations_mdot} panels \textit{v-viii.}).

\end{enumerate}

\begin{acknowledgments}
We thank the referee for their thoughtful and constructive feedback. Their contributions significantly improved the quality of the manuscript. OO acknowledges early support in this project from the National Science Foundation Graduate Research Fellowship and ALMA Student Observing Support program. The Flatiron Institute is supported by the Simons Foundation.

\software{
Astropy \citep{robitaille_astropy_2013, collaboration_astropy_2018, collaboration_astropy_2022},
IPython \citep{perez_ipython_2007},
Matplotlib \citep{caswell_matplotlibmatplotlib_2022},
NumPy \citep{van_der_walt_numpy_2011},
SciPy \citep{virtanen_scipy_2020}
}

\textit{Data Availability:} The data underlying this article will be shared on reasonable request to the corresponding author.

\end{acknowledgments}

\begin{contribution}

All authors contributed equally to the manuscript.

\end{contribution}

\appendix

\section{Building Blocks of the TNG SAM}
\label{app:tngsam_creation}

\subsection{Comparison to the published SC SAM}
\label{subapp:scsam_mod}

\begin{figure}
    \includegraphics[width=0.4\linewidth]{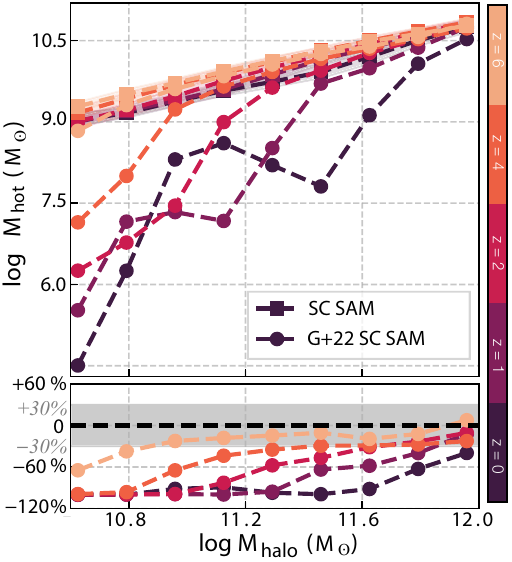}
    \centering
    \caption{Comparison of the Santa Cruz SAM used in this work (dashed squares) to the Santa Cruz SAM presented in \citealt{gabrielpillai_galaxy_2022} (dashed circles) for \mhot. The original SC SAM's cooling model depleted the CGM by up to $\sim120\%$ relative to the updated model presented here. While the revised cooling prescription produces a substantially less depleted hot halo, the updated SC SAM still underpredicts \mhot\ relative to TNG100 by up to $\sim80\%$ (see Figure \ref{fig:scsam_halo}).}
    \label{fig:new_csam_v_asam_halo}
\end{figure}

\begin{deluxetable*}{ccccc}
\tabletypesize{\scriptsize}
\tablecaption{Correspondence between the parameters studied in this paper, their fields in the SAM, and their corresponding fields in the IllustrisTNG simulations. Columns (1)–(5) list, respectively, the physical scale considered (halo or galaxy), the parameter name used throughout this paper, the corresponding SAM field, the matching IllustrisTNG field, and a brief description of the parameter.}
\label{tab:tng_sam_def}
\tablewidth{0pt}
\tablehead{
\colhead{Scale} &
\colhead{Parameter} &
\colhead{SAM Label} &
\colhead{TNG Label} &
\colhead{Description} \\
\colhead{(1)} &
\colhead{(2)} &
\colhead{(3)} &
\colhead{(4)} &
\colhead{(5)} 
}
\startdata
\hline
\textit{Halo} & \mhalo & \texttt{GalpropMvir} & \texttt{Group\_M\_TopHat\_200['SubhaloGrNr']} & Total virial mass \\
 & $M_{\text{DM}}$ & \texttt{GalpropMvir} $(1 - f_{\text{baryon}})$ & \texttt{SubhaloMassType Part 1 (DM)} & Total dark matter mass \\
 & $M_{\text{CGM}}$ & \texttt{HalopropMhot} & \texttt{GroupMassType part 0 ['SubhaloGrNr']}  & Gas mass of the circumgalactic halo \\
  &  &  & $- \sum\limits_{i}^{n}$ \texttt{SubhaloMassInRadType  part 0}  &  \\
\hline
\textit{Galaxy} & $M_*$ & \texttt{GalpropMstar} & \texttt{SubhaloMassType part 4} & Stellar mass \\
 & $M_{\text{cold}}$ & \texttt{GalpropMcold} & \texttt{SubhaloMassInRadType part 0} & Gas mass of the interstellar medium \\
 & SFR & \texttt{GalpropSfr} & \texttt{SubhaloSFR} & Star formation rate \\
 & $Z_*$ & \texttt{GalpropZstar} & \texttt{SubhaloStarMetallicity} & Stellar metallicity \\
 & $Z_{\text{cold}}$ & \texttt{GalpropZcold} & \texttt{SubhaloGasMetallicity} & Cold gas metallicity \\
 & $Z_{\text{hot}}$ & \texttt{HalopropZhot} & \texttt{GroupGasMetallicity['SubhaloGrNr']} & Halo gas metallicity \\ 
 &  &  & $- \Big(\sum\limits_{i}^{n}$ \texttt{SubhaloGasMetallicity} $\cdot$  &  \\ 
  &  &  & \texttt{SubhaloMassInRadType part 0} \Big) &  \\
\enddata
\end{deluxetable*}

\cite{pandya_first_2020} showed that the SC SAM substantially underpredicts the mass of the hot gas relative to the FIRE simulations,  and \citet{gabrielpillai_galaxy_2022} (G+22) showed that the same discrepancy appears when the SC SAM is compared with TNG. These results motivated the revision to the SC SAM's cooling model described in Section \ref{subsubsec:scsam_modifications}, which is implemented in the SC SAM used throughout this work. As shown in Figure \ref{fig:new_csam_v_asam_halo}, the updated cooling model proposed by \cite{pandya_first_2020} leads to a significant improvement in \mhot\  relative to the original prescription used in G+22. Despite this major improvement at the halo-scale, the impact on galaxy-scale gas properties and overall metal content remains modest, with deviations mostly within 0.5 dex relative to the G+22 SC SAM.

\subsection{Analytic Scaling relations}
\label{subapp:analytic_functions}

Table \ref{tab:tng_sam_def} lists all galaxy- and halo-scale properties compared between the SAM and TNG. For each property, the corresponding field names in the SAM and TNG outputs are provided.

We calibrated the SAM to TNG using analytic scaling relations as a function of halo mass and redshift. To flexibly capture the mass and redshift evolution of the physical processes governing gas cooling, star formation, stellar feedback, and metal enrichment in TNG, we adopt the three following functional forms:

\begin{align}
f(M_{\rm halo}, z, K, \alpha_0, \alpha_z, A, \beta, \delta) = & \nonumber \\
K \cdot \Big[ A \cdot (1 + z)^{\beta}
\cdot \left(\frac{M_{\rm halo}}{10^{12}\,M_\odot}\right)^{\alpha_0 + \alpha_z (1+z)^2}
+ \delta\Big] &
\label{eqn:tuned_function_A}
\end{align}

\begin{align}
\log(f(M_{\rm halo}, z, K, b, c, d, \alpha_0, \alpha_z)) = & \nonumber \\ -K \cdot \left(\arctan \left[b \cdot (\log M_{\rm halo} - c)\right] + d \right)
\cdot \left[\alpha_0 + e^{\alpha_z} (1 + z)\right] &
\label{eqn:tuned_function_B}
\end{align}

\begin{align}
\log(f(M_{\rm halo}, z, K, b, c, d, \alpha_0, \alpha_z)) = & \nonumber \\
-K \cdot \left(\arctan \left[b \cdot (\log M_{\rm halo} - c)\right] + d \right)
\cdot \left[\alpha_0 (1 + z)^{\alpha_z}\right] &
\label{eqn:tuned_function_C}
\end{align}

All tuned parameters, the functional form they correspond to (Equations \ref{eqn:tuned_function_A} - \ref{eqn:tuned_function_C}), and their best-fit coefficients are listed in Table \ref{tab:app_tuned_params}. 

{
\begin{table*}
\renewcommand{\arraystretch}{1.2}
 \begin{tabular}{c|c|c|c|c}
 \hline
 \hline
Model Update & Parameter & Equation & Coefficients & Comment \\
(1) & (2) & (3) & (4) & (5) \\
  \hline
  \hline
Gas Cooling & \fin & \ref{eqn:tuned_function_A} & 1.0, 0.0, 0.0, 1033.2389, -0.0004, -1031.1689 & $z \le 1.8$\\ 
& & \ref{eqn:tuned_function_A} & 1.5, 0.0, 0.0, 1033.2389, -0.0004, -1031.1689 & $1.8 < z \le 3.5$\\ 
& & \ref{eqn:tuned_function_A} & 1.0, 0.0, 0.0, 1033.2389, -0.0004, -1031.1689 & $3.5 < z \le 4.5$\\ 
& & \ref{eqn:tuned_function_A} & 0.9, 0.0, 0.0, 1033.2389, -0.0004, -1031.1689 & $4.5 < z \le 5.5$\\ 
& & \ref{eqn:tuned_function_A} & 1.6, 0.0, 0.0, 1033.2389, -0.0004, -1031.1689 & $5.5 < z \le 6.5$\\ 
& & \ref{eqn:tuned_function_A} & 1.0, 0.0, 0.0, 1033.2389, -0.0004, -1031.1689 & $z > 6.5$\\ 
  \hline
  
  \nodata & $t_{\rm cool}$ & \ref{eqn:tuned_function_C} & 1.05, -234.56, 10.578, 14.995, -0.052522, -0.82487 & $z \le 1.55$\\ 
& & \ref{eqn:tuned_function_B} & 1.26, 0.52727, 9.9922, 1.1905, -0.33785, -2.3922 & $1.55 < z \le 2.55$\\ 
& & \ref{eqn:tuned_function_B} & 0.84, 0.52727, 9.9922, 1.1905, -0.33785, -2.3922 & $z > 2.55$\\

  \hline

  Star Formation & $\tau_*$ & \ref{eqn:tuned_function_C} & 1.0, 0.70584, 10.879, -8.1082, -1.2408, -0.049007 & $z \le 1.5$\\ 
& & \ref{eqn:tuned_function_C} & 1.0, 3.0988, 16.725, 1.3864, -80.13, -0.077247 & $1.5 < z \le 3.5$\\ 
& & \ref{eqn:tuned_function_C} & 1.2, 11.288, 14.382, 1.4563, -120.65, -0.095178 & $3.5 < z \le 5.5$\\ 
& & \ref{eqn:tuned_function_C} & 1.0, 1.1836, 11.565, -6.6394, -1.5893, -0.13023 & $z > 5.5$\\ 

  \hline

  Stellar Feedback & $\eta_{\rm launch}/\eta_{ISM}$ &  \ref{eqn:tuned_function_C} & 0.85, -2.1682, 11.277, 3.2989, -0.36799, -0.10655 & $z \le 2.05$\\ 
& & \ref{eqn:tuned_function_C} & 0.9, -2.5067, 11.6902, 1.3298, -0.7013, -0.2257 & $2.05 < z < 5.5$\\ 
& & \ref{eqn:tuned_function_C} & 0.85, -4.4457, 11.477, 2.0363, -0.56883, -0.24664 & $z > 5.5$\\ 

\hline

  \nodata & $\eta_{\rm halo}$ & \ref{eqn:tuned_function_B} & 0.65, 2.4603, 10.505, -1.9555, 0.85904, -20.155 & $z \le 1.4$\\ 
& & \ref{eqn:tuned_function_C} & 0.688, -0.90685, 10.768, 1.015, -1.6071, -0.23343 & $1.4 < z \le 3.05$\\ 
& & \ref{eqn:tuned_function_C} & 0.7, -2.3859, 11.113, 0.97005, -1.4023, -0.38535 & $3.05 < z \le 5.05$\\ 
& & \ref{eqn:tuned_function_C} & 0.6, -2.0278, 11.203, 0.52378, -0.2216, 0.70667 & $5.05 < z \le 6.5$\\ 
& & \ref{eqn:tuned_function_C} & 0.7, -1.9895, 11.128, 0.0737, -4.4154, -0.56573 & $z > 6.5$\\ 

  \hline

  Metal Circulation & $\zeta_{\rm halo}^{\rm in}$ & \ref{eqn:tuned_function_C} & 1.0, 1.4293, 10.022, -1.7885, -1.8285, 0.10172 & $z < 1.1$\\ 
& & \ref{eqn:tuned_function_B} & 1.0, -1.1446, 9.2119, 1.7167, 2.6688, -2.2636 & $1.1 \le z \le 3.5)$\\ 
& & \ref{eqn:tuned_function_C} & 0.8, -1.5531, 10.7872, 4.2776, 0.421, 0.1029 & $3.5 < z \le 5.5$\\ 
& & \ref{eqn:tuned_function_C} & 1.0, -48.756, 9.7789, 1.8536, 5.8717, 0.064435 & $z > 5.5$\\ 
     
\hline
  
  \nodata & $\zeta_{\rm halo}^{\rm out}$ & \ref{eqn:tuned_function_C} & 1.0, -10.584, 11.184, 1.4135, 0.057361, -4.1385 & $z < 0.75$\\ 
& & \ref{eqn:tuned_function_C} & 1.2, 1.0759, 11.482, -0.3432, 0.082956, 0.69927 & $ 0.75 \le z \le 3.5$\\ 
& & \ref{eqn:tuned_function_C} & 1.2, -3.67, 11.375, 0.029179, -0.018041, 1.4446 & $3.5 < z \le 5.1$\\ 
& & \ref{eqn:tuned_function_C} & 0.8, -3.67, 11.375, 0.029179, -0.018041, 1.4446 & $5.1 < z \le 6.5$\\ 
& & \ref{eqn:tuned_function_C} & 0.3, -3.67, 11.375, 0.029179, -0.018041, 1.4446 & $z > 6.5$\\ 

\hline
  
  \nodata & $\zeta_{\rm ISM}^{\rm in}$ & \ref{eqn:tuned_function_C} & 1.0, -25.462, 10.554, 2.6799, -0.11964, 0.27383 & $z < 1.6$\\ 
& & \ref{eqn:tuned_function_B} & 0.95, 2.1292, 11.067, -1.9443, 0.15012, -4.3082 & $1.6 \le z < 2.5$\\ 
& & \ref{eqn:tuned_function_B} & 1.0, 2.1292, 11.067, -1.9443, 0.15012, -4.3082 & $2.5 \le z < 5.2$\\ 
& & \ref{eqn:tuned_function_B} & 1.0, 2.257, 11.213, -1.8659, 0.21026, -5.6133 & $z \ge 5.2$\\ 

   \hline
  
  \nodata & $\zeta_{\rm ISM}^{\rm out}$ & \ref{eqn:tuned_function_C} & 1.0, 4.0357, 10.309, 0.34813, 0.23049, 0.063895 & $z < 1.1$\\ 
& & \ref{eqn:tuned_function_C} & 1.0, 7.3556, 6.9872, -1.5214, 21.999, -0.12318 & $z \ge 1.1$\\
 \end{tabular}

  \caption{Parameters calibrated to emulate TNG and the coefficients used in Equations \ref{eqn:tuned_function_A}-\ref{eqn:tuned_function_C}. The columns list: (1) the model component updated, (2) the parameter calibrated, (3) the functional form adopted, (4) the best-fit coefficients, and (5) the domain over which the fit is applied.  } 
 \label{tab:app_tuned_params}
\end{table*}
}

\newpage 

\begin{figure*}
    \includegraphics[width=\linewidth]{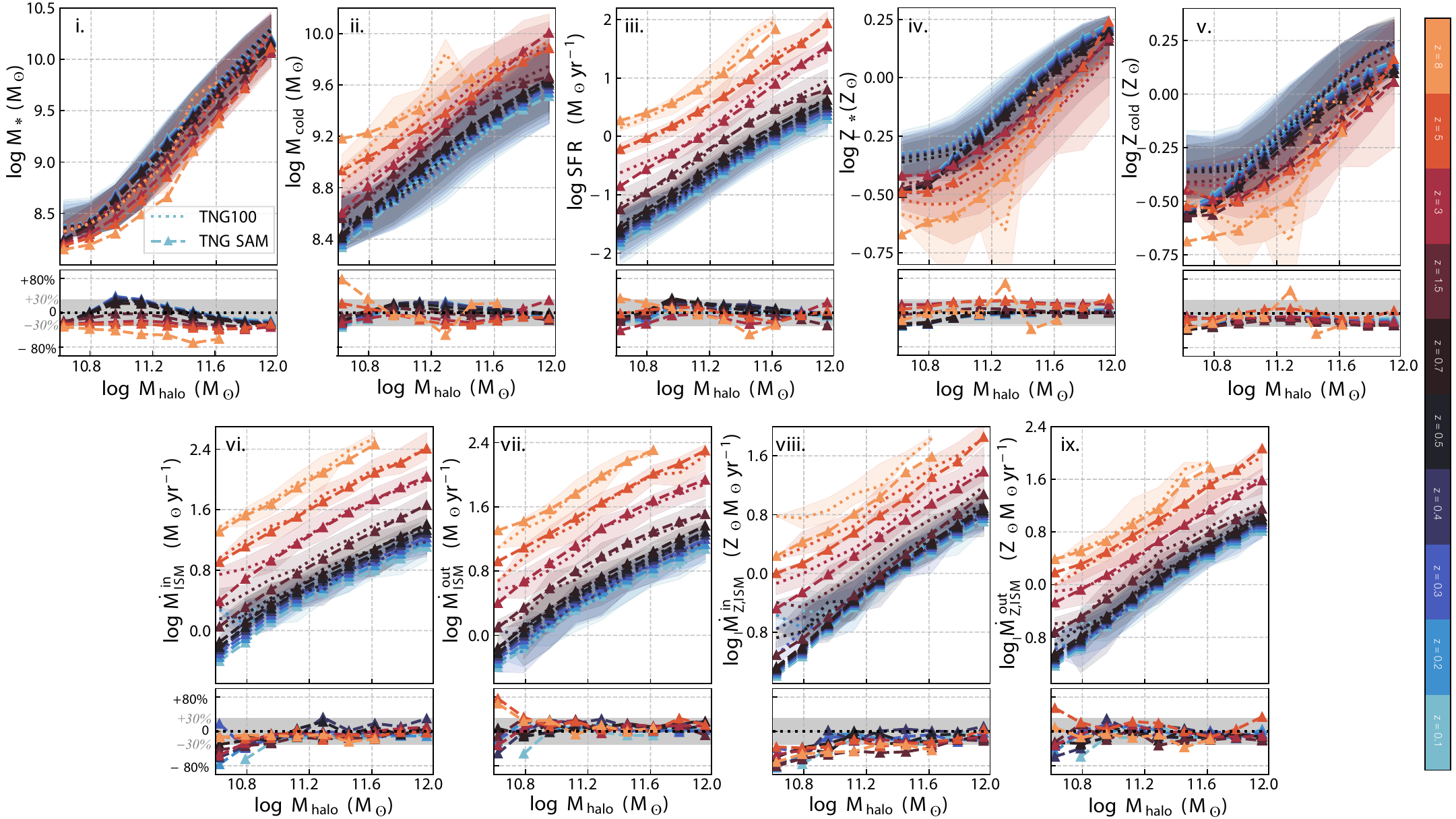}
    \caption{Extension of Figure \ref{fig:tngsam_gal} showing the evolution of galaxy-scale properties (panels \textit{i–iv.}) and baryon flow rates (panels \textit{v–ix.}) for the redshifts not displayed in Figure \ref{fig:tngsam_gal}. Low-redshift relations are mostly shown in blues $(z<0.5)$, high-redshift relations in reds $(1.5\leq z \leq 8)$, and intermediate redshifts in darker colors $(0.5\leq z \leq 0.7)$. Across most halo masses and redshifts, the TNG SAM continues to reproduce TNG100’s global galaxy properties and flow rates to within $\sim30\%$, with larger deviations appearing primarily at the highest redshift ($z=8$) where the number of well-resolved systems is limited.} 
    \label{fig:tngsam_gal_morez}
\end{figure*}

\begin{figure*}
    \includegraphics[width=\linewidth]{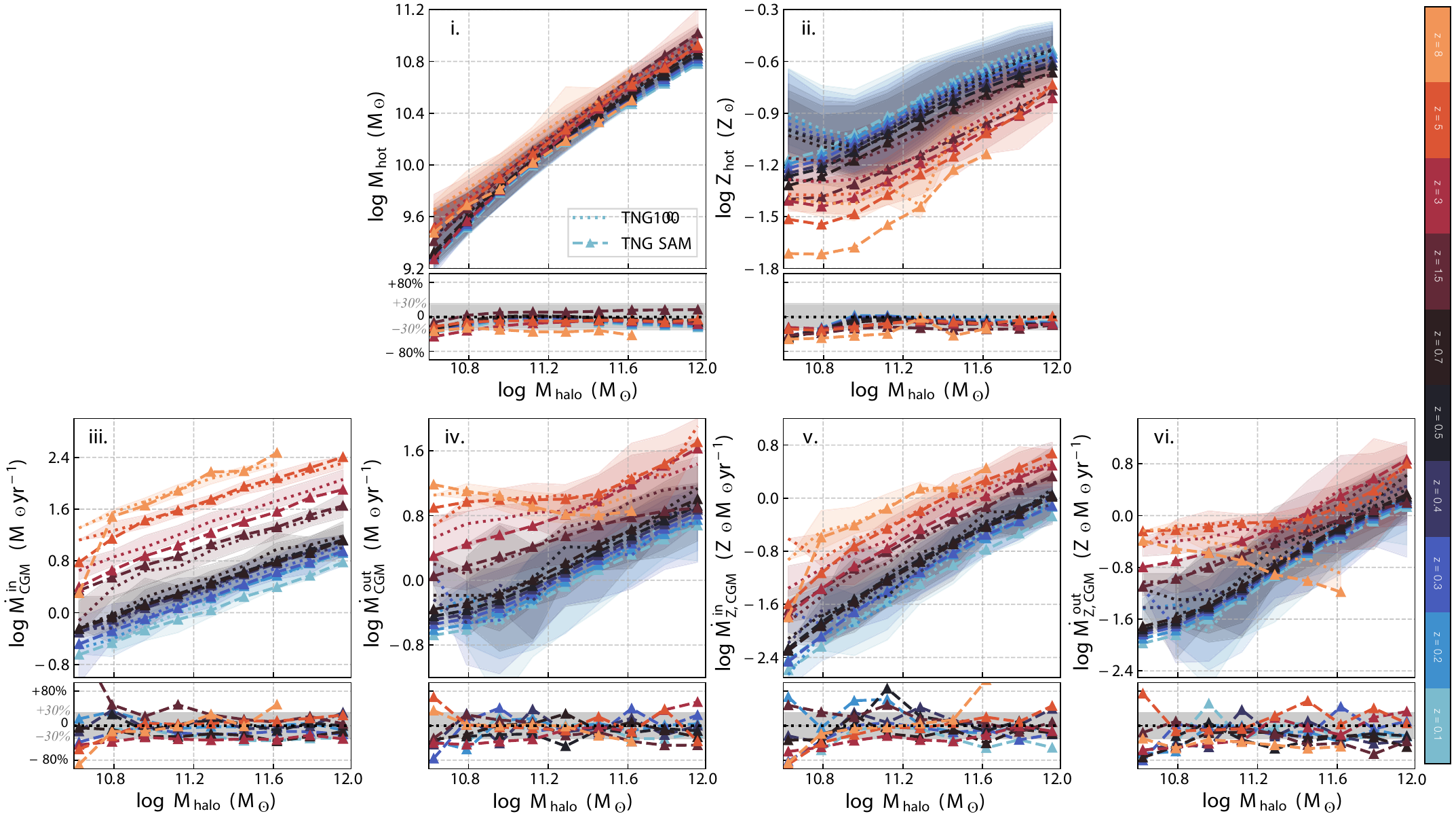}
    \caption{Extension of Figure \ref{fig:tngsam_halo} showing halo-scale quantities (panels \textit{i–ii.}) and the corresponding gas and metal flow rates (panels \textit{iii–vi.}) for the redshifts not displayed in Figure \ref{fig:tngsam_halo}. Median relations are shown using the same redshift color-coding as in Figure \ref{fig:tngsam_gal_morez}. The TNG SAM maintains agreement with TNG100 at the $\sim30\%$ level for most halo masses and redshifts, including for halo-scale gas and metal inflows and outflows, with discrepancies increasing toward the higher redshift.}
    \label{fig:tngsam_halo_morez}
\end{figure*}

\begin{figure*}
    \includegraphics[width=\linewidth]{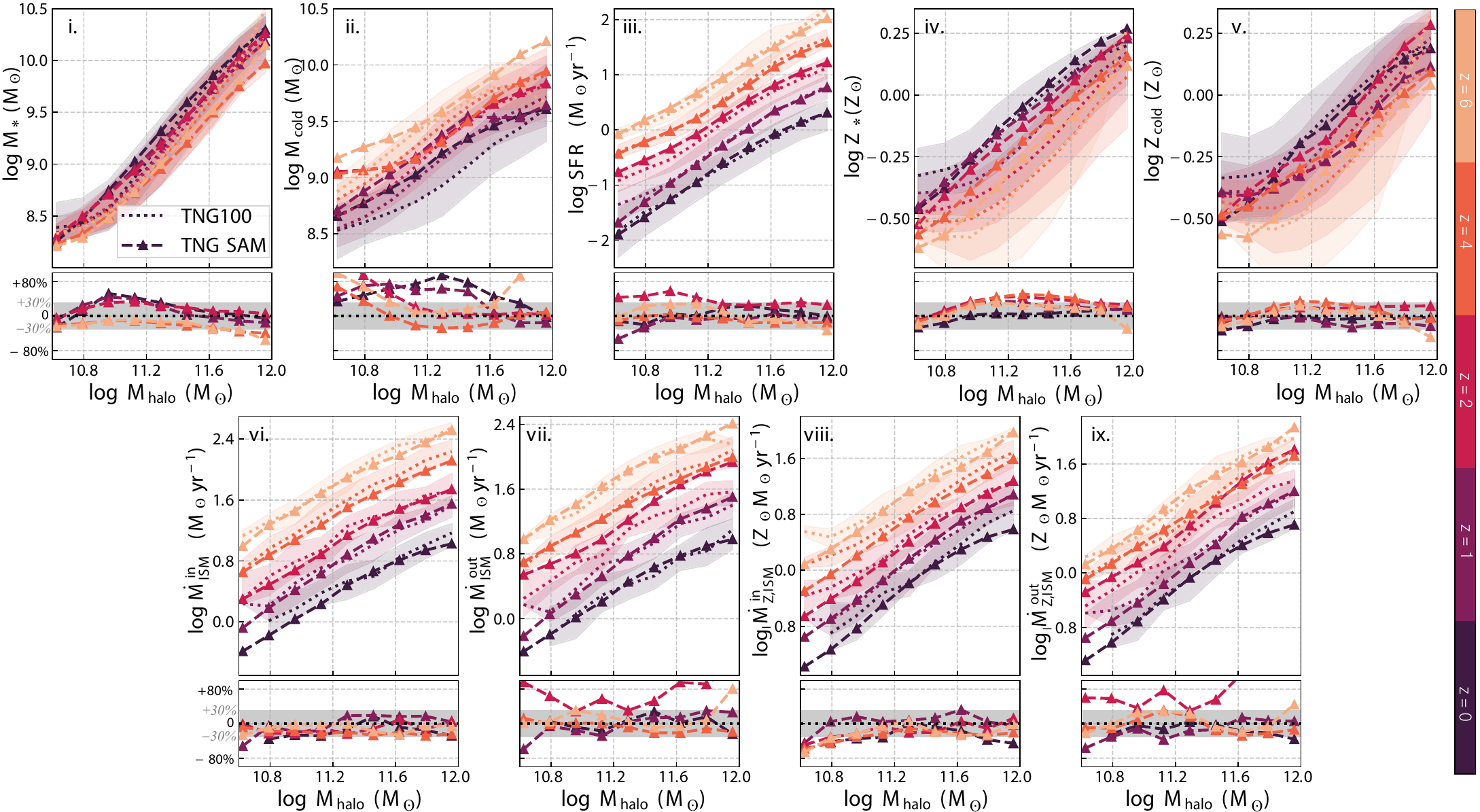}
    \caption{Galaxy-scale properties and baryon flow rates for the TNG SAM calibrated using only $\sim100$ randomly selected galaxies per redshift. The figure follows the same layout, quantities, and redshift color-coding as Figure~\ref{fig:tngsam_gal}, but for the reduced calibration sample. Despite the limited number of galaxies used for calibration, the TNG SAM continues to reproduce aggregate galaxy-scale properties to mostly within $\sim40\%$ accuracy across redshift. The cold gas mass shows the weakest agreement, with deviations reaching up to $\sim80\%$ in some regimes.}
    \label{fig:app_tngsam_100gal}
\end{figure*}

\begin{figure*}
    \includegraphics[width=\linewidth]{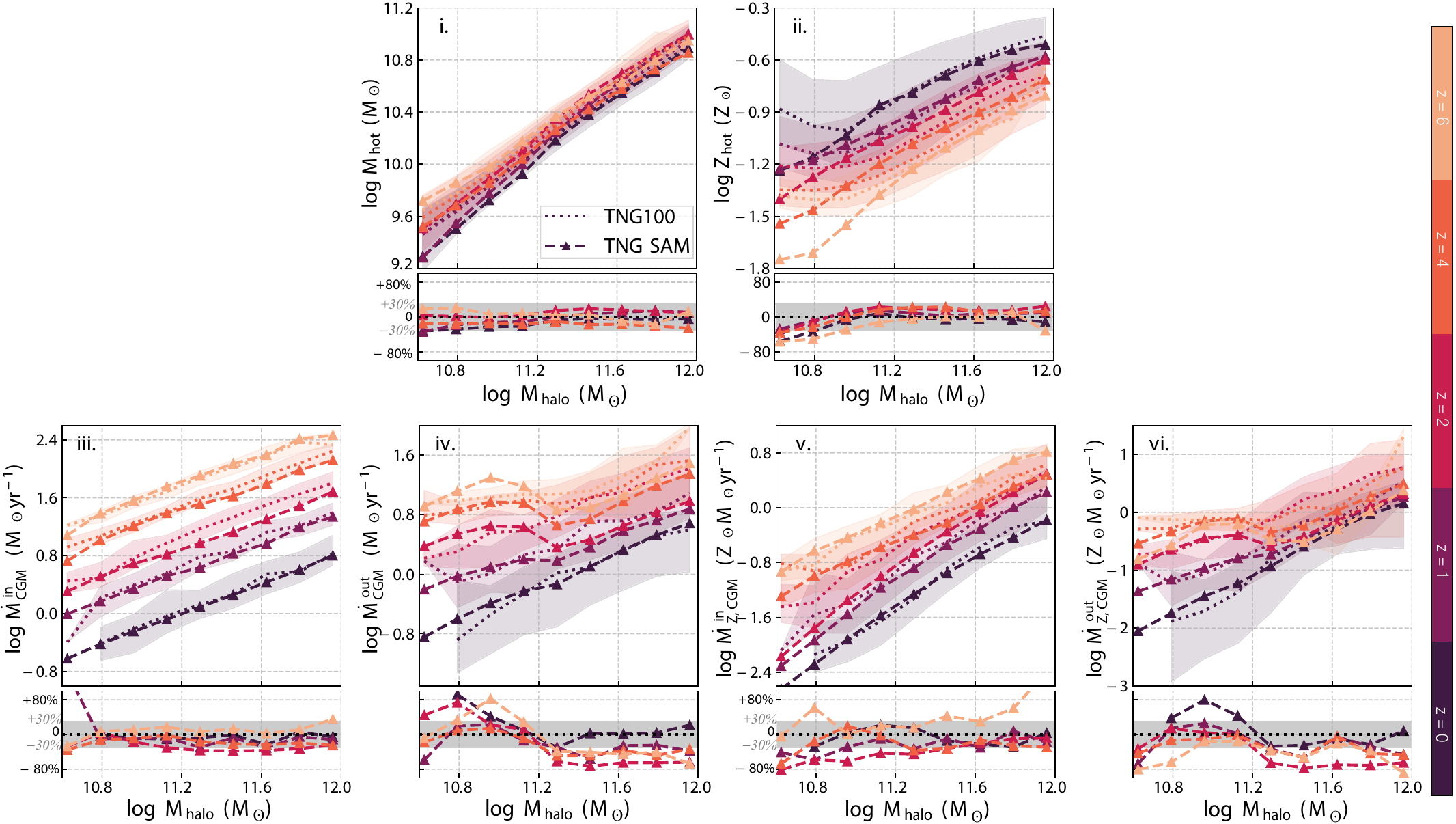}
    \caption{Halo-scale quantities and baryon flow rates for the TNG SAM calibrated using only $\sim100$ randomly selected galaxies per redshift, shown using the same layout and redshift color scheme as Figure~\ref{fig:tngsam_halo}. While aggregate halo properties remain mostly consistent with TNG100 within $\sim40\%$, halo-scale flow rates are more sensitive to the reduced calibration set. The largest discrepancies occur for gas and metal outflows from the halo, which can differ by up to $\sim80\%$.}
    \label{fig:app_tngsam100_halo}
\end{figure*}

\subsection{TNG SAM Results at Intermediate Redshifts}
\label{subapp:tngsam_redshift}

Figures \ref{fig:tngsam_gal} and \ref{fig:tngsam_halo} show the results of the TNG SAM at select redshifts to avoid visual clutter. For completeness, Figures \ref{fig:tngsam_gal_morez} and \ref{fig:tngsam_halo_morez} show the remaining intermediate redshifts. Across the full redshift range explored, the TNG SAM generally reproduces TNG’s predictions to within $\pm30\%$. The agreement degrades primarily at $z=8$, where the number of well-resolved galaxies in TNG is substantially lower ($\lesssim 10$).

\newpage

\subsection{How Few Galaxies Are Enough? Calibration with 100 Galaxies}
\label{subapp:100gals}

To evaluate how sensitive the TNG SAM’s performance is to the size of the calibration sample, we constructed a version of the model using only $\sim 100$ randomly selected galaxies per redshift, spanning the halo mass range $10^{10} - 10^{12}\ M_{\odot}$. Despite the substantially reduced calibration set, the TNG SAM matches TNG's predictions at the galaxy-scale mostly within $\sim40\%$ across redshift. Figures \ref{fig:app_tngsam_100gal} and \ref{fig:app_tngsam100_halo} show that aggregate quantities such as stellar mass, star formation rate, and halo-scale gas content generally remain within this range, although the cold gas mass shows the weakest agreement, with deviations reaching up to $\sim80\%$ in some regimes. 

Figures \ref{fig:app_tngsam_100gal} and \ref{fig:app_tngsam100_halo} also show that the baryon flow rates are more sensitive to the reduced calibration set. Gas and metal inflow and outflow rates are typically recovered to within $\sim40\%$, while the largest discrepancies—up to $\sim80\%$—occur for gas and metals leaving the halo. Nevertheless, the overall redshift and halo-mass evolution of these flows is captured reasonably well.

\newpage

\bibliography{tng_bib_filtered.bib}
\bibliographystyle{aasjournal}

\end{document}